\documentclass[prb,twocolumn,showpacs,amsmath,amssymb,superscriptaddress,letterpaper]{revtex4}
\usepackage{multirow}
\usepackage{bigstrut}
\usepackage{amssymb}
\usepackage{amsmath}
\usepackage{graphicx}
\usepackage{dcolumn}
\usepackage{bm}
\usepackage{CJK}
\usepackage{relsize}
\usepackage{cancel}
\usepackage{color}

\begin{document}
\title{Analysis of phonon-induced spin relaxation processes in silicon}

\author{Yang Song ËÎÑï}\email{yangsong@pas.rochester.edu}\affiliation{Department of Physics and Astronomy,
University of Rochester, Rochester, New York, 14627}
\author{Hanan Dery}\affiliation{Department of Physics and Astronomy,
University of Rochester, Rochester, New York, 14627}\affiliation{Department of Electrical and Computer Engineering, University of Rochester, Rochester, New York, 14627}

\begin{abstract}
We study all of the leading-order contributions to spin relaxation of \textit{conduction} electrons in silicon due to the electron-phonon interaction. Using group theory, $k\cdot p$ perturbation method and rigid-ion model, we derive an extensive set of matrix element expressions for all of the important spin-flip transitions in the multi-valley conduction band. The scattering angle has an explicit dependence on the electron wavevectors, phonon polarization, valley position and spin orientation of the electron. Comparison of the derived analytical expressions with results of empirical pseudopotential and adiabatic band charge models shows excellent agreement.
\end{abstract}
\maketitle

\section{Introduction}

Silicon is an ideal material choice for semiconductor spin-based devices.\cite{Zutic_RMP04,Dery_Nature07,Behin_NatureNano10} It has a  relatively weak spin-orbit coupling which leads to a negligible probability of flipping the electron's spin during a scattering event. Furthermore, the Dyakonov-Perel spin relaxation mechanism is absent in bulk silicon due to its inversion symmetry (i.e., no intrinsic magnetic field around which the electron spin precesses).\cite{Dyakonov_JETP73} Finally, the zero nuclear spin of its naturally abundant isotope suppresses spin relaxation by hyperfine interactions.\cite{Jelezko_PRL04,Fodor_JPCM06,Tyryshkin_NatureMater12} These characteristics have motivated a wide interest in silicon spintronics.\cite{Zutic_PRL06,Appelbaum_Nature07,Jonker_NaturePhysics07,Mavropoulos_PRB08,Dash_Nature09,Sasaki_APE11,Ando_APL11b,Dery_APL11,Jansen_NatureMater12}

\begin{figure}[t]
\includegraphics[width=8cm]{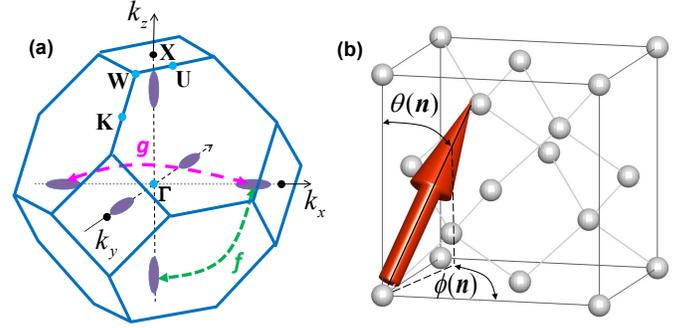} 
\caption{(Color online) (a) Valley positions and high symmetry points in the Brillouin zone. Valley centers are about $0.15\times 2\pi/a$ from the $X$ points where $a$ is the lattice constant. Also marked are representative intervalley transitions by $g$ and $f$ processes. (b) The spin orientation ($\mathbf{n}$) with respect to the crystallographic axes.} \label{fig:brillouin}
\end{figure}

In cases where spin dephasing by precession is weak, the spin relaxation rate is set by spin-flip scattering. For conduction electrons in crystals, this scattering is described by a matrix element of the form $\langle \mathbf{k}_2, \Downarrow_{\mathbf{n}} | H_{\rm{sf}} | \mathbf{k}_1 , \Uparrow_{\mathbf{n}} \rangle$  where the states are identified with wavevector $\mathbf{k}$ due to the translational symmetry. For a given spin-flip mechanism ($H_{\rm{sf}}$), this matrix element depends on the initial and final state wavevectors $\mathbf{k}_1$ and $\mathbf{k}_2$, as well as on the spin orientation $\mathbf{n}$. In silicon, conduction electrons reside in six valleys near the edges of the Brillouin zone as shown in Fig.~\ref{fig:brillouin}(a). The valley centers are positioned on the $\Delta$-axis connecting the $\Gamma$ and $X$ points. In multivalley  semiconductors, the electron remains in the same valley after an intravalley scattering and it switches valleys after an intervalley scattering. Figure~\ref{fig:brillouin}(a) shows examples of intervalley scattering in Si, where $g$ and $f$ processes refer to electron scattering between opposite valleys and between valleys that reside on perpendicular crystal axes, respectively. When dealing with spin-flip scattering, the spin orientation ($\mathbf{n}$) further breaks the scattering symmetry. The resulting anisotropy in spin relaxation depends on the projections of $\mathbf{n}$ on crystallographic axes [Fig.~\ref{fig:brillouin}(b)]. 

Spin-flip mechanisms can be classified into Yafet and Elliott processes. The former involves spin-dependent interaction whereas the states are viewed as pure spin states.\cite{Yafet_SSP63} Examples of spin-dependent interactions include the spin-orbit coupling of the host crystal via electron-phonon interaction ($H_{\rm{sf}} \sim \delta\mathbf{R} \cdot \bm\nabla V_{\rm{so}}$), spin-orbit coupling of defects via electron-impurity scattering ($H_{\rm{sf}} \sim \bm\nabla V_{\rm{imp}}\times \mathbf{p} \cdot \sigma$), and electron-nuclear hyperfine interaction ($H_{\rm{sf}} \sim \lambda \mathbf{s}\cdot\mathbf{I}$). The \textit{Elliott} processes are governed by spin mixing in the electron states due to the crystal spin-orbit coupling, whereas the interaction is spin independent (e.g., $\delta\mathbf{R} \cdot \bm\nabla V$).\cite{Elliott_PR54a} In this paper the focus is on spin relaxation due to the electron-phonon interaction where the interplay between Elliott and Yafet processes plays a key role in setting the intrinsic spin lifetime. In Sec.~\ref{sec:phases} we provide a general overview of the spin relaxation mechanisms in $n$-type silicon and we show that the electron-phonon interaction dominates the spin relaxation of conduction electrons over a wide range of temperature and doping conditions.


The spin relaxation due to electron-phonon interaction in silicon was studied using different theoretical approaches. The seminal works of Overhauser, Elliott and Yafet enabled an approximate quantitative connection between the spin and momentum relaxation times via the shift of the $g$-factor.\cite{Overhauser_PR53,Elliott_PR54a,Yafet_SSP63} Yafet also derived a general form of the \textit{intravalley} scattering by \textit{acoustic} phonon modes.\cite{Yafet_SSP63} \textit{Intervalley} scattering  was included in three recent works.\cite{Cheng_PRL10,Pengke_PRL11,Tang_PRB12} Numerical analysis of the measured spin relaxation times above 50~K in intrinsic silicon was carried out by Cheng, Wu and Fabian.\cite{Cheng_PRL10} The analysis included calculation of electron energies and states using an empirical pseudopotential model (EPM),\cite{Chelikowsky_PRB76} of phonon dispersion and polarization vectors using an adiabatic bond charge model (ABCM),\cite{Weber_PRB77} and of a rigid-ion model approximation to describe the electron-phonon interaction.\cite{Allen_PRB81} An analytical approach was then developed by Li and Dery using a compact spin-dependent $\mathbf{k}\cdot\mathbf{p}$ Hamiltonian model around the zone edge $X$ point.\cite{Pengke_PRL11} This model was then used to derive the dominant spin-flip matrix elements where the spin-orbit coupling signature appeared only in the expansion of electronic states. The electron-phonon interaction, on the other hand, was mimicked by a phenomenological connection with deformation potential and scattering parameters. In another recent theory, Tang, Collins and Flatte have used a tight-binding model to calculate the spin relaxation in strained silicon and germanium.\cite{Tang_PRB12}

In this paper we present a comprehensive theory of phonon-induced spin-flip mechanisms of conduction electrons in bulk silicon. Matrix elements are derived using two different approaches based on whether they are wavevector dependent or not. Table~\ref{tab:symmetry_argument} presents the power-law dependence for spin-flip and spin-conserving matrix elements as well as the theoretical approaches to derive their forms. When transitions between valley-center states do not vanish [i.e., $M(0,\mathbf{s}_1\,;\,0,\mathbf{s}_2)\neq0$], group theory can be used alone to derive explicit forms of their matrix elements including the spin orientation dependence. As shown in the right column of the table, spin-conserving scattering is wavevector independent other than with long-wavelength acoustic phonons. On the other hand, spin-flip scattering is wavevector independent only for the $f$-process whereas other processes have higher power-law dependence than in spin-conserving scattering. Accordingly, the  $f$-process is dominant in spin relaxation whereas its weight is comparable to other processes in momentum relaxation. In the next three paragraphs we discuss key aspects of the present work including explanation of the approach to derive wavevector-dependent matrix elements.

\begin{table}
\caption{\label{tab:symmetry_argument}
Power-law dependence of leading order matrix elements for all types of electron-phonon scattering. $\mathbf{k}_{1,2}$ are the initial and final electron wavevectors and they are measured from their respective valley centers. $\mathbf{q}$ and $\mathbf{K}$ denote their difference and average, respectively. Intravalley scattering is divided to interaction with long-wavelength optical (OP) and acoustic (AC) phonons. Also mentioned are theoretical approaches to derive explicit forms of the matrix elements (beyond the power-law dependence). See text for further details.
}
\renewcommand{\arraystretch}{2.0}
\tabcolsep=0.07 cm
\begin{tabular}{p{0.62in}|p{1.22in}|p{1.22in}}%
 \hline\hline
            &   $M(\mathbf{k}_1,\mathbf{s}\,;\,\mathbf{k}_2,-\mathbf{s})$   &  $M(\mathbf{k}_1,\mathbf{s}\,;\,\mathbf{k}_2,\mathbf{s})$ \\
               \hline
$f$-process & $\mathbf{k}$-independent, $\quad$ $\qquad\quad$ double group. & $\mathbf{k}$-independent, $\quad$ $\qquad\quad$ single group.  \\
\hline
$g$-process &  linear in $\mathbf{K}$, $\qquad\qquad$ $\qquad\quad$ `$\mathbf{k}\!\cdot\!\mathbf{p}$'+single group. & $\mathbf{k}$-independent, $\quad$$\qquad$  single group.\\
 \hline
intravalley & AC: quadratic in $\mathbf{q}$, $\,\,\,$  OP: linear in $\mathbf{q}$, $\,\,\,\,\,\,\,\,\,\,\,\,$  `$\mathbf{k}\!\cdot\!\mathbf{p}$'+single group.   & AC: linear in $\mathbf{q}$, $\,\,\,\,\,$ OP: $\mathbf{k}$-independent, $\qquad$ `$\mathbf{k}\!\cdot\!\mathbf{p}$'+single group. \\
\hline\hline
\end{tabular}
\end{table}

For every type of electron-phonon scattering, we express the leading-order spin-flip matrix elements as functions of the electron wavevector ($\mathbf{k}$), phonon polarization ($\bm{\xi}$), valley position of the electron and its spin orientation ($\mathbf{n}$). For the important $f$-process we derive a complete set of selection rules by rendering double group representation matrices in conjunction with time reversal symmetry. This approach is more informative than the common practice of using the character table since it distinguishes between \textit{momentum} and \textit{spin} scattering processes, it unambiguously identifies parameters for both processes, and it is generalized for any spin orientation.  The latter is important in analyzing experimental measurements where the orientation of injected spins is dictated by the shape and magnetocrystalline anisotropy of ferromagnetic contacts or by the propagation and helicity of a circularly-polarized light beam.

In the analysis of intravalley and $g$-process spin flips we employ a different approach. Derivation of selection rules in these cases cannot be carried out solely by group theory because the spin-flip matrix elements are wavevector dependent. Instead, we utilize a combined approach that involves spin-dependent $\mathbf{k}\cdot\mathbf{p}$, rigid-ion and group theories. \textit{Ad~hoc} selection rules are derived for electron-phonon interaction between basis functions that appear in the $\mathbf{k}\!\cdot\!\mathbf{p}$ expansion of states around the valley center. The selection rules are derived from single group theory whenever multiple bands have to be included in the $\mathbf{k}$$\,$$\cdot$$\,$$\mathbf{p}$ expansion.  Double group bares no advantage in this case. This theoretical procedure allows us to include the crystal symmetry not only in the electronic \textit{states} but also in the electron-phonon \textit{interactions}. One outcome is that new scattering-angle symmetries are revealed. We derive appealing matrix element forms by employing an elastic continuum approximation for diamond crystal structures. A second outcome is that the out-of-phase motion of atoms in the primitive cell is quantified and shown to play a role in all types of intravalley spin-flip processes including scattering with acoustic phonons.

In this paper we thoroughly study the proximity effect between the lowest pair of conduction bands in silicon. In his seminal work, Yafet showed that the intravalley spin-flip matrix element due to scattering with acoustic phonon modes is proportional to the general form $C \Delta_{\rm{so}} q^2 / E_g^2$ where $C$ is a deformation potential constant, $E_g$ is the energy gap from bands where the spin-orbit coupling is considered (upper valence bands), and $\Delta_{\rm{so}}$ is the strength of the coupling.\cite{Yafet_SSP63} The quadratic dependence on the phonon wavevector ($q^2$) is a consequence of time reversal and space inversion symmetries. This quadratic dependence allowed Yafet to predict the $T^{-5/2}$ temperature dependence of the spin lifetime due to intravalley electron scattering with acoustic phonon modes. Yafet's general matrix element, however, misses two important features. First, the scalar form of this matrix element does not provide details of the dominant phonon modes and cannot capture the dependence on directions of $\mathbf{q}$ and $\mathbf{n}$. Second, the interband deformation potential between the lowest pair of conduction bands plays a crucial role in intravalley spin relaxation of silicon. As a result, $\Delta_{\rm{so}}/E_g^2$ should be replaced by $\Delta_{\rm{so}}/(E_g \Delta_C)$ where $\Delta_C$ denotes the band gap between the two conduction bands at the valley center.\cite{Pengke_PRL11} This interband coupling is a unique feature of the conduction band in \textit{silicon}. It results from the proximity of the valley center to the two-band degeneracy at the $X$ point. This coupling also explains the discrepancy between theoretical predictions of the spin lifetimes in strained silicon.\cite{Dery_APL11,Tang_PRB12} This discrepancy can be resolved by inclusion of $d$-orbitals in the tight-binding model to correctly capture the electronic behavior near the $X$ point.\cite{Jancu_PRB98}

\subsection*{Paper organization and outline of central results}

This paper is organized as follows. Section~\ref{sec:phases} surveys various spin relaxation processes in \textit{n}-type silicon. Section~\ref{sec:e-ph} provides a theoretical framework for the electron-phonon interaction. The dominant $f$-process is studied in Sec.~\ref{sec:f-process}. Using four scattering constants, Eq.~(\ref{eq:Sigma_Sn}) lists explicit forms of the $f$-process matrix elements as a function of the phonon symmetry and of the spin orientation.  In section~\ref{sec:intra_g} we present detailed derivations and extensive results of intravalley and $g$-process spin flips. This section begins with an introduction of the special features and tabulation of representative results (Table~\ref{tab:intra_g_M_Sn}). Selection rules and Hamiltonian forms are introduced in section~\ref{sec:kp hamilton}, as well as accurate state vectors up to quadratic-in-$\mathbf{k}$ terms.   Sections~\ref{sec:intravalley}-\ref{sec:valley-spin dependence} include derivations of the core spin-flip matrix element expressions with spin orientation dependence [Eqs.~(\ref{eq:ac_Sn})-(\ref{eq:g_Sn}) with their deformation potential constants defined in Eqs.~(\ref{eq:D_xy}), (\ref{eq:Dop}) and (\ref{eq:Dgs})]. Some insights about the relation between analytical derivation and EPM  are discussed in Sec.\ref{sec:EPM}. Spin lifetimes from integrating analytical matrix elements [Eqs.~(\ref{eq:tau_ac})-(\ref{eq:tau_f})] as well as different levels of numerical calculations are presented and compared in Section~\ref{sec:lifetime}.  Section~\ref{sec:conclusion} includes a summary of results and an outlook on future work. Appendices \ref{app:f}-\ref{app:details_intra_g} include technical details from different derivation phases.

\section{Spin Relaxation Processes} \label{sec:phases}

\begin{figure}
\includegraphics[width=8.5cm, height=7.5cm]{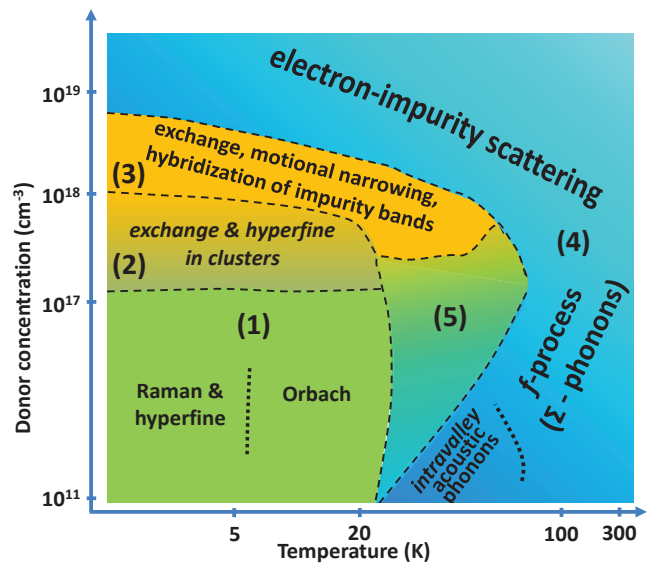}
\caption{(Color online) Diagram of dominant spin relaxation mechanisms in $n$-type silicon as a function of temperature and donor concentration. It is applicable at or near equilibrium conditions. In region 1 electrons are localized on isolated impurity sites. In regions 4(3) they populate the conduction (impurity) band. Region 2 is a precursor of the impurity band (donor clusters). Region 5 includes more than a single phase. Transitions between regions are generally gradual and colors as well as dashed border lines are added for ease of illustration. The findings in this paper are relevant for non-degenerate conditions ($<$10$^{17}$~cm$^{-3}$) in region 4.} \label{fig:phase}
\end{figure}

Spin lifetime of electrons in \textit{n}-type bulk silicon varies by more than 12 orders of magnitude with changing temperature and doping conditions.  \cite{Fletcher_PR54,Honig_PR54,Slichter_PR55,Feher_PR55,Honig_PR56,Feher_PR59,Castner_PR63,Lancaster_PPSL64,Jerome_PR64,Maekawa_JPSJ65,Kodera_JPSJ66,Lepine_PRB70,Ue_PRB71,Quirt_PRB72,Gershenzon_PSS72,Pifer_PRB75,Ochiai_PSS76,Paalanen_PRL86,Sachdev_PRB86,Sachdev_PRB87,Zarifis_PRB87,Huang_PRL07} We summarize the important mechanisms using the map in Fig.~\ref{fig:phase}.

In the low-temperature and non-degenerate regime (region 1), electrons freeze on isolated donor sites. The ground states is split into 6 states (without spin) due to valley-orbit coupling.\cite{Kohn_SSP57} For typical shallow donors such as phosphorus the singlet (nondegenerate) state is located $\sim$45~meV below the conduction band edge, while the doublet (two-fold degenerate) and triplet (three-fold degenerate) states are only  slightly split and located $\sim$35~meV below the conduction band edge. The extremely long spin lifetime of localized electrons in region 1 is governed by electron-phonon Raman processes and by hyperfine interactions with the non-zero nuclear spin of the impurity or Si$^{29}$ isotopes.\cite{Pines_PR57,Hasegawa_PR60,Roth_PR60,Castner_PR63,Castner_PR67} As an example, we mention the Orbach process which was shown by Castner to dominate between 5~K and 20~K.\cite{Castner_PR67} Spin flips at the low-energy singlet state are caused by phonon-induced virtual transitions to intermediate triplet states at which the spin-orbit coupling admixes spin-up and spin-down components. This process requires an initial absorption of a $\sim$$10$~meV phonon to mediate a singlet-to-triplet transition followed by phonon emission to transfer the electron back to the singlet state but with opposite spin.

The spin relaxation in region 3 of Fig.~\ref{fig:phase} is due to the formation of an impurity band. At these low temperatures and intermediate donor concentrations, the impurity band is populated and separated from the nearly empty conduction band. Compared with region 1, the spin relaxation is enhanced by orders of magnitude due to the overlap of wave functions in different impurity centers.\cite{Feher_PR55} On the insulator side (region 2), the spin relaxation is governed by exchange interactions between localized electrons and by rich hyperfine configurations of donor clusters.\cite{Slichter_PR55,Jerome_PR64,New_PR84} In barely metallic samples (region 3) the spin admixture is increased by broadening of the triplet and singlet bands.\cite{Zarifis_PRB87,Zarifis_PRB98} Spin relaxation is then governed by exchange and motional narrowing when electrons travel across the (random) potential.\cite{Ue_PRB71,Ochiai_PSS76,Paalanen_PRL86,Sachdev_PRB86,Zarifis_PRB87,Sachdev_PRB87,Zarifis_PRB98}

Spin relaxation in region 4 of Fig.~\ref{fig:phase} is of conduction electrons. Transition into region 4 is made either by increasing the temperature (electrons are thermally excited to the conduction band), or when the impurity band is merged into the conduction band (heavily degenerate doping). Region 4 is pertinent to spintronic devices in which electrons are swept away from the magnetic junction.\cite{Huang_JAP07,Huang_PRL07,Li_APL08,Huang_APL08,Li_APL09a,Kioseoglou_APL09,Li_APL09b,Grenet_APL09,Jansen_PRB10,Huang_PRB10} Then, the relaxation is governed by impurity scattering (degenerate doping) or by electron-phonon interactions (non-degenerate doping). In lateral spintronic devices, on the other hand, electrons are kept at the vicinity of the contact,\cite{Erve_APL07,Erve_IEEE09,Jang_PRL09,Ando_APL09,Sasaki_APE09,Breton_Nature11,Jeon_APL11,Ando_PRB12} and the transport is affected by the geometry and properties of the contact.\cite{Min_NatureMaterials06,Dery_PRB06a,Dery_PRB06b,Cywinsky_APL06,Song_PRB10,Dash_PRB11,Li_PRB11,Li_nature_com11}  Spin lifetimes that exceed the 1~ns scale were recently reported in lateral devices that incorporate a highly degenerate $n$-type silicon channel.\cite{Sasaki_APL11,Sasaki_APL10,Ando_APL11a,Shiraishi_PRB11,Sasaki_IEEE10,Ando_APE10} These reports suggest that impurity scattering in the bulk silicon channel plays an important role in heavily degenerate lateral devices that rely on spin accumulation.

Electron-phonon interaction plays a key role in the non-degenerate doping regime of region 4. In nearly intrinsic conditions, it dominates the spin relaxation already above $\sim$30~K,\cite{Huang_PRL07} while at $\sim$10$^{17}$~cm$^{-3}$ it dominates above $\sim$100~K. The increase in temperature is brought by the extended range at which freeze-out conditions persist, by exchange between free and localized electrons, and by scattering from ionized impurities.\cite{Lepine_PRB70} For transport at large applied electrical fields (where Fig.~\ref{fig:phase} is no longer valid), the electron-phonon scattering dominates the spin relaxation of conduction electrons at all lattice temperatures since the effective temperature of drifting electrons is significantly higher than the lattice temperature.\cite{Li_PRL12,Jang_PRB08}

Electron-impurity scattering dominates the spin relaxation of conduction electrons in highly degenerate doping concentrations (top part of region 4). An important aspect is that at intermediate and degenerate concentrations the impurity type influences the spin lifetime (irrespective of the temperature).\cite{Pifer_PRB75,Ochiai_PSS76,Zarifis_PRB98} The original Elliott-Yafet theory, on the other hand, suggests that the spin-orbit coupling of the host crystal (rather than of the impurity) determines the spin relaxation of conduction electrons.\cite{Elliott_PR54a,Yafet_SSP63} Finally, in region 5 several processes coexist and provide measurable spin relaxation. At donor concentrations of 10$^{15}$~-~10$^{17}$~cm$^{-3}$, L\'{e}pine showed that the spin relaxation is governed by exchange between free and localized electrons, by the larger spin-orbit coupling of triplet states, and by the modulation of the hyperfine field (when electrons make thermal transitions between localized states).\cite{Lepine_PRB70} At lower donor concentrations, the role of electron-phonon scattering increases since the \textit{relative} fraction of electrons in the conduction band increases.\cite{Lepine_PRB70,Lu_PRL11}


\section{Electron-phonon interaction}\label{sec:e-ph}

Within the harmonic approximation framework the electron-phonon interaction is generally described by,\cite{Kittel1963,Maradudin1971}
\begin{eqnarray}
\mathcal{H}_{\rm{ep}}\!\! & =&\!\! \sum\limits_{j,\alpha}\delta \mathbf{R}_{j\alpha}\cdot \bm{\nabla}_{\mathbf{R}_{j\alpha}} \mathcal{V}_{\rm{at}}(\mathbf{r}-\mathbf{R}_{j\alpha})\label{eq:Hep} \,,
\end{eqnarray}
where $j$ sums over $N$ primitive cells and $\alpha=\{A,B\}$ labels the two atoms of a primitive cell with the origin chosen at their mid-point position, $\bm\tau=\bm\tau_{A}=-\bm\tau_{B}=-(a/8)(1,1,1)$. An atom position is then denoted by $\mathbf{R}_{j\alpha}=\mathbf{R}_{j}+\bm\tau_{\alpha}$, and its potential by
\begin{eqnarray}
\mathcal{V}_{\rm{at}}(\mathbf{r}) = V_{\rm{at}}(\mathbf{r}) \mathcal{I} + \frac{\hbar}{4m^2_0 c^2}\left[\bm{\nabla}V_{\rm{at}}(\mathbf{r})\!\times\!\mathbf{p} \right]\cdot \bm\sigma\,. \label{eq:potential}
\end{eqnarray}
$\mathcal{I}$ is the 2$\times$2 identity matrix and $\bm\sigma$ is the vector of Pauli matrices. The first (second) term is the bare potential (spin-orbit coupling), and it corresponds to the Elliott (Yafet) part of $\mathcal{H}_{\rm{ep}}$.   The atom vibration in Eq.~(\ref{eq:Hep}) is expressed as
\begin{eqnarray}
\delta \mathbf{R}_{j\alpha} =  \sum_{\mathbf{q}}\sqrt{2\hbar/[\rho\omega(\mathbf{q}) N a^3 ]} [a_\mathbf{q} \bm\xi_\alpha (\mathbf{q}) e^{i \mathbf{q} \mathbf{R}_{j\alpha}} + \rm{h.c.}]\,,\,\,\, \label{eq:Rja}
\end{eqnarray}
where $\rho$ is the material density and $\omega(\mathbf{q})$ is phonon frequency. $a_\mathbf{q}$ is the annihilation operator, $\bm{\xi}$ is the normalized phonon polarization vector,\cite{footnote_wtype} and `h.c.' stands for Hermitian conjugate. The square-root prefactor is written in accordance with $\sum_{\alpha}|\bm\xi_{\alpha}|^2=2$ (the number of atoms in the primitive cell). Using crystal momentum conservation  together with identities of phonon creation, $\langle  n_2|a^{\dagger}|n_1\rangle=\sqrt{n_1+1}\delta_{n_2,n_1+1}$, and annihilation, $\langle  n_2|a| n_1\rangle=\sqrt{n_1}\delta_{n_2,n_1-1}$, we write the transition amplitude of an electron from wavevector $\mathbf{k}_1$ to $\mathbf{k}_2$ by,
\begin{eqnarray}
&& \!\!\!\!\!\!\!\! \langle \mathbf{k}_2,\mathbf{s}_2; n(\mathbf{q})\pm 1 | \mathcal{H}_{\rm{ep}} | \mathbf{k}_1,\mathbf{s}_1;n(\mathbf{q})\rangle = \label{eq:k2Hepk1} \\ && \!\!\!\!\!\!\!\!  -\sqrt{\frac{2\hbar}{\rho \omega(\mathbf{q}) N a^3}} \sqrt{n(\mathbf{q})+\frac{1}{2} \pm \frac{1}{2}}    \times\langle \mathbf{k}_2,\mathbf{s}_2| {H}_{\rm{ep}}| \mathbf{k}_1,\mathbf{s}_1\rangle\,,\,\,\,\,\, \nonumber
\end{eqnarray} 
where
\begin{eqnarray}
H_{\rm{ep}}=\sum\limits_{j,\alpha}\bm\xi_\alpha (\mathbf{q}) e^{i \mathbf{q} \mathbf{R}_{j\alpha}}\cdot \bm{\nabla}_{r} \mathcal{V}_{\rm{at}}(\mathbf{r}-\mathbf{R}_{j\alpha}),\label{eq:Hep_atom}
\end{eqnarray}
$\mathbf{q}=\mathbf{k}_2-\mathbf{k}_1$, $n(\mathbf{q})$ is phonon occupation (Bose-Einstein distribution at thermal equilibrium), and $\mathbf{s}=\{\Uparrow,\Downarrow\}$ denotes the spin-up and spin-down states. Due to time reversal and space inversion symmetries each band at wavevector $\mathbf{k}$ is spin degenerate and we can define its states with respect to the spin orientation $\mathbf{n}$, such that
\begin{eqnarray}
 &\langle \mathbf{k}, \mu, \Uparrow |  \bm{\sigma}\cdot \hat{\mathbf{n}} | \,\mathbf{k}, \mu, \Uparrow \rangle  \equiv -\langle \mathbf{k}, \mu, \Downarrow |  \bm{\sigma}\cdot \hat{\mathbf{n}} |\, \mathbf{k}, \mu, \Downarrow \rangle \geq 0&\,,\,\nonumber  \\
 & \langle \mathbf{k}, \mu, \Uparrow |  \bm{\sigma}\cdot \hat{\mathbf{n}} | \,\mathbf{k}, \mu, \Downarrow \rangle \equiv 0\, ,&\label{eq:spin_up_down_states}
\end{eqnarray}
where $\mu$ is the band index. Mixed by spin-orbit interaction, the defined spin-up and spin-down states ($\Uparrow, \Downarrow$) are not pure spin states ($\uparrow, \downarrow$).

Based on the symmetry of electron and phonon states we will derive selection rules for $M(\mathbf{k}_1,\mathbf{s}_1;\mathbf{k}_2,\mathbf{s}_2)=\langle \mathbf{k}_2,\mathbf{s}_2 |H_{\rm{ep}}|\mathbf{k}_1,\mathbf{s}_1 \rangle$. It is often convenient to work with interaction that has a fixed parity under space inversion operation of the crystal. We convert Eq.~(\ref{eq:Hep}) into in-phase and out-of-phase parts,
\begin{eqnarray}
\!\!\!\! \mathcal{H}_{\rm{ep}}\!\!\! & =& \!\!\!  -\!\!\sum\limits_{j,\pm}\!\delta \mathbf{R}^{\pm}_{j}\!\cdot\!\! \bm{\nabla}_{\mathbf{r}} \mathcal{V}_{\pm}(\mathbf{r}\!-\!\mathbf{R}_{j}),  \nonumber\\
\delta \mathbf{R}^{\pm}_{j}\!\! & =& \!\! \tfrac{1}{2}\left(\delta \mathbf{R}_{jA}\pm \delta \mathbf{R}_{jB}\right),\nonumber  \\
\mathcal{V}_{\pm}(\mathbf{r}) \!\!\! & =& \!\!\! \mathcal{V}_{\rm{at}}(\mathbf{r}-\bm\tau_A) \pm \mathcal{V}_{\rm{at}}(\mathbf{r}-\bm\tau_B)\label{eq:Hep1}.
\end{eqnarray}
The (anti)symmetrized cell potential $\mathcal{V}_{+(-)}$ is associated with the in(out-of)-phase motion. Using the crystal translational symmetry,\cite{footnote_translational} $\sum_{j} e^{i \mathbf{q} \mathbf{R}_{j}} \bm{\nabla} \mathcal{V}_{\pm}(\mathbf{r}-\mathbf{R}_j)$ coupled  between states at $\mathbf{k}_1$ and $\mathbf{k}_2$ is further reduced to $N \bm{\nabla} \mathcal{V}_{\pm}(\mathbf{r}) \delta_{\mathbf{q}+\mathbf{k}_1-\mathbf{k}_2,\mathbf{g}}$. With Eq.~(\ref{eq:Hep_atom}), the matrix element becomes
\begin{eqnarray}
&&M(\mathbf{k}_1,\mathbf{s}_1; \mathbf{k}_2,\mathbf{s}_2) = \langle \mathbf{k}_2,\mathbf{s}_2| H_{\rm{ep}} |\mathbf{k}_1,\mathbf{s}_1 \rangle \nonumber\\
&=&N\sum\limits_{\pm}\bm\xi^{\pm} (\mathbf{q})\cdot \langle \mathbf{k}_2,\mathbf{s}_2| \bm{\nabla}_{\mathbf{r}} \mathcal{V}_{\pm}(\mathbf{r}) |\mathbf{k}_1,\mathbf{s}_1 \rangle\;, \label{eq:M}
\end{eqnarray}
where in-phase and out-of-phase polarization vectors
\begin{eqnarray}
 \bm\xi^{\pm}(\mathbf{q})=[\bm\xi_A (\mathbf{q}) e^{i \mathbf{q} \bm{\tau}_A} \pm  \bm\xi_B (\mathbf{q}) e^{i \mathbf{q} \bm{\tau}_B}]/2\,,\label{eq:inout_pol_vec}
\end{eqnarray}
satisfy $|\bm\xi^{+}|^2+|\bm\xi^{-}|^2=1$. In what follows we abbreviate the notation and absorb the scaling factor into the matrix element definition (i.e., $N\langle...\rangle \rightarrow \langle...\rangle$).

\begin{table}
\tabcolsep=0.1cm
\renewcommand{\arraystretch}{2.0}
\caption{\label{tab:elastic_continuum}
In-phase and out-of-phase phonon polarization vectors in the long-wavelength limit ($\mathbf{q} \ll 2\pi/a$). We have used the elastic continuum approximation for diamond crystal structures. $\lambda=\{\rm{TA}_1,\rm{TA}_2,\rm{LA}\}$ and $\Gamma_{xyz} \approx a/16$. Two TA or TO polarizations can be linearly combined into any other orthonormal ones. See text for further explanation.
 }
\begin{tabular}{cc}
\hline \hline
$\bm\xi^+_{\rm{TA_1}}, \bm\xi^-_{\rm{TO_1}}(\mathbf{q})$  & $ (q_y,-q_x,0)/\sqrt{q^2_x+q^2_y} $ \\
$\bm\xi^+_{\rm{TA_2}}, \bm\xi^-_{\rm{TO_2}}(\mathbf{q})$  & $ (q_x q_z,q_y q_z,-q^2_x-q^2_y) /(\sqrt{q^2_x+q^2_y}|\mathbf{q}|)$ \\
$\bm\xi^+_{\rm{LA}}, \bm\xi^-_{\rm{LO}}(\mathbf{q})$  & $\mathbf{q}/|\mathbf{q}|$\\
$\xi^-_{\lambda,l}(\mathbf{q})$  & $ i\mathbf{q}\cdot\bm\tau \xi^+_{\lambda,l}(\mathbf{q}) +i \Gamma_{xyz} [q_m\xi^+_{\lambda,n}(\mathbf{q}) +q_n\xi^+_{\lambda,m}(\mathbf{q})]$\\
\hline \hline
\end{tabular}
\end{table}

In the analysis of intravalley scattering, explicit forms of polarization vectors are utilized. To gain further insight of their symmetries we invoke the elastic continuum approximation for diamond crystal structures (incorporating internal displacement).\cite{Born_book, Smith_1948, Ehrenreich1956} Table~\ref{tab:elastic_continuum} lists forms of $\bm\xi^{\pm}_{\lambda}(\mathbf{q})$ in the {\it{long-wavelength}}  limit where $\lambda$ is the phonon mode. The in-phase polarization vectors of acoustic modes (first three rows) are of the order of unity while their out-of-phase vectors (fourth row) are of the order of $qa/2\pi \ll 1$. They flip roles for optical modes (i.e. order of unity for out-of-phase and the order of $qa/2\pi$ for in-phase vectors). Components of out-of-phase vectors relate to the in-phase components and embody the structure of the two-atom primitive cell,
 \begin{eqnarray}
 \xi^-_{\lambda,\ell}(\mathbf{q}) \!=\! \tfrac{1}{2}{\xi}_{\lambda,\ell}^+(\mathbf{q}) (e^{i\mathbf{q}\cdot\bm\tau_A}\!\! - \! e^{i\mathbf{q}\cdot\bm\tau_B}) + i\! \sum_{m,n} \Gamma_{mn\ell}q_m\xi_{\lambda,n}^-\!(\mathbf{q}). \nonumber 
\end{eqnarray}
Here, $\lambda$ denotes any of the acoustic modes and ${l,m,n}$ are cyclic permutation of the coordinates. In writing the expression in the fourth row of the table we have used the following approximations. The exponential term, brought by the macroscopic strain, is replaced by $i\mathbf{q}\cdot\bm\tau\bm{\xi}^+(\mathbf{q})$ due to the long-wavelength nature ($\mathbf{q}\rightarrow0$). For the other term, brought by internal displacement, the only non-vanishing components of the third-rank tensor $\Gamma_{ijk}$ are $\Gamma_{xyz}=\Gamma_{yzx}=\Gamma_{zxy}$ in a diamond-crystal structure.\cite{Bir_Pikus_Book} Their value is $0.5a/8$.\cite{Weber_PRB77,Nielsen_PRB85}

It is emphasized that in this work we use the rigid-ion model to derive general results of intravalley spin-flip matrix elements [Eqs.~(\ref{eq:ac_Sn})-(\ref{eq:op_Sn})]. These results do \textit{not} depend on the approximate forms of $\bm\xi^{\pm}$ in Table~\ref{tab:elastic_continuum}. However, since these approximations are fairly accurate and become exact along high symmetry directions, we can make use of them to derive appealing spin-flip matrix elements of long-wavelength phonon modes (Table~\ref{tab:intra_g_M_Sn}).

\section{Intervalley $f$-process spin flips}\label{sec:f-process}

At elevated temperatures, spin relaxation in unstrained silicon is largely dominated by the intervalley $f$-process.\cite{Pengke_PRL11,Tang_PRB12,Dery_APL11,Li_PRL12} Therefore, it is crucial to have a complete set of selection rules for $f$-process spin scattering. Selection rules of the intervalley momentum relaxation (neglecting spin-orbit coupling)  were worked out by Lax and Hopfield and by Streitwolf with single group theory.\cite{Lax_PR61,Streitwolf_PSS70} In this section we present a direct and detailed application of double group theory to derive spin-conserving and spin-flip matrix elements along arbitrary spin orientation directions.

\begin{figure}[t]
\includegraphics[width=8cm]{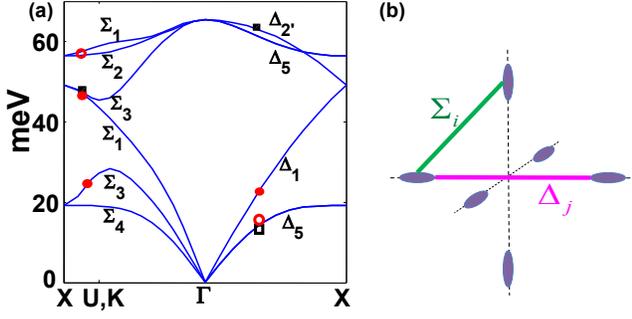} 
\caption{(Color online) (a) Phonon dispersion curves in silicon. Phonon modes at the vicinity of the red circles (black squares) take part in spin (momentum) \textit{intervalley} scattering. The $f$-process involves the zone-edge phonons near the $\Sigma$-symmetry axis (connecting the $\Gamma$ and $K$ points). The $g$-process involves phonons near the $\Delta$-symmetry axis (between $\Gamma $ and $X$ points). It will be shown that leading contributions are associated with phonon modes with the solid symbols. Intravalley scattering relates to all branches of long-wavelength phonons around the $\Gamma$ point. (b) Relating phonons with $\Delta_j$ and $\Sigma_i$ symmetries to representative intervalley electron transitions.
} \label{fig:BZ_phonon}
\end{figure}


Rendering double group theory to explain the $f$-process spin flips is particularly simple since electron states involve only the two-dimensional (2D) irreducible representation (IR) $\Delta_6$,\cite{Elliott_PR54b} and no perturbation between other symmetry states is needed. For convenience, the $f$-process is studied by changing the spin orientation ($\mathbf{n}$) while fixing the scattering to be  between the $+x$ and $+y$ valleys.  While the total number of independent (scattering) constants is not changed, the matrix elements depend on the chosen spin orientation. As such, the character table is not sufficient in mapping the dependence on the spin orientation. We bypass this limitation and use the explicit IR matrices. In this approach spin-conserving (in momentum scattering) and spin-flip transitions are treated on equal footing. In the rest of this section we first study the case of spin orientation along a valley axis. Then we will remove this restriction and provide general expressions for the $f$-process spin-flip matrix elements [Eqs.~(\ref{eq:Mm_Sigma1_Sn})-(\ref{eq:Ms_Sigma3_Sn}) and Table~\ref{tab:fM_Sn}].

\subsection{Spin orientation parallel to a valley axis}\label{sec:f_para}
The $f$-process between opposite spin states in the $+x$ and $+y$ valleys depends on whether $\mathbf{n}$ is perpendicular to both valleys ($\mathbf{n} \| \mathbf{z}$) or parallel to one of them ($\mathbf{n} \| \mathbf{x}$ or $\mathbf{n} \| \mathbf{y}$). For either case, we should consider four phonon symmetries represented by 1D IRs, $\Sigma_{1-4}$. These symmetries dictate how each of the phonon-induced interactions, $H_{\Sigma_i}$ [$H_{\rm{ep}}$ in Eq.~(\ref{eq:Hep_atom}) with $\Sigma_i$ phonon], transforms under symmetry operations of the $\Sigma$ group. In the left part of Fig.~\ref{fig:BZ_phonon}(a) we designate the phonon symmetries with the six phonon branches and in row 4 of Table \ref{tab:f_character} we associate these symmetries with phonon modes. Row 5 of the table lists the energies of these modes at the phonon wavevector that connects the valley centers.


The nonvanishing matrix elements $\langle \mathbf{k}_2| H_{\Sigma_{i}} |\mathbf{k}_1\rangle$ are obtained from selection rules that involve very few transformations.\cite{Elliott_PR54b,Lax_PR61} These transformations include common symmetry operations, $g_c\in\{ (\epsilon|0), (\bar{\epsilon}|0),(\rho_z|\tau), (\bar{\rho}_z|\tau) \}$, of the little groups at $\mathbf{k_1}$, $-\mathbf{k_2}$ and $\mathbf{q}$. They also include  operations $g_e\in \{ (\rho_{xy}|0), (\bar\rho_{xy}|0),(\delta_{2\bar{x}y}|\tau),(\bar\delta_{2\bar{x}y}|\tau) \}$ that exchange $\mathbf{k_1}$ and $-\mathbf{k_2}$. The bar over operations denotes an additional $2\pi$ rotation. The character table of the nontrivial operations is shown at the upper three rows of Table~\ref{tab:f_character}. The number of independent real constants involved in the matrix elements for each phonon symmetry is provided by consideration of these operations and time reversal symmetry,
\begin{eqnarray}
\mathcal{N}_{\Sigma_i}=\frac{1}{2h_0}\left[\sum\limits_{g_c} \chi^{\mathbf{-k}_2}_{\Delta_6}(g_c) \chi^{\mathbf{k}_1}_{\Delta_6}(g_c) \chi^{\mathbf{q}}_{\Sigma_i}(g_c) \right.\nonumber\\
\left.- \sum\limits_{g_e} \chi^{\mathbf{k}_1}_{\Delta_6}(g^2_e) \chi^{\mathbf{q}}_{\Sigma_i}(g_e) \right], \label{eq:indep_num}
\end{eqnarray}
where $h_0=4$ is the number of $g_c$ or $g_e$ operations and $\chi_{\Delta_6}=\chi_{\Delta_1} \times \chi_{1/2}$. 
The second sum in Eq.~(\ref{eq:indep_num}) represents the effect of time reversal symmetry and the minus sign takes into account the parity from the spinor basis and interaction $H_{\rm{ep}}$ (see, for example, Refs.~[\onlinecite{Bir_Pikus_Book}] and ~[\onlinecite{doni_JPC73}] for more details).  By straightforwardly plugging the characters of Table~\ref{tab:f_character} into Eq.~(\ref{eq:indep_num}) we get
\begin{eqnarray}
\mathcal{N}_{\Sigma_1}=2,\, \mathcal{N}_{\Sigma_2}=1, \,\mathcal{N}_{\Sigma_3}=1,\, \mathcal{N}_{\Sigma_4}=0.\label{eq:N_0_Sigma}
\end{eqnarray}

\begin{table}
\caption{\label{tab:f_character}
Non-trivial relevant IR matrices in a $f$-process between $+x$ and $+y$ valleys [$\Sigma$ evaluated at $(-k_0,k_0,0)$, $\Delta$ at $(k_0,0,0)$ or $(0,k_0,0)$, $k_0= 0.85 \times 2\pi/a$]. $\chi^{-\mathbf{k}_2}_{\Delta_1}=(\chi^{\mathbf{k}_2}_{\Delta_1})^*$. Also shown is the effect of exchange operations on $\Delta$ star. Basis states in $D_{1/2}$ is along $\pm z$ in spin space. The $\|$ and $\bot$ directions in the mode subscript are relative to $\mathbf{z}$.  $E^f_q$ is in unit of meV and $D_{\Sigma_i s}$ of $\rm{meV}\cdot 2\pi/a$.
}
\renewcommand{\arraystretch}{2.0}
\tabcolsep=0.07 cm
\begin{tabular}{c|c|c|c|c|c|c}
\hline \hline
               & $\Sigma_1$ &$\Sigma_2$ & $\Sigma_3$ & $\Sigma_4$ & $\Delta_1$ &$D_{1/2}$\\
               \hline
 $(\rho_z|\tau)$ & 1           & -1         &  -1       &  1         &  $\displaystyle  e^{ \frac{-ik_0 a}{4}}$ & $ \!\!\!\left(\!\!\!\renewcommand{\arraystretch}{1.0}\begin{array}{cc}-i&0\\0&i\end{array}\!\!\!\right)$\\
 \hline
 $(\rho_{xy}|0)$ & 1           & -1         &  1       &  -1         &  $\mathbf{k}_1\!\!\!\leftrightarrow\!\! \mbox{-}\mathbf{k}_2$\! &$\displaystyle e^{\mbox{-}\tfrac{3\pi i}{4}}\!\! \left(\!\!\!\renewcommand{\arraystretch}{1.0}\begin{array}{cc}0&1\\i&0\end{array}\!\!\!\right)$\\
 \hline
 $(\delta_{2\bar{x}y}|\tau)$ & 1   & 1  & -1      &  -1         &  $\mathbf{k}_1\!\!\!\leftrightarrow\!\! \mbox{-}\mathbf{k}_2$\!  &$\displaystyle
 e^{\mbox{-}\tfrac{3\pi i}{4}}\!\! \left(\!\!\!\renewcommand{\arraystretch}{1.0}\begin{array}{cc}0&i\\1&0\end{array}\!\!\!\right)$\\
 \hline\hline
Mode      &  LA,\,$\rm{TO}_{\| }$    & $\rm{TO}_{\perp }$          &  $\rm{TA}_{\| }$,\,LO   & $\rm{TA}_{\perp }$               \\
 \hline
$E^f_q$     &  46.6,\,58    & 57         & 23,\,46.8    &19              \\
  \hline
$D_{\Sigma_i s}$      &   6.5,\,2.9   & 3.7         & 1.7,\,1.1   &         \\
 \hline \hline
\end{tabular}
\end{table}

\begin{table*} [t]
\renewcommand{\arraystretch}{1.5}
\tabcolsep=0.7cm
\caption{\label{tab:fM_Sn}
$|M_{\Sigma_i} (\mathbf{k}_1,\Uparrow_\mathbf{n}; \mathbf{k}_2,\Downarrow_\mathbf{n})/ D_{\Sigma_i s}|^2$ for $f$-process spin flips between $+x$ and $+y$ valleys. The values of the scattering constants $D_{\Sigma_i s}$ are given in Table~\ref{tab:f_character}. For each of the non-vanishing modes, $\Sigma_i$, the relative amplitude is provided for spin orientation ($\mathbf{n})$ along any of the inequivalent high-symmetry crystal directions. Results between other valleys can all be obtained by trivial symmetry arguments.}
\begin{tabular}{cccccccc}
\hline \hline
$\mathbf{n}$    &   $[0\;0\;1]$  & $[1\;0\;0]$ & $[1\;1\;0]$ & $[1\;\bar{1}\;0]$ &  $[1\;0\;1]$     &  $[1\;1\;1]$    &   $[1\;\bar{1}\;1]$     \\ \hline
$\Sigma_1$      &       0        &      1      &        1    &      1            &  $\frac{1}{2}$   &  $\frac{2}{3}$  &   $\frac{2}{3}$         \\
$\Sigma_2$      &       2        &      1      &        0    &      2            &  $\frac{3}{2}$   &  $\frac{2}{3}$  &       2                 \\
$\Sigma_3$      &       2        &      1      &        2    &      0            &  $\frac{3}{2}$   &      2          &   $\frac{2}{3}$         \\ \hline \hline
\end{tabular}
\end{table*}

Next, the interaction matrix elements $\langle \mathbf{k}_2,\mathbf{s}_2 |H_{\Sigma_i}| \mathbf{k}_1,\mathbf{s}_1 \rangle$ between specific spin species are expressed in terms of $\mathcal{N}_{\Sigma_i}$ independent scattering constants. To reach this goal, we first write down the IR matrix $D_{\Delta_6}$ for a given spin orientation. We choose the spin orientation conveniently along $z$ direction, and then $D_{\Delta_6}=D_{\Delta_1} \times D_{1/2}$ using Table~\ref{tab:f_character}.\cite{footnote_rot_spin} Via appropriate group operations (see details in Appendix~\ref{app:f}), we obtain
\begin{subequations}\label{eq:Sigma}
\begin{eqnarray}
\langle \mathbf{k}_2, \Uparrow_z |H_{\Sigma_1}| \mathbf{k}_1, \Uparrow_z \rangle &=& D_{\Sigma_1m} + i D_{\Sigma_1s}, \label{eq:Sigma_1}\\
\langle \mathbf{k}_2, \Downarrow_z |{H}_{\Sigma_2}| \mathbf{k}_1, \Uparrow_z \rangle &=& D_{\Sigma_2s} - i D_{\Sigma_2s},\label{eq:Sigma_2}\\
\langle \mathbf{k}_2, \Downarrow_z |{H}_{\Sigma_3}| \mathbf{k}_1, \Uparrow_z \rangle &=& D_{\Sigma_3s} + i D_{\Sigma_3s},\label{eq:Sigma_3}
\end{eqnarray}
\end{subequations}
and
\begin{subequations}\label{eq:TS}
\begin{eqnarray}
\langle \mathbf{k}_2, \Uparrow_{\mathbf{n}} |H_{\Sigma_i}| \mathbf{k}_1, \Uparrow_{\mathbf{n}} \rangle & = & \langle \mathbf{k}_2, \Downarrow_{\mathbf{n}} |{H}_{\Sigma_i}| \mathbf{k}_1 \Downarrow_{\mathbf{n}} \rangle^*, \label{eq:TS_m}\\
\langle \mathbf{k}_2, \Downarrow_{\mathbf{n}} |{H}_{\Sigma_i}| \mathbf{k}_1, \Uparrow_{\mathbf{n}} \rangle & = & -\langle \mathbf{k}_2, \Uparrow_{\mathbf{n}} |{H}_{\Sigma_i}| \mathbf{k}_1, \Downarrow_{\mathbf{n}} \rangle^*,\quad\label{eq:TS_s}
\end{eqnarray}
\end{subequations}
up to a phase freedom for each matrix element. The scattering constants are all real numbers with $D_{\Sigma_1m}$ much larger (about three orders of magnitude) than the rest. $m$ denotes momentum and $s$ denotes spin, with the reason more obvious in the general spin orientation case. Eq.~(\ref{eq:TS}) holds for all phonon modes and spin orientations, and it expresses the effects of time reversal and space inversion (diamond structure).

When the spin orientation is parallel to the axis of one of the valleys that participate in the $f$-process,  then instead of changing the $\Delta_6$ basis states and the matrix form, we write the new spin states in terms of the previous $\Delta_6$ basis (we call it the `original basis'). As an example, for spin orientation along the $x$ valley we use the rotation matrix $[1,-1;1, 1]/\sqrt{2}$ as a unitary transformation matrix and get\cite{footnote_rot_Delta6}
\begin{eqnarray}
|\mathbf{k}, \Uparrow_x \rangle &\simeq&\left(|\mathbf{k}, \Uparrow_z \rangle +|\mathbf{k} , \Downarrow_z, \rangle\right)/\sqrt{2},\nonumber\\
|\mathbf{k}, \Downarrow_x \rangle &\simeq&\left(-|\mathbf{k}, \Uparrow_z \rangle +|\mathbf{k}, \Downarrow_z \rangle\right)/\sqrt{2}.\label{eq:xz_spin_rot}
\end{eqnarray}
Using Eqs.~(\ref{eq:Sigma}) and (\ref{eq:TS}) we can obtain the results for spin orientation along one of the involved valleys ($\mathbf{n}$$\,$$\|$$\,$$\mathbf{x}$; the case of $\mathbf{n}$$\,$$\|$$\,$$\mathbf{y}$ is equivalent). For $\Sigma_1$,
\begin{subequations}\label{eq:fspinx}
\begin{eqnarray}
\langle \mathbf{k}_2, \Uparrow_x \!\!\!&\!|\!&\!\!{H}_{\Sigma_1}| \mathbf{k}_1, \Uparrow_x \rangle \nonumber \\
\!\!\!\!\!\!\!\!\!&&\!\!\!\!=\!\! (\!\!\renewcommand{\arraystretch}{1.6}\begin{array}{cc} \frac{1}{\sqrt{2}}\!\! & \frac{1}{\sqrt{2}} \end{array}\!\!) \left(\!\!\begin{array}{cc} \!D_{\Sigma_1m}\!\!+\!iD_{\Sigma_1s} \!& 0\\0 & \!\! D_{\Sigma_1m}\!\!-\!iD_{\Sigma_1s}\!\! \end{array}\!\!\right) \!\!\!\left(\!\!\begin{array}{c} \frac{1}{\sqrt{2}}\!\! \\ \frac{1}{\sqrt{2}} \end{array}\!\!\right) \nonumber\\
&&=D_{\Sigma_1m}\,.\label{eq:fspinx1}
\end{eqnarray}
Repeating the same procedure for the spin-flip case and for other phonon modes, we get
\begin{eqnarray}
\langle \mathbf{k}_2, \Downarrow_x |{H}_{\Sigma_1}| \mathbf{k}_1, \Uparrow_x \rangle &=& -i D_{\Sigma_1s},\\
\langle \mathbf{k}_2, \Uparrow_x |{H}_{\Sigma_2}| \mathbf{k}_1, \Uparrow_x \rangle &=& -i D_{\Sigma_2s},\\
\langle \mathbf{k}_2, \Downarrow_x |{H}_{\Sigma_2}| \mathbf{k}_1, \Uparrow_x \rangle &=& D_{\Sigma_2s},\\
\langle \mathbf{k}_2, \Uparrow_x |{H}_{\Sigma_3}| \mathbf{k}_1, \Uparrow_x \rangle &=& i D_{\Sigma_3s},\\
\langle \mathbf{k}_2, \Downarrow_x |{H}_{\Sigma_3}| \mathbf{k}_1, \Uparrow_x \rangle &=& D_{\Sigma_3s}.
\end{eqnarray}
\end{subequations}
Together with Eq.~(\ref{eq:TS}), the set of expressions in Eqs.~(\ref{eq:Sigma}) and (\ref{eq:fspinx}) complete the results of inequivalent types of $f$-process scattering when the spin orientation is set along a valley axis. Values of the scattering constants, $D_{\Sigma_is}$, are obtained by numerical calculations and they are listed in the last row of Table~\ref{tab:f_character}.  Preferably, their values are extracted from experiments which cover some parameter range (e.g., temperature, stress and external fields).

\subsection{General dependence  on  spin orientation}\label{sec:f_rot}
So far we have restricted the spin orientation during an $f$-process spin flip to be along a valley axis (main crystallographic axis). Removing this restriction adds to the anisotropy of spin relaxation processes and it allows one to make a direct comparison to a wide range of spin injection experiments.

We generalized the previous derivation to an arbitrary spin orientation direction. We define $\mathbf{n}$ in terms of polar and azimuthal angles $\theta$ and $\phi$ with respect to the $+z$ and $+x$ directions. The new spin states relate to the original ones by an `active' rotation matrix in spin space,\cite{Sakurai}
\begin{eqnarray}
|\mathbf{k}, \Uparrow_\mathbf{n} \rangle &\simeq&\cos\frac{\theta}{2}|\mathbf{k}, \Uparrow_z \rangle +\sin\frac{\theta}{2} e^{i\phi}|\mathbf{k}, \Downarrow_z \rangle,\nonumber\\
|\mathbf{k}, \Downarrow_\mathbf{n} \rangle &\simeq&-\sin\frac{\theta}{2} e^{-i\phi}|\mathbf{k}, \Uparrow_z \rangle + \cos\frac{\theta}{2}|\mathbf{k}, \Downarrow_z \rangle.\label{eq:k_n}
\end{eqnarray}
Using the explicit matrix for interaction ${H}_{\Sigma_i}$ between $|\mathbf{k}, \Uparrow(\Downarrow)_z \rangle$ basis states, as the example in Eq.~(\ref{eq:fspinx1}), the spin-flip matrix elements under the new spin orientation are
\begin{subequations}\label{eq:Sigma_Sn}
\begin{eqnarray}
\langle \mathbf{k}_2, \Uparrow_\mathbf{n}\! |{H}_{\Sigma_1}| \mathbf{k}_1, \Uparrow_\mathbf{n} \rangle \!\!&=&\!\! D_{\Sigma_1m}+ i \cos\theta D_{\Sigma_1s},\label{eq:Mm_Sigma1_Sn}
\\
\langle \mathbf{k}_2, \Downarrow_\mathbf{n}\! |{H}_{\Sigma_1}| \mathbf{k}_1, \Uparrow_\mathbf{n} \rangle \!\!&=&\!\! -i \sin\theta e^{i\phi} D_{\Sigma_1s}, \label{eq:Ms_Sigma1_Sn}
\\
\langle \mathbf{k}_2, \Uparrow_\mathbf{n}\! |{H}_{\Sigma_2}| \mathbf{k}_1, \Uparrow_\mathbf{n} \rangle \!\!&=&\!\! -i\sqrt{2} \sin\theta \sin(\phi+\frac{\pi}{4})D_{\Sigma_2s},\qquad\label{eq:Mm_Sigma2_Sn}
\\
\langle \mathbf{k}_2, \Downarrow_\mathbf{n}\! |{H}_{\Sigma_2}| \mathbf{k}_1, \Uparrow_\mathbf{n} \rangle \!\!&=&\!\! \left[(1+i) \sin^2\frac{\theta}{2} e^{2i\phi} \right. \nonumber\\
 &&\quad\left. + (1-i) \cos^2\frac{\theta}{2}\right] D_{\Sigma_2s},\label{eq:Ms_Sigma2_Sn}
\\
\langle \mathbf{k}_2, \Uparrow_\mathbf{n}\! |{H}_{\Sigma_3}| \mathbf{k}_1, \Uparrow_\mathbf{n} \rangle \!\!&=&\!\! i\sqrt{2} \sin\theta \sin(\phi+\frac{3\pi}{4})D_{\Sigma_3s},\label{eq:Mm_Sigma3_Sn}
\\
\langle \mathbf{k}_2, \Downarrow_\mathbf{n}\! |{H}_{\Sigma_3}| \mathbf{k}_1, \Uparrow_\mathbf{n} \rangle \!\!&=&\!\! \left[(1-i) \sin^2\frac{\theta}{2} e^{2i\phi} \right. \nonumber\\
 &&\quad\left. + (1+i) \cos^2\frac{\theta}{2}\right] D_{\Sigma_3s}\,,\label{eq:Ms_Sigma3_Sn}
\end{eqnarray}
\end{subequations}
with Eq.~(\ref{eq:TS}) for other matrix elements. Table~\ref{tab:fM_Sn} lists the relative magnitudes of the squared spin-flip matrix elements for $\mathbf{n}$ along all inequivalent high-symmetry directions of the crystal (they are often the spin orientations of injected electrons). Analysis of the spin relaxation time due to $f$-process spin flips will be given in Sec.~\ref{sec:lifetime}.



\section{intravalley and intervalley $g$-process spin flips}\label{sec:intra_g}

In this section we will present a rigorous procedure to reach at compact intravalley and $g$-process spin-flip matrix elements. Before embarking on the theory, we discuss key considerations that underly the analysis and exemplify their outcomes via representative results (Table~\ref{tab:intra_g_M_Sn}).  This choice allows one to understand the most important physical parameters without delving into details of the analysis (which are provided in Secs.~\ref{sec:kp hamilton} through ~\ref{sec:valley-spin dependence}).

Spin-flip matrix elements vanish for intravalley ($g$-process) scattering if wavevectors of the initial and final states are the same (opposite). The wavevector power-law dependence was mentioned in Table~\ref{tab:symmetry_argument} along with the fact that suppression of zeroth-order terms in these processes leads to relative slow spin relaxation compared with the $f$-process. However, intravalley and $g$-process become more important when shear strain is applied since different valley minima are split thus suppressing the scattering by the $f$-process.\cite{Dery_APL11,Tang_PRB12}
The intravalley spin flips are also important at low temperature due to the larger population of long wavelength acoustic phonons [smaller energies; See Fig.~\ref{fig:BZ_phonon}(a)].



It is instrumental to compare spin flips with {\it momentum} scattering of which the study is more established. Following the ingenious connection with deformation potential parameters by Bardeen and Shockley,\cite{Bardeen_PR50} Herring and Vogt derived a detailed angle dependence of intravalley momentum scattering due to interaction with  acoustic phonon.\cite{Herring_PR56} In spin flips, the connection with deformation potential is more subtle and complicated by the dependence on high-order wavevector components. Later, we will derive explicit forms while making no {\it a prior} assumptions about the form of possible deformation potential parameters. This approach allows us to identify the crucial role of the coupling between the lowest pair of conduction bands in setting the intravalley spin relaxation. The dependence of spin-flip matrix elements on high-order wavevector components makes this coupling effective in spin relaxation (while being marginal in momentum relaxation).  The relaxation rate becomes inversely proportional to the square of the energy gap between the conduction bands at the valley center, $\Delta_C\approx0.5$~eV. 

\begin{table} [h]
\renewcommand{\arraystretch}{2.5}
\tabcolsep=0.10 cm
\caption{\label{tab:intra_g_M_Sn}
Squared spin-flip matrix element $ |M_{\rm{sf}}|^2$, induced by all types of phonon modes in intravalley scattering, and by the LA phonon mode in $g$-process scattering. $\mathbf{q}=\mathbf{k}_2-\mathbf{k}_1$ and $\mathbf{K}=(\mathbf{k}_2+\mathbf{k}_1)/2$. Valley centers are set along $z$ axis. Spin orientations $\mathbf{n}$ are taken along all inequivalent crystal symmetry directions.  Results in other valleys can be obtained by trivial symmetry arguments. See text for related parameters and further explanations.
}
\begin{tabular}{cc|cc}
\hline \hline
\multicolumn{4}{c}{$|M(\mathbf{k}_1\Uparrow_\mathbf{n},\mathbf{k}_2\Downarrow_\mathbf{n})|^2$\;= $\left\{\begin{array}{ccc} \left(\displaystyle\frac{|\eta|}{\Delta_C}\right)^2 D^2_{\lambda} S_{\mathbf{n}}(\mathbf{q}) I_{\lambda}(\mathbf{q})\!\!\!\!&,\!\! &\rm{intra} \\
D^2_{gs} S_\mathbf{n}(\mathbf{K})&,\!\!&g\end{array}\right.$}
\\ \hline\hline
$\mathbf{n}$ & $S_\mathbf{n}(\mathbf{q})$ & $\lambda$ & $I_{\lambda}(\mathbf{q})$ \\ \hline
$[0\;0\;1]$ & $q^2_x+q^2_y$ & TA & $\displaystyle \frac{(q^2_x-q^2_y)^2}{q^2_x+q^2_y}+\frac{4q^2_xq^2_yq^2_z}{(q^2_x+q^2_y)|\mathbf{q}|^2}$ \\
$[1\;0\;0]$ & $q^2_y$ & LA & $\displaystyle \frac{4 q^2_xq^2_y}{|\mathbf{q}|^2}$ \\
$[1\;1\;0]$ & $\displaystyle \frac{ q^2_x+q^2_y}{2}$ & TO & $\displaystyle \frac{q^2_x+q^2_y}{|\mathbf{q}|^2}$\\
$[1\;0\;1]$ & $\displaystyle \frac{ q^2_x}{2}+q^2_y $ & LO & $\displaystyle \frac{q^2_z}{|\mathbf{q}|^2}$\\
$[1\;1\;1]$ & $\displaystyle \frac{ 2}{3}(q^2_x+q^2_y+q_x q_y) $ && \\
\hline \hline
\end{tabular}
\end{table}

Implications of the above considerations are manifested in the spin-flip expressions of intravalley and $g$-process scattering. The matrix element of intravalley spin flips will be shown to consist of four factors that represent different aspects of the above considerations
\begin{eqnarray}
|M_{\lambda}^{\rm{intra}}(\mathbf{k}_1\Uparrow_\mathbf{n},\mathbf{k}_2\Downarrow_\mathbf{n})|^2 = \left(\frac{|\eta|}{\Delta_C}\right)^2 \!\! D^2_{\lambda} S_{\mathbf{n}}(\mathbf{q}) I_{\lambda}(\mathbf{q}).\,\,\,\,\,\, \nonumber 
\end{eqnarray}
The factor $|\eta|/\Delta_C$$\,$$\approx$$\,$0.03~$\AA$ is calculated from band structure parameters that originate from spin-dependent $\mathbf{k}\!\cdot\!\mathbf{p}$ perturbation terms in the Hamiltonian. The second factor, $D_{\lambda}$, depends on the phonon mode. Scattering with long-wavelength acoustic modes is governed by a deformation potential constant, $D_{TA/LA}=D'_{xy}$, that couples the two lowest conduction bands. Its explicit integral expression will be given in Eq.~(\ref{eq:D_xy}) and its numerical solution (via EPM calculation) yields a value of 6~eV. This value agrees well with the measured energy splitting, $4D'_{xy}\epsilon_{xy}$, of the lowest conduction bands at the $X$ point when applying a shear strain.\cite{Hensel_PR65,Laude_PRB71} Scattering with long-wavelength optical modes is governed by the constant $D_{TO/LO}=D_{\rm{op}}\approx 5~\text{eV}\cdot 2\pi/a$ which originates from the out-of-phase motion of atoms in the primitive cell. This parameter is also calculated from the coupling between the lowest pair of conduction bands [Eq.~(\ref{eq:Dop})]. The remaining two factors in the above expression depend on the valley position. Their forms for electrons that reside in the $z$ valley are listed in Table~\ref{tab:intra_g_M_Sn}. $I_{\lambda}$ and $S_\mathbf{n}$ reflects the dependence on phonon properties and on the electron spin orientation, respectively. The right column of the table lists the values of $I_{\lambda}$ for all long-wavelength phonon modes and the left column lists the values of $S_\mathbf{n}$ for five inequivalent high-symmetry directions in the $z$~valley.

The $g$-process spin-flip will be shown to share several properties with the intravalley case. For electron transition between the $\pm z$ valleys, the dominant $g$-process spin-flip mechanism will be shown to originate from scattering with LA phonon modes,
\begin{eqnarray}
|M_{LA}^{g}(\mathbf{k}_1\Uparrow_\mathbf{n},\mathbf{k}_2\Downarrow_\mathbf{n})|^2 = D^2_{gs} S_\mathbf{n}(\mathbf{K}),\nonumber
\end{eqnarray}
where its prefactor $D_{gs} \approx 0.1 \rm{eV}$ is dominated by the sum of two scattering constants. The first is related to the deformation potential of the lowest conduction band [Eqs.~(\ref{eq:mom_matrix_element})-(\ref{eq:Djk})] and the second couples the lowest conduction band with upper valence band by spin-orbit modulated electron-phonon interaction [Eq.~(\ref{eq:D_so})]. 

The rest of this section is organized in the following logical order. In Sec.~\ref{sec:kp hamilton} we derive selection rules that pertain to both the band structure Hamiltonian and the electron-phonon interaction. The obtained electron state vectors are used in Secs.~\ref{sec:intravalley} and \ref{sec:g-process} where we present the core derivation for intravalley and $g$-process spin-flip matrix elements. In these parts we derive exact forms of the various scattering constants.  Finally, the dependence on  spin orientation ($\mathbf{n}$) is given in Sec.~\ref{sec:valley-spin dependence}.
Readers who are not interested in the full derivation may directly skip to the most general expressions in Eqs.~(\ref{eq:intra_AC})-(\ref{eq:Dop}) and Fig.~\ref{fig:intra_M_q} for intravalley spin-flips in the $z$ valley with $\mathbf{n}$$\,$$\|$$\,$$\mathbf{z}$; in Eqs.~(\ref{eq:Mg})-(\ref{eq:Dgs}) and Fig.~\ref{fig:g_M} for $g$-process spin-flips between the $\pm z$ valleys with $\mathbf{n}$$\,$$\|$$\,$$\mathbf{z}$; and in Eqs.~(\ref{eq:ac_Sn})-(\ref{eq:g_Sn}) for general spin orientations ($\hat{\mathbf{n}}=[\cos{\phi}\sin{\theta}, \sin{\phi}\sin{\theta},\cos{\theta}]$).

\subsection{$X$-point selection rules and spin-dependent eigenstates of the $\mathbf{k}\cdot \mathbf{p}$ Hamiltonian \label{sec:kp hamilton}}

\begin{figure}
\includegraphics[width=8cm, height=4.5cm]{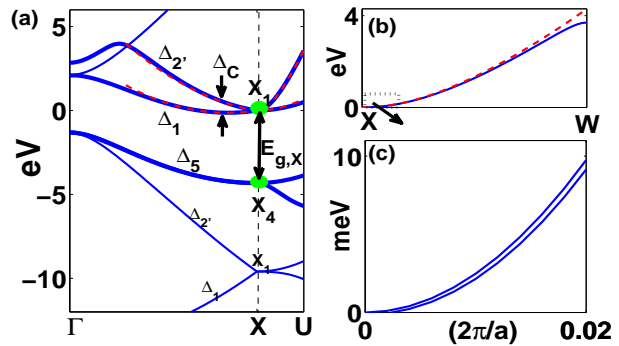}
\caption{(Color online) (a) Energy band structure along the $\Delta$-axis between the $X=(0,0,1)2\pi/a$ and $\Gamma=(0,0,0)$ points and along the $\Sigma$-axis between $X$ and $U=(1/4,1/4,1)2\pi/a$ points [see Fig.~\ref{fig:brillouin}(a)]. Solid lines are the results of EPM and dash lines are taken from Eq.~(\ref{eq:eigen_energy}) of the Appendix. $E_{X_1}-E_{X_4}=E_{g,\scriptscriptstyle{X}}$ and we set $E_{X_1}=0$. $\Delta_C$ is the energy gap between the two conduction bands at the valley center $\mathbf{k}_0\simeq (0,0,0.85)2\pi/a$. (b) Conduction band dispersion along the $Z$-symmetry axis between $X$ and $W=(0,1/2,1)2\pi/a$ points. Along this direction the two-band conduction degeneracy can only be lifted by spin-orbit coupling (spin hot-spot). (c) The induced energy splitting by zooming-in closer to the $X$ point.} \label{fig:band_structure}
\end{figure}

In this part we present group theory results and analytically quantify the signature of spin-orbit coupling on electronic states. Since the bottom of the conduction band is at the vicinity of the $X$ point [see Fig.~\ref{fig:band_structure}(a)], we employ a compact $\mathbf{k}\cdot \mathbf{p}$ model with a small set of basis states that pertain to the symmetry of this point. Findings of this model will be used to derive spin-flip matrix elements. These findings will also be benchmarked against numerical calculations of an EPM that includes spin-orbit coupling.\cite{Chelikowsky_PRB76}

For intravalley scattering, working with $X$-point space group is as effective as working with the group of the $\Delta$ axis (position of the valley center), and bares favorable features over the latter choice.\cite{footnote_Xpoint_choice} The initial and final states of a $g$-process scattering can be expanded to the same $X$ point, making it possible to relate the matrix element with a deformation potential parameter, though with modifications (to be discussed).

We first mention spin independent features.\cite{Hensel_PR65} Due to the symmetry of the crystal, only one of the six conduction band valleys is studied and we arbitrary identify it as the valley along the $+z$ crystallographic axis. We choose a basis of four eigenstates of the Hamiltonian $H_0=p^2/2m_0 + V(r)$ and denote them by $\{|X_1^{2'}\rangle, |X_1^1\rangle, |X_4^x\rangle, |X_4^y\rangle\}$, where $\psi_{X}(\mathbf{r})=\langle \mathbf{r}| X \rangle$.  These $X$-point states are associated with the lowest pair of conduction bands and upper pair of valence bands at $\mathbf{k}=(0,0,k_X)$, where $k_X=2\pi/a$ [see Fig.~\ref{fig:band_structure}(a)]. Inclusion of the $X_4$ valence states is imperative since they bring the mass anisotropy of conduction electrons.  The $X_1$ and $X_4$ nomenclature denotes 2D IRs of the space group $G^2_{32}$ which describes the symmetry of the $X$ point.\cite{Hensel_PR65,Jones1960,Dresselhaus_PR67,Yu_Cardona_Book,Bradley_Cracknell2010,Dresselhauses_Jorio_Book} The dimensionality complies with the two-band degeneracies at the $X$ point of diamond crystal structures (due to time reversal and glide reflection symmetries). The indexing of the basis states implies of their compatibility relations when going from the $X$ point into the $\Delta$ axis ($\{|\Delta_{2'}\rangle, |\Delta_1\rangle, |\Delta^x_5\rangle, |\Delta^y_5\rangle\}$).\cite{Hensel_PR65} For example, $\psi_{\Delta_{2'},k_z}(\mathbf{r})\simeq e^{i (k_z-k_X)z} \psi_{X_1^{2'}}(\mathbf{r})$ . Finally,  we mention that for the following derivations, the symmetry properties rather than explicit functional forms of the basis states are important.

When adding the spin degree of freedom the new basis set reads $\{$ $|X_1^{2'},\uparrow\rangle$, $|X_1^{2'},\downarrow\rangle$, $|X_1^1,\uparrow\rangle$, $|X_1^1,\downarrow\rangle$, $|X_4^x,\uparrow\rangle$, $|X_4^x,\downarrow\rangle$, $|X_4^y,\uparrow\rangle$, $|X_4^y,\downarrow\rangle$ $\}$. Two remarks on this basis are in place. First, inclusion of the $X_4$ states in the basis set enables us to capture the salient spin-dependent properties of conduction electrons (originate from spin-orbit coupling with the upper pair of valence bands).\cite{Pengke_PRL11} Second, the spin-orbit coupling is treated as a perturbation and, accordingly, the new basis states are not eigenstates of the spin-dependent Hamiltonian at the $X$ point.  In Eqs.~(\ref{eq:spin_up_state})-(\ref{eq:state_structure}) of this section we find specific forms of these states for electrons around the bottom of the $+z$ conduction valley for spin orientation  $\mathbf{n}\|\mathbf{z}$ [see Eq.~(\ref{eq:spin_up_down_states})]. In sections~\ref{sec:g-process} we discuss the needed changes when considering electron states from opposite valleys. In Section~\ref{sec:valley-spin dependence} we show the anisotropy of spin-flip processes by varying $\mathbf{n}$.

Group theory is invoked to construct the band structure Hamiltonian matrix and matrix elements of electron-phonon interactions.\cite{Bir_Pikus_Book,Dresselhauses_Jorio_Book,Winkler_Book,Winkler_PRB10} Both depend on couplings of $X_1$ basis states, of $X_4$ basis states, and of $X_1$ with $X_4$ basis states. Using the character table of $G^2_{32}$ (Appendix~\ref{app:G32}), these couplings are clarified by the following decompositions\cite{footnote_realIR},
\begin{subequations}\label{eq:XX}
\begin{eqnarray}
H_{cc}:\,\,\,\,\, X_1\otimes X_1 &=& M_1 \oplus M_4 \oplus M_2' \oplus M_3' \,, \label{eq:X1X1}\\
H_{\upsilon\upsilon}:\,\,\,\,\, X_4\otimes X_4 &=& M_1\oplus M_4 \oplus M_1' \oplus M_4' \,,\label{eq:X4X4} \\
H_{\upsilon c}:\,\,\,\,\, X_4\otimes X_1&=& M_5\oplus M_5'\,\,. \label{eq:X1X4}
\end{eqnarray}
\end{subequations}
$M_5$ and $M_5'$ are 2D IRs and the remaining $M_i$ are 1D IRs.  Next we relate these decomposition with specific linear combination of state products.\cite{Bir_Pikus_Book}  For example, one of the $X_1$ states coupling [Eq.~(\ref{eq:X1X1})], $\psi^*_{X^{2'}_1}(\mathbf{r}) \psi_{X^1_1}(\mathbf{r}) + \psi_{X^{2'}_1}(\mathbf{r}) \psi^*_{X^1_1}(\mathbf{r})$ is found to transform as $M_4$ (apply the operations of $G^2_{32}$ on it). We associate this linear combination with $[0,1;1,0]$, or $\rho_x$, in $H_{cc}$ block (In this paper, $\rho_i$ and $\sigma_i$ identify, respectively, Pauli matrices in the 2D IRs product and spin space). We find the following associations for product combinations in intraband coupling,
\begin{subequations}\label{eq:rep}
\begin{eqnarray}
H_{cc}\!\!&:&\,M_1  \! \leftrightarrow \! \mathcal{I}\,,        \,\,   M_4 \!\leftrightarrow  \! \mathcal{\rho}_x \,, \,\,   M_2'\! \leftrightarrow \!\mathcal{\rho}_y\,,     \, \, M_3' \! \leftrightarrow \!\mathcal{\rho}_z\,,\,\, \label{eq:rep_cc} \\
H_{\upsilon\upsilon}\!\!&:&\,M_1  \! \leftrightarrow \! \mathcal{I}\,,  \,\,M_4 \!\leftrightarrow  \!\mathcal{\rho}_x \,, \,\,   M_1' \! \leftrightarrow \! \mathcal{\rho}_y \,, \,\,     M_4' \!\leftrightarrow \!\mathcal{\rho}_z \,. \,\,    \quad \label{eq:rep_vv}
\end{eqnarray}
In case of interband coupling where the decomposed IRs ($M_5$ and $M_5'$) are 2D, we classify pairs of product combinations. For example, $\{\psi^*_{X^x_4} \psi_{X^{2'}_1}+ \psi^*_{X^y_4} \psi_{X^1_1}, \psi^*_{X^x_4} \psi_{X^1_1}+ \psi^*_{X^y_4} \psi_{X^{2'}_1}\}$ belongs to $M_5'$. We find
\begin{equation}
H_{\upsilon c}:\,\,\,\,\,M_5  \leftrightarrow  \{\mathcal{\rho}_y\,, \mathcal{\rho}_z\} \,, \,\,\,\, M_5' \leftrightarrow \{\mathcal{I}\,, \mathcal{\rho}_x\} \,. \label{eq:rep_2D}
\end{equation}
\end{subequations}

Non-vanishing matrix elements of any interaction (i.e., selection rules) are readily identified once we specify how terms in that interaction transform under symmetry operations. These terms usually can be made into parts that transform as components of vectors ($\mathbf{r}$)  or axial vectors ($\mathbf{R}$). The longitudinal component of the vector (axial vector) transforms as $M_3'$ ($M_3$), and transverse components as $M_5$ ($M'_5$):
\begin{subequations}\label{eq:operator_IR}
\begin{equation}
\!z\sim M'_3,  \,\,  \{x,y\}\sim  M_5,  \,\, R_z\sim M_3,  \,\, \{R_x,R_y\}\sim  M'_5\,.\!\! \!\! \! \label{eq:vectorIR}
\end{equation}
where the longitudinal component is along the valley axis ($z$ in our choice). Besides, the symmetric and antisymmetric potentials [Eq.~(\ref{eq:Hep1})] are even and odd under space inversion and transform as
\begin{equation}
\mathcal{V}_+\sim M_1,  \qquad  \mathcal{V}_-\sim  M'_2. \label{eq:potentialIR}
\end{equation}
\end{subequations}
By the fundamental theorem of group theory, a non-vanishing matrix element results when the interaction operator and the states product decomposition belong to the same\cite{footnote_realIR} IR(s). The integrand of the non-vanishing matrix element as a whole then belongs to the identity IR (in this case, $M_1$).  From Eqs.~(\ref{eq:X1X4}) and (\ref{eq:vectorIR}), the transverse components of a vector ($M_5$) or axial vector ($M'_5$) can couple conduction band with valence band. However, since $M_1$ only appears once in either $M_5\otimes M_5$ or $M_5'\otimes M_5'$, only one of the linear combinations of the operator and state product can survive and the unique linear combination read
\begin{equation}
ix\rho_y - y \rho_z\sim M_1\,\,\, \text{and}\,\,\, R_x \mathcal{I}-R_y \rho_x\sim M_1\,, \label{eq:possible_couplings}
\end{equation}
for $M_5\otimes M_5$ or $M_5'\otimes M_5'$, respectively. The IR product decompositions and the IR assignments with (axial) vector are derived from $G^2_{32}$ character table (see Appendix~\ref{app:G32}).

Transformation under time reversal operation ($\mathcal{T}$) is equally essential in obtaining selection rules (in addition to spatial operations). Time reversal symmetry connects matrix elements by,\cite{Bir_Pikus_Book}
\begin{subequations}\label{eq:TR}
\begin{eqnarray}
\langle X_j | \mathcal{O} |X_i \rangle = \langle \mathcal{T}X_i| \mathcal{T}\mathcal{O}^\dag\mathcal{T}^{-1}|\mathcal{T}X_j \rangle \;.\label{eq:time_reversal}
\end{eqnarray}
Physical operators $\mathcal{O}$ have different parities under time reversal operation,
\begin{equation}
\mathbf{r}\rightarrow \mathbf{r},\; \bm\nabla V \!\rightarrow \! \bm\nabla V,\; \mathbf{p}\rightarrow -\mathbf{p}, \;\bm\nabla V \!\times \mathbf{p}\rightarrow-\bm\nabla V \!\times \mathbf{p}, \!\! \label{eq:vectorTR}
\end{equation}
and this feature distinguishes between vectors such as $\bm\nabla{V}$ and $\mathbf{p}$.  $\mathcal{T}$ acting on $|X_i \rangle $, by our convention, exchanges the basis states,\cite{Hensel_PR65}
\begin{equation}
|{X^{2'}_1}\rangle \leftrightarrow |X^1_1\rangle,\quad |X^x_4\rangle \leftrightarrow | X^y_4\rangle. \label{eq:X_TR}
\end{equation}
Therefore, for the diagonal blocks ($H_{cc}$ and $H_{vv}$)
\begin{equation}
\mathcal{I} \rightarrow \mathcal{I}, \quad\rho_x\rightarrow\rho_x, \quad\rho_y\rightarrow \rho_y, \quad \rho_z\rightarrow -\rho_z \label{eq:XX_TR}
\end{equation}
\end{subequations}
under time reversal operation in light of Eq.~(\ref{eq:time_reversal}). A matrix element vanishes if under time reversal operation the changes of sign are opposite for the state product [Eq.~(\ref{eq:XX_TR})] and the interaction [Eq.~(\ref{eq:vectorTR})].

The spin-dependent $\mathbf{k}\cdot\mathbf{p}$ Hamiltonian matrix can be prescribed using the above group theory analysis, with perturbation Hamiltonian ($H=H_0+H_1$)
\begin{eqnarray}
H_1\!=\! \left(\!\frac{\hbar^2 k'^2}{2 m_0}\! + \! \frac{\hbar \mathbf{k}'\cdot \mathbf{p}}{m_0}\!\right)\mathcal{I}\!+ \frac{\hbar[\bm{\nabla}V(\mathbf{r})\times(\mathbf{p}+\hbar \mathbf{k}')]\cdot \bm{\mathcal{\sigma}}}{4m_0^2c^2}\;,\,\,\,\label{eq:kp_perturb}
\end{eqnarray}
where $\mathbf{k}'=(k_x,k_y,k_z-k_X)$ is measured from the $X$ point and $V(\mathbf{r})$ is the crystal potential $\left[\sum_j \mathcal{V}_+(\mathbf{r}-\mathbf{R}_j)\right]$. The Hamiltonian matrix blocks are thus, respectively
\begin{subequations}\label{eq:H_block}
\begin{eqnarray}
H_{1cc} &=& \frac{\hbar^2}{2m_0} (k'^2 \mathcal{I}\otimes \mathcal{I} + 2 k'_0 k'_z \mathcal{\rho}_z \otimes \mathcal{I}),\label{eq:Hcc}\\
H_{1v v} &=& \frac{\hbar^2}{2m_0} k'^2 \mathcal{I}\otimes \mathcal{I},\label{eq:Hvv}\\
H_{1v c} &=& -i P(k_x \mathcal{\rho}_y +i k_y \mathcal{\rho}_z)\otimes\mathcal{I} + i \Delta_X ( \mathcal{\rho}_x \otimes \mathcal{\sigma}_y \nonumber\\&&
- \mathcal{I}\otimes \mathcal{\sigma}_x  ) +\alpha[k'_z(i \mathcal{\rho}_z\otimes\mathcal{\sigma}_x -\mathcal{\rho}_y\otimes\mathcal{\sigma}_y)\nonumber\\&&+(k_y \mathcal{\rho}_y-i k_x \mathcal{\rho}_z)\otimes \mathcal{\sigma}_z ]\label{eq:Hvc}\,.
\end{eqnarray}
\end{subequations}
We stress that $\partial V(\mathbf{r})/\partial z$ in $H_{1cc}$ block vanishes considering time reversal symmetry.The spin-independent constants in Eq.~(\ref{eq:H_block}) are empirically known from experiments.\cite{Hensel_PR65} They relate to the position of the conduction band minima $\mathbf{k}_0=(0,0,k'_0+k_X)$, and to the momentum matrix element that sets the mass anisotropy ($P$). The spin-dependent parameters ($\Delta_X$ and $\alpha$) are calculated by the empirical pseudopotential model. Table~\ref{tab:params} lists these and other parameters that we use in this paper. In Appendix~\ref{app:partitioning}, we provide their integral expressions and we also apply a partitioning technique to analytically diagonalize the total Hamiltonian matrix
\begin{eqnarray}
\left(
\begin{array}{cc}
H_{1cc}&H^\dagger_{1vc}\\H_{1vc} & H_{1vv}\!-\! E_{g,\scriptscriptstyle{X}}
\end{array}\!
\right)  ,
\label{eq:8 by 8}
\end{eqnarray}
where $E_{g,\scriptscriptstyle{X}}$ is the energy-gap between the conduction and valence band at the $X$ point [see Fig.~\ref{fig:band_structure}(a)].

\begin{table}
\caption{\label{tab:params}
Parameters of bulk silicon. The lattice constant is $a$=5.43~$\AA$, the $X$-point energy gap is $E_{g,\scriptscriptstyle{X}} \approx 4.3$~eV, and the free electron mass is $m_0$.}
\renewcommand{\arraystretch}{1.2}
\tabcolsep=0.2 cm
\begin{tabular}{c|ll|l}
\hline \hline 
 $k'_0$ & -0.15 & $2\pi/a$ &   Eq.~(\ref{eq:k0}) \\
 $P$   & 10 & eV$\cdot$$a/2\pi$ &  Eq.~(\ref{eq:P}) \\
 $\Delta_X$ & 3.6 & meV & Eq.~(\ref{eq:Delta_X}) \\
 $\alpha$   & -3.1 & meV$\cdot$$a/2\pi$ &  Eq.~(\ref{eq:alpha}) \\ \hline
 $\Delta_C$ & 0.5 & eV &  $2\hbar^2k'^2_0/m_0$ \\
 $|\eta|$ & 16.7 & meV$\cdot$$a/2\pi$ & $\eta=2iP\Delta_X/E_{g,\scriptscriptstyle{X}}$ \\
 $\Delta_X'$ & 4.1 & meV & $\Delta_X + \alpha k'_0$ \\
 $|\eta'|$ & 18.9 & meV$\cdot$$a/2\pi$ & $\eta'=2iP\Delta_X'/E_{g,\scriptscriptstyle{X}}$ \\ \hline \hline
\end{tabular}
\end{table}

Appendix~\ref{app:spin_align} provides a general procedure to align the resulting degenerate states to spin states that satisfy Eq.~(\ref{eq:spin_up_down_states}). For spin orientation along the valley axis ($\mathbf{n}\|\mathbf{z}$), degenerate states in a conduction band valley read
\begin{eqnarray}
| \mathbf{k}, \Uparrow   \rangle &=& \exp{(i\mathbf{k}'\cdot\mathbf{r})}\left[ \mathbf{A}(\mathbf{k})|\uparrow\rangle + \; \mathbf{B}(\mathbf{k})|\downarrow \rangle\right]|\mathbf{X}\rangle \,, \nonumber \\
| \mathbf{k}, \Downarrow \rangle &=& \exp{(i\mathbf{k}'\cdot\mathbf{r})}\left[ \mathbf{A}^{\!\!\ast}\!(\mathbf{k})|\downarrow \rangle - \mathbf{B}^{\ast}\!(\mathbf{k})|\uparrow\rangle\right]|\mathbf{X}\rangle \,  \label{eq:spin_up_state}
\end{eqnarray}
where
\begin{eqnarray}
|\mathbf{X}\rangle \!\! &=& \!\! \left[ |X_1^{2'}\rangle, |X_1^{1}\rangle, |X_4^x\rangle, |X_4^y\rangle \right]^T   \nonumber \\
\mathbf{A}(\mathbf{k}) \!\! &=& \!\!  \left[ \frac{2P^2k_xk_y}{E_{g,\scriptscriptstyle{X}}\Delta_C},\,\, 1 - \frac{P^2k_xk_y}{2E_{g,\scriptscriptstyle{X}}^2},\,\,  -\frac{Pk_x}{E_{g,\scriptscriptstyle{X}}},\,\,-\frac{Pk_y}{E_{g,\scriptscriptstyle{X}}} \right] \nonumber \\
\mathbf{B}(\mathbf{k}) \!\! &=& \!\!  \bigg[ \frac{ \eta (k_x\!-\!ik_y)}{\Delta_C} \!\left( \! 1 \!-\!\frac{k_z-k_0}{k'_0}\right),\,\, \frac{i\eta'(k_x\!+\!ik_y)}{2E_{g,\scriptscriptstyle{X}}}, \nonumber \\ && \,\,\! -\!\frac{\Delta_X'}{E_{g,\scriptscriptstyle{X}}}\! -\!\!\frac{i\eta k_y}{\Delta_C}\frac{Pk_y}{E_{g,\scriptscriptstyle{X}}},\,\, - \frac{i\Delta_X'}{E_{g,\scriptscriptstyle{X}}} \!+\!\frac{\eta k_x}{\Delta_C}\frac{Pk_x}{E_{g,\scriptscriptstyle{X}}} \bigg].  \label{eq:state_structure}
\end{eqnarray}
As shown in Appendix.~\ref{app:momentum}, momentum scattering is governed only by the zeroth- and first-order wavevector components of $\mathbf{A}(\mathbf{k})$. For spin relaxation, on the other hand, the `negligible' spin-orbit coupling coefficients become crucial [$\mathbf{B}(\mathbf{k})$ vector]. In writing the coefficients of $\mathbf{A}(\mathbf{k})$ and $\mathbf{B}(\mathbf{k})$, we have kept only those terms that are relevant to the spin relaxation analysis of the following sections. Terms that scale with  $\Delta_C/(2 E_{\scriptscriptstyle{g,X}})\approx 1/17$ are omitted due to their negligible effect. The energy gap between the two conduction bands at the valley center $\Delta_C \approx 0.5$~eV [see Fig.~\ref{fig:band_structure}(a)] will be extensively used when dealing with intravalley spin relaxation. The main difference of the above solution from Ref.~[\onlinecite{Pengke_PRL11}] is that we have identified all quadratic-wavevector terms that play important roles in intravalley spin-flip processes. We have checked the accuracy of these expressions with numerical EPM solutions.\cite{footnote_check_state} The main difference between them comes from the omission of the lower valence band states [lower $X_1$ point in Fig.~\ref{fig:band_structure}(a)]. These states are responsible for making the longitudinal mass slightly less than $m_0$. In spite of missing some $k_z$ dependent components, the error amounts to a few percent and is irrelevant in \textit{intravalley} spin relaxation (to be shown later).

\subsection{Intravalley spin flips}\label{sec:intravalley}
In this subsection we present the theoretical procedure for deriving intravalley spin-flip matrix elements for {\it all} types of phonon modes. Scattering with long-wavelength TA phonon modes will be shown to dominate the intravalley spin relaxation at low temperature. This property is largely set by the phonon dispersion around the $\Gamma$ point. Figure~\ref{fig:BZ_phonon}(a) shows that the energy of long-wavelength acoustic phonons is linear in $\mathbf{q}$ with a smaller slope for TA, and approaches zero at $\Gamma$ point. For long-wavelength optical modes, the phonon energy is almost wavevector independent ($\approx$63~meV). However, when the temperature increases, optical modes become more important and together with the LA mode they provide about half of the intravalley spin relaxation at room temperature. The reason is twofold. Spin-flip matrix elements due to scattering with optical modes are linear in $\mathbf{q}$ while being quadratic with acoustic modes (to be proven below). Second, the rise of phonon population with temperature is much faster for optical modes.

In Appendix \ref{app:momentum}, by invoking selection rules we have rederived the leading-order matrix elements of intravalley $momentum$ scattering, which is typically described in terms of deformation potential constants.\cite{Bardeen_PR50,Herring_PR56} When extended to analyze the intravalley spin-flip matrix elements, this group theory approach will be appreciated by its ability to reach essential terms efficiently. It also provides flexibility in the sense that no \textit{a priori} knowledge of the connection between deformation potential quantities and spin-flip processes is needed.

\subsubsection*{long-wavelength acoustic phonon modes}\label{sec:intra_ac}
Spin-flip matrix elements due to electron interaction with long-wavelength acoustic phonon modes have a quadratic-wavevector dependence. We show it by starting with the most general form,
\begin{eqnarray}
&& M_{\lambda}\left(\mathbf{k}_1=\mathbf{K}-\tfrac{\mathbf{q}}{2},\Uparrow \;\; ;\;\;  \mathbf{k}_2=\mathbf{K}+\tfrac{\mathbf{q}}{2},\Downarrow \right) = \label{eq:spin_flip_general}\\
&& \langle \mathbf{K}\!+\!\tfrac{\mathbf{q}}{2},\Downarrow | \bm\xi^+_{\lambda}(\mathbf{q})\cdot \bm{\nabla} \!\mathcal{V}_{+}(\mathbf{r}) + \bm\xi^-_{\lambda}(\mathbf{q})\cdot\bm{\nabla} \!\mathcal{V}_{-}(\mathbf{r})    | \mathbf{K}\!-\!\tfrac{\mathbf{q}}{2},\Uparrow \rangle \nonumber \\
&& \equiv \bcancel{M}_{{\rm{sf}},\lambda}^{(0)} + \bcancel{M}_{{\rm{sf}},\lambda}^{(1)} + M_{{\rm{sf}},\lambda}^{(2)} + \mathcal{O}(q^3)\;.\nonumber
\end{eqnarray}
$M_{{\rm{sf}},\lambda}^{(j)}$ denotes terms of $j^{th}$ order in $\mathbf{q}$ due to $\lambda$ phonon mode. In the next step we expand the states in Eq.~(\ref{eq:spin_flip_general}) in increasing orders of $\mathbf{q}$. We show how the zeroth- and first-order terms vanish (${M}_{{\rm{sf}},\lambda}^{(0)} = M_{{\rm{sf}},\lambda}^{(1)} = 0$), and then we find the form of the dominant quadratic terms ($M_{{\rm{sf}}}^{(2)}$). In this wavevector-order analysis, the $\mathbf{q}$ dependence of the in-phase and out-of-phase polarization vectors has to be taken into account. In Table~\ref{tab:elastic_continuum} we showed these dependencies in the long-wavelength regime which is relevant for intravalley scattering ($\mathbf{q} \ll 2\pi/a$). The out-of-phase vector, $\bm\xi^-_{\lambda}(\mathbf{q})$, is linear in $\mathbf{q}$ while $\bm\xi^+_{\lambda}(\mathbf{q})$ has a zeroth-order dependence (e.g., $q_i/q$ terms). Hereafter, we abbreviate the notation $\bm\xi^{\pm}_{\lambda}(\mathbf{q})$ and $M_{{\rm{sf}},\lambda}^{(j)}$ as $\bm\xi^{\pm}$ and $M_{{\rm{sf}}}^{(j)}$.

\textbf{Zeroth-order}: The lowest-order spin-flip matrix element has the form
\begin{eqnarray}
M_{{\rm{sf}}}^{(0) } = \langle \mathbf{K},\Downarrow | \bm{\xi}^+ \cdot \bm{\nabla} \!\mathcal{V}_{+} | \mathbf{K},\Uparrow \rangle .
\end{eqnarray}
We exclude the out-of-phase part since the acoustic interaction at the infinite wavelength limit is governed solely by the in-phase motion of atoms in the primitive cell [$\bm{\xi}^-_{\rm{ac}}(\mathbf{q}=0)=0$ and $|\bm{\xi}^+_{\rm{ac}}(\mathbf{q}=0)|=1$]. In this limit, the phonon-induced interaction reduces to displacement of the entire crystal, $\sum_{j} \bm\nabla_{\mathbf{r}} \mathcal{V}_+(\mathbf{r}-\mathbf{R}_j)=  \bm\nabla \mathcal{V}_{\rm{crystal}}$. Based on the relation
\begin{eqnarray}
\bm{\nabla}\mathcal{V}_{\rm{crystal}}= i[\mathbf{p},H]/\hbar,\label{eq:commutator}
\end{eqnarray}
the coupling of the in-phase part between spin-degenerate eigenstates of $H$ vanishes. When breaking this vanishing matrix element into parts that come from interaction with the bare symmetrical potential and with its spin-orbit coupling part,
\begin{eqnarray}
\langle \mathbf{K},\Downarrow |  \bm{\nabla} \!V_+ | \mathbf{K},\Uparrow \rangle +  \langle \mathbf{K},\Downarrow |  \bm{\nabla} \!V^{\rm{so}}_{+} | \mathbf{K},\Uparrow  \rangle = 0 \;,\label{eq:EYcancel}
\end{eqnarray}
the sum is zero but each of these two contributions is finite. This result was first pointed out by Elliott.\cite{Elliott_PR54a} Later we will make use of this property.

\textbf{First-order}: To write the spin-flip matrix elements of this order we begin by linearizing the state,
\begin{eqnarray}
| \mathbf{K}\pm \tfrac{\mathbf{q}}{2},s \rangle =   | \mathbf{K},s \rangle   \pm  \tfrac{1}{2}\mathbf{q}\cdot \bm{\mathcal{L}} | \mathbf{K},s \rangle + \mathcal{O}(q^2)\,, \label{eq:stat_gen}
\end{eqnarray}
where $\bm{\mathcal{L}}$ denotes the derivative of the state in $\mathbf{k}$-space,
\begin{eqnarray}
\mathcal{L}_i | \mathbf{k},s \rangle  \equiv  \underset{\delta \mathbf{k} \rightarrow 0}{\text{lim}} \frac{| \mathbf{k}+\delta k_i,s \rangle - | \mathbf{k},s \rangle}{\delta k_i} \;. \label{eq:Lop}
\end{eqnarray}
Substituting Eq.~(\ref{eq:stat_gen}) into Eq.~(\ref{eq:spin_flip_general}) and considering the wavevector dependence of $\bm\xi^{\pm}$, the first-order spin-flip matrix elements are,
\begin{eqnarray}
M_{{\rm{sf}}}^{(1) } &=&  \langle \mathbf{K},\Downarrow | \bm\xi^-\cdot\bm{\nabla} \!\mathcal{V}_{-} | \mathbf{K},\Uparrow \rangle \label{eq:first_order_sf} \\
&+& \tfrac{1}{2}\langle \mathbf{K},\Downarrow | \left(\mathbf{q}\cdot \bm{\mathcal{L}}^{\dagger}\right) \left(\bm\xi^+\cdot\bm{\nabla} \!\mathcal{V}_{+} \right)  | \mathbf{K},\Uparrow \rangle \nonumber \\
&-& \tfrac{1}{2}\langle \mathbf{K},\Downarrow | \left(\bm\xi^+\cdot\bm{\nabla} \!\mathcal{V}_{+} \right) \left(\mathbf{q}\cdot \bm{\mathcal{L}}\right) | \mathbf{K},\Uparrow \rangle \;. \nonumber
\end{eqnarray}
Combining space inversion and time-reversal symmetries, the out-of-phase term can be shown to vanish (first line on the right-hand side), and the in-phase terms to cancel each other (second and third lines). These facts follow the relations
\begin{eqnarray}
\langle \psi_1 , \mathcal{O}\psi_2 \rangle = \langle \mathcal{S}\psi_1 , \mathcal{S}\mathcal{O}\mathcal{S}^{-1}\mathcal{S}\psi_2 \rangle = \langle \mathcal{T}\psi_2 , \mathcal{T}\mathcal{O}^\dag\mathcal{T}^{-1}\mathcal{T}\psi_1 \rangle \;,  \nonumber
\end{eqnarray}
where for the case in hand space inversion provides
\begin{eqnarray}
\!\!\mathcal{S}| \mathbf{k},\Uparrow \rangle =| \mbox{-}\mathbf{k},\Uparrow   \rangle    \;&,&\; \;  \mathcal{S}| \mathbf{k},\Downarrow \rangle =| \mbox{-}\mathbf{k},\Downarrow \rangle     \nonumber \\
\!\!\mathcal{S}\bm{\nabla} \!\mathcal{V}_{\pm} \mathcal{S}^{-1} = \mp \bm{\nabla} \!\mathcal{V}_{\pm} \;&,&\;\;   \mathcal{S} \bm{\mathcal{L}} \mathcal{S}^{-1} = -\bm{\mathcal{L}} \;,  \label{eq:SI_Prop}
\end{eqnarray}
and time reversal provides
\begin{eqnarray}
\!\!\mathcal{T}| \mathbf{k},\Uparrow \rangle =| \mbox{-}\mathbf{k},\Downarrow \rangle    \;&,&\;\;   \mathcal{T}| \mathbf{k},\Downarrow \rangle =-| \mbox{-}\mathbf{k},\Uparrow \rangle      \nonumber \\
\!\!\mathcal{T}\bm{\nabla} \!\mathcal{V}_{\pm}^{\dagger} \mathcal{T}^{-1} = \bm{\nabla} \!\mathcal{V}_{\pm}     \;&,&\;\;   \mathcal{T} (\bm{\mathcal{L}}^{\dagger})^\dag \mathcal{T}^{-1} = -\bm{\mathcal{L}} \;. \label{eq:TR_Prop}
\end{eqnarray}
Yafet separated the in-phase terms into two parts to prove them vanishing.\cite{Yafet_SSP63} Together with Eq.~(\ref{eq:EYcancel}), it is the celebrated Elliott-Yafet cancellation. We emphasize that the first-order (linear-in-$\mathbf{q}$) Elliott and Yafet terms vanish separately, rather than interfere destructively.  All in all, the zero- and first-order spin-flip matrix element identically vanish, $M_{{\rm{sf}}}^{(0) } = M_{{\rm{sf}}}^{(1) }  = 0$.

\textbf{Second-order}: It is the lowest order at which spin-flip matrix elements due to electron interaction with acoustic phonon modes do not vanish. At this order, states are expanded by
\begin{eqnarray}
| \mathbf{K}\pm \tfrac{\mathbf{q}}{2},s \rangle \simeq   | \mathbf{K},s \rangle   \pm  \tfrac{1}{2}\mathbf{q}\!\cdot\!\bm{\mathcal{L}} | \mathbf{K},s \rangle +   \tfrac{1}{8}\mathbf{q}^{\otimes 2}\!\cdot\!\bm{\mathcal{L}}^{\otimes 2} | \mathbf{K},s \rangle,\;\;\; \label{eq:gen_state_2nd_order}
\end{eqnarray}
where the vector components of $\bm{\mathcal{L}}$ were formally defined in Eq.~(\ref{eq:Lop}), and $\mathbf{q}^{\otimes 2}\!\cdot\!\bm{\mathcal{L}}^{\otimes 2}$ denotes the scalar product of two second-rank tensors. An explicit form of this state was derived by a spin-dependent $\mathbf{k}\!\cdot\!\mathbf{p}$ expansion of the $X$-point basis states [Eqs.~(\ref{eq:spin_up_state})-(\ref{eq:state_structure})]. Using this basis, the general spin-flip matrix element is converted to
\begin{eqnarray}
\!\!\!\!\!&& \!\!\!\!\! M\left(\mathbf{k}_1=\mathbf{K}-\tfrac{\mathbf{q}}{2},\Uparrow;  \mathbf{k}_2=\mathbf{K}+\tfrac{\mathbf{q}}{2},\Downarrow \right) = M_{{\rm{sf}}}^{(2)} + \mathcal{O}(q^3) \approx \nonumber \\
\!\!\!\!\!&& \!\!\!\!\! \sum_{\mu,\nu} \langle X_\mu| e^{\mbox{-}\frac{i\mathbf{q}\cdot\mathbf{r}}{2}}\!\left[ \!\Gamma_{\mu,\nu}^+(\mathbf{k}_1,\mathbf{k}_2) \!+\! \Gamma_{\mu,\nu}^-(\mathbf{k}_1,\mathbf{k}_2) \!\right]\! e^{\mbox{-}\frac{i\mathbf{q}\cdot\mathbf{r}}{2}} |X_\nu \rangle.\,\,\, \label{eq:gen_sf_ac_2nd_X}
\end{eqnarray}
$\Gamma_{\mu,\nu}^{\pm}(\mathbf{k}_1,\mathbf{k}_2)$ are scalars formed by products of column and row spinors with the potential matrix,
\begin{eqnarray}
\!\!\!\!&& \!\!\!\! \Gamma_{\mu,\nu}^{\pm}(\mathbf{k}_1,\mathbf{k}_2) =  \label{eq:gamma_form_1} \\  \!\!\!\! && \!\!\!\!  \left[-B_{\scriptscriptstyle{X_{\mu}}}(\mathbf{k}_1) ,A_{\scriptscriptstyle{X_\mu}}(\mathbf{k}_1)\right] \left(\bm\xi^{\pm}\cdot \bm\nabla \mathcal{V}_{\pm,\mathbf{K}}(\mathbf{r})\right)\left[\begin{array}{c} A_{\scriptscriptstyle{X_{\nu}}}(\mathbf{k}_2) \\ B_{\scriptscriptstyle{X_{\nu}}}(\mathbf{k}_2)  \end{array} \right], \nonumber
\end{eqnarray}
where the coefficients of $\mathbf{A}(\mathbf{k})$ and $\mathbf{B}(\mathbf{k})$ are taken from Eq.~(\ref{eq:state_structure}), and the $\mathbf{K}$-dependent potential has the form
\begin{eqnarray}
\mathcal{V}_{\pm,\mathbf{K}} &=& V_{\pm} \mathcal{I}+ V^{\rm{so}}_{\pm,\mathbf{K}}\nonumber\\
&= &V_{\pm} \mathcal{I} + \frac{\hbar}{4m^2_0 c^2}\left[\bm{\nabla}V_{\pm}\!\times\!(\mathbf{p} +\hbar\mathbf{K}')\right]\cdot \bm\sigma\,. \label{eq:potential_mod}
\end{eqnarray}
The bare potential, $V_{\pm}$, is diagonal and generates Elliott products of the type $A_{\mu}B_\nu$. The spin-orbit coupling potential, $V_{\pm,\mathbf{k}}^{\rm{so}}$, generates Yafet products of all types but the dominant signature comes from $A_mA_n$ terms due to the smallness of the coefficients in  $\mathbf{B}(\mathbf{k})$.

We identify the general forms of second-order matrix elements between the $X$-point basis states. After expanding the exponential and $\Gamma$ terms in Eqs.~(\ref{eq:gen_sf_ac_2nd_X})-(\ref{eq:gamma_form_1}) into power series, quadratic terms in $\mathbf{q}$ are classified by six integrals that read
\begin{eqnarray}
M_{{\rm{sf}}}^{(2)} &=& \sum_{\mu,\nu} \sum_{n=1}^6 I_{\mu,\nu;n} \;, \label{eq:Msf2}
\end{eqnarray}
\begin{widetext}
\begin{subequations} \label{eq:spin_flip_int}
\begin{align}
\!\!\!\!\! I_{\mu,\nu;1} &= -\tfrac{i}{2} \left\langle X_\mu \left| \tilde{\mathbf{C}}_{\mu}(\mathbf{k})  \left\{ \mathbf{q}\cdot\mathbf{r},\bm\xi^{-}\cdot \bm\nabla \mathcal{V}_{-,\mathbf{k}}(\mathbf{r})\right\} \mathbf{C}_{\nu}(\mathbf{k})  \right| X_{\nu} \right\rangle_{\mathbf{k}=\mathbf{k}_0}  \label{eq:Msf2_out1} \\
\!\!\!\!\! I_{\mu,\nu;2} &= \left\langle X_\mu \left| \left( \mathbf{q}\cdot \bm\nabla_{\mathbf{k}}\tilde{\mathbf{C}}_{\mu}(\mathbf{k})\right) \left(\bm\xi^{-}\cdot \bm\nabla \mathcal{V}_{-,\mathbf{k}}(\mathbf{r})\right) \mathbf{C}_{\nu}(\mathbf{k})  \right| X_{\nu} \right\rangle_{\mathbf{k}=\mathbf{k}_0}  \label{eq:Msf2_out2} \\
\!\!\!\!\! I_{\mu,\nu;3} &= -\tfrac{1}{4} \left\langle X_\mu \left| \left(\mathbf{q}\cdot \bm\nabla_{\mathbf{k}}\tilde{\mathbf{C}}_{\mu}(\mathbf{k}) \right) \left(\bm\xi^{+}\cdot \bm\nabla \mathcal{V}_{+,\mathbf{k}}(\mathbf{r})\right)  \left(\mathbf{q}\cdot \bm\nabla_{\mathbf{k}} \mathbf{C}_{\nu}(\mathbf{k}) \right) \right| X_{\nu} \right\rangle_{\mathbf{k}=\mathbf{k}_0} \label{eq:Msf2_in1} \\
\!\!\!\!\! I_{\mu,\nu;4} &= -\tfrac{i}{2} \left\langle X_\mu \left| \left(\mathbf{q}\cdot \bm\nabla_{\mathbf{k}}\tilde{\mathbf{C}}_{\mu}(\mathbf{k}) \right) \left\{ \mathbf{q}\cdot\mathbf{r},\bm\xi^{+}\cdot \bm\nabla \mathcal{V}_{+,\mathbf{k}}(\mathbf{r})\right\} \mathbf{C}_{\nu}(\mathbf{k})  \right| X_{\nu} \right\rangle_{\mathbf{k}=\mathbf{k}_0}  \label{eq:Msf2_in2} \\
\!\!\!\!\! I_{\mu,\nu;5} &= +\tfrac{1}{4} \left\langle X_\mu \left| \left( \mathbf{q}^{\otimes 2}\cdot\bm\nabla_{\mathbf{k}}^{\otimes 2}\tilde{\mathbf{C}}_{\mu}(\mathbf{k}) \right) \left(   \bm\xi^{+}\cdot \bm\nabla \mathcal{V}_{+,\mathbf{k}}(\mathbf{r})\right)\mathbf{C}_{\nu}(\mathbf{k})  \right| X_{\nu} \right\rangle_{\mathbf{k}=\mathbf{k}_0}         \label{eq:Msf2_in3} \\
\!\!\!\!\! I_{\mu,\nu;6} &= -\tfrac{1}{2} \left\langle X_\mu \left|  \tilde{\mathbf{C}}_{\mu}(\mathbf{k}) \mathbf{q}\cdot\mathbf{r} \left(\bm\xi^{+}\cdot \bm\nabla \mathcal{V}_{+,\mathbf{k}}(\mathbf{r}) \right) \mathbf{q}\cdot\mathbf{r} \mathbf{C}_{\nu}(\mathbf{k})  \right| X_{\nu} \right\rangle_{\mathbf{k}=\mathbf{k}_0}   ,       \label{eq:Msf2_in4}
\end{align}
\end{subequations}
\end{widetext}
where $\{ U,W \} = UW +WU$ and
\begin{eqnarray}
\mathbf{C}_{\nu}(\mathbf{k})\!\!&=&\!\left[  A_{\scriptscriptstyle{X_{\nu}}}(\mathbf{k}),  B_{\scriptscriptstyle{X_{\nu}}}(\mathbf{k})  \right]^T\!\!  ,\! \nonumber\\ \tilde{\mathbf{C}}_{\mu}(\mathbf{k})\!&=&\!\left[-B_{\scriptscriptstyle{X_{\mu}}}(\mathbf{k}) , A_{\scriptscriptstyle{X_\mu}}(\mathbf{k})\right]. \label{eq:C_k}
\end{eqnarray}
In writing these matrix elements we have used time reversal and space inversion to unify terms.  The mean wavevector $\mathbf{K}$ is replaced with $\mathbf{k}_0$, which brings approximation of $\mathcal{O}(q^3)$.

At this point, the integrals of Eq.~(\ref{eq:spin_flip_int}) can be numerically calculated in a straightforward manner. Given the four basis states and the above integral forms, this numerical procedure involves a few hundreds of different types of space integrations. Performing selection rules for all of them also demands a large amount of work. However, we can avoid these labors and come to an accurate and compact matrix element [Eq.~(\ref{eq:intra_AC})]. This possibility is enabled by three observations that greatly simplify the analysis of Eq.~(\ref{eq:spin_flip_int}). These observations will also allow us to connect the derived matrix element with an experimentally known deformation potential quantity.

\textit{Observation 1}: Integration of bare potential related terms such as $\langle X_\mu |\bm\nabla V_{\pm}| X_\nu \rangle$, $\langle X_\mu | r_i \bm\nabla V_{\pm} | X_\nu \rangle$ and $\langle X_\mu | r_i r_j \bm\nabla V_{\pm} | X_\nu \rangle$ yield numbers of the same order of magnitude given that $r_i$ is measured in units of $a/2 \pi$. Moreover, these numbers are comparable for both the in-phase ($+$) and out-of-phase ($-$) parts. The same applies to the integrals of the spin-orbit coupling potential, $\langle X_\mu |\bm\nabla V_{\pm,\mathbf{k}}^{\rm{so}} | X_\nu \rangle$, $\langle X_\mu | \{r_i, \bm\nabla V_{\pm,\mathbf{k}}^{\rm{so}}\} | X_\nu \rangle$ and $\langle X_\mu | r_i \bm\nabla V_{\pm,\mathbf{k}}^{\rm{so}}  r_j| X_\nu \rangle$. These conjectures are backed by explicit numerical calculations. The physical rationale is that these potentials are significant within the size of a primitive cell where $r_i$ is of the order of 1. In what follows we keep consistency and measure length in units of $a/2\pi$. Accordingly, band-structure parameters such as $P$, $\eta$ or $\alpha$ are approached as energy scales, and compared directly with $E_{\scriptscriptstyle{g,X}}$ or $\Delta_C$ (see Table~\ref{tab:params}).

\textit{Observation 2}: The number of matrix elements is greatly reduced by estimating the amplitude of their coefficients. The largest Elliott and Yafet products scale, respectively, with $|\eta|/\Delta_C$ (or $|\eta|P/\Delta_CE_{\scriptscriptstyle{g,X}}$)  and  $2 P^2/(E_{\scriptscriptstyle{g,X}} \Delta_C)$. For both cases, other products are significantly smaller. These products relate, respectively, to the bare and spin-orbit coupling potentials. These amplitudes are evaluated by inspection after substituting the explicit coefficients of $\mathbf{A}(\mathbf{k})$ and $\mathbf{B}(\mathbf{k})$ [Eq.~(\ref{eq:state_structure})] into Eq.~(\ref{eq:spin_flip_int}). Notice that the terms $\{\mathbf{C}(\mathbf{k}),\bm\nabla_{\mathbf{k}}\mathbf{C}(\mathbf{k}),\bm\nabla^2_{\mathbf{k}}\mathbf{C}(\mathbf{k})\}|_{\mathbf{k}=\mathbf{k}_0}$
are all constants without $\mathbf{k}$ dependence (and similar for $\tilde{\mathbf{C}}$).

\textit{Observation 3}: Prior to the application of selection rules, contributions of Elliott and Yafet processes to intravalley spin relaxation are conceivably comparable. To understand this physics we first recall the zeroth-order Elliott-Yafet cancelation [Eq.~(\ref{eq:EYcancel})]. For states at the valley center ($\mathbf{k}=\mathbf{k_0}$) this cancellation means
\begin{eqnarray}
\left| \frac{\langle X_1 |\bm\nabla V_{+,\mathbf{k}_0}^{\rm{so}}| X_1 \rangle}{\langle X_1 |\bm\nabla V_{+}| X_4 \rangle } \right| = \frac{\Delta_X}{E_{\scriptscriptstyle{g,X}}} . \label{eq:EYratio}
\end{eqnarray}
Taking into account observation 1, we can generalize this order of magnitude to ratios between all Yafet-related integrals and all Elliott-related ones. Dominant Elliott processes relate to matrix elements of the type that appears in the denominator of Eq.~(\ref{eq:EYratio}) multiplied by $\eta/\Delta_C$ (see observation 2). Similarly, dominant Yafet processes relate to matrix elements of the type that appear in the numerator multiplied by $2 P^2/(E_{\scriptscriptstyle{g,X}} \Delta_C)$.  The overall ratio between Elliott and Yafet processes is thus of the order of unity ($\sim P/E_{\scriptscriptstyle{g,X}}$).

Application of observations 1-3 in Eq.~(\ref{eq:spin_flip_int}) results in a handful of matrix elements that are worth examination. For example, the integral classes  $I_{\mu,\nu;1}$ and $I_{\mu,\nu;6}$ are eliminated on ground of their small coefficients [$\Delta'_X/E_{\scriptscriptstyle{g,X}}$ and 1 rather than $\eta/\Delta_C$ and $2 P^2/(E_{\scriptscriptstyle{g,X}} \Delta_C)$ in observation 2]. Therefore,
\begin{eqnarray}
I_{\mu,\nu;1} \sim  0 \;\; \text{and} \;\;  I_{\mu,\nu;\,6} \sim  0  \;, \;\nonumber
\end{eqnarray}
for any possible basis state combination ($X_\mu$ and  $X_\nu$).

We invoke group theory and evaluate the remaining integrals in  Eq.~(\ref{eq:spin_flip_int}). Group theory results [Eqs.~(\ref{eq:XX})-(\ref{eq:TR}) and their discussion] are extensively utilized, coupling integrals are expressed analytically by reasonable approximations, and the internal displacement of silicon structure are carefully accounted. The detailed procedure is shown in Appendix \ref{app:details_intra_g}. We reach at the intravalley spin-flip matrix elements due to interaction with acoustic phonon modes,
\begin{eqnarray}
\!\!\!\!\! && M_{\lambda}^{\text{intra}} \left(\mathbf{k}_1,\Uparrow \;\; ;\;\;  \mathbf{k}_2,\Downarrow \right) = \frac{2\eta }{\Delta_C} D'_{xy}\epsilon_{xy,\scriptscriptstyle{\lambda}}(\mathbf{q})(q_x-iq_y) \nonumber\\
&\,&\,\,\,\, = \frac{i\eta D'_{xy}}{\Delta_C} (q_x\!-\!iq_y) \left[q_x\xi_{\lambda,y}^+(\mathbf{q})+q_y\xi_{\lambda,x}^+(\mathbf{q})\right], \label{eq:intra_AC}
\end{eqnarray}
where $\lambda$=$\{$TA$_1$, TA$_2$, LA$\}$, with a physical deformation potential constant $D'_{xy}$ whose complete integral expression reads
\begin{eqnarray}
D'_{xy}\!&=&\!\frac{-  P^2 m_0 }{ \hbar^2} + \Gamma_{xyz} \left\langle X^{2'}_1\left|\frac{\partial V_-}{\partial z} \right|X^1_1\right\rangle \label{eq:D_xy} \\
\!\!\!\!\! &-& \!\!\left\langle  \!\! X^{2'}_1\left| \sum^{A,B}_\alpha (y-\tau_{\alpha,y}) \frac{\partial V_{\rm{\!at}} (\mathbf{r}-\bm\tau_\alpha)}{\partial x}  \right|X^1_1\right\rangle \approx 6~\text{eV}. \nonumber
\end{eqnarray}

\begin{figure}
\includegraphics[width=8.4cm, height=7.5cm]{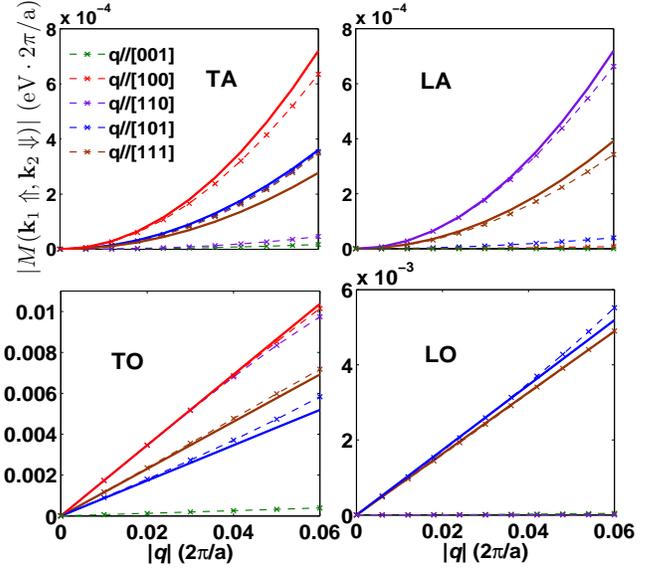}
\caption[intravalley M vs q] {(Color online) Analytical (solid lines) and numerical results (-x- lines) of spin-flip matrix element as a function of $\mathbf{q}=\mathbf{k}_1-\mathbf{k}_2$ along five different scattering directions. Each direction is represented by a different line color. The electron resides in the $z$-valley and its spin is orientated along the valley axis. The analytical results for TA, LA, TO and LO modes are taken from Table ~\ref{tab:intra_g_M_Sn} for $\mathbf{n}$$\,$$\|$$\,$[001].  Numerical results rely on empirical pseudopotential and adiabatic bond charge models (to be discussed in Sec.~\ref{sec:numerical_procedures}). Analytical and numerical curves agree very well for all cases where the quadratic dependence of the acoustic cases holds best for small $q$ values. Note that analytical results for different $\mathbf{q}$ directions may overlap each other. } \label{fig:intra_M_q}
\end{figure}

\subsubsection*{long-wavelength optical phonon modes}\label{sec:intra_op}
Using the gained knowledge of spin-flip processes with acoustic phonon modes, we can readily derive the optical case. The optical phonon modes have a dominant out-of-phase polarization vector, $\bm\xi_{\rm{op}}^-$, with magnitude of about unity at the long-wavelength regime (Table~\ref{tab:elastic_continuum}). From the detailed analysis of the out-of-phase acoustic phonon modes we could have recognized the fact that the antisymmetric interaction, $\bm\nabla V_{-}$, has a leading-order matrix element that is linear in $\mathbf{q}$. Separating the interaction with optical modes into in-phase and out-of-phase parts, we find in this case that the out-of-phase part dominates the in-phase by two orders of magnitude. This result is a consequence of the $\mathbf{q}$ dependence of the polarization vectors as well as of the interaction integrals. The leading integral term is simply
\begin{eqnarray}
M_{\chi}^{\text{intra}} \!\left(\mathbf{k}_1,\Uparrow \; ;\;  \mathbf{k}_2,\Downarrow \right) =
\frac{-\eta}{\Delta_C} D_{\rm{op}}\left(q_x-iq_y\right)\xi^-_{\chi,z}(\mathbf{q})\,,\label{eq:intra_OP}
\end{eqnarray}
where $\chi$=$\{$TO$_1$, TO$_2$, LO$\}$, with the associated scattering integral
\begin{eqnarray}
D_{\rm{op}} = \langle X^{2'}_1 | \partial V_{\!-} / \partial z |X^1_1\rangle \approx 5~\text{eV}\cdot2\pi/a \text .\label{eq:Dop}
\end{eqnarray}

The ability to investigate for each individual mode is desired under conditions of anisotropic fields or stress. Then, the relative importance of different modes may vary  depending on the symmetry breaking mechanism.

We render the elastic continuum approximation in order to achieve complete and analytical $\mathbf{q}$-dependent matrix elements and to facilitate analytical integrations. The polarization vectors, $\bm{\xi}^{\pm}(\mathbf{q})$, in Eqs.~(\ref{eq:intra_AC}) and (\ref{eq:intra_OP}) are then replaced by the expressions of Table~\ref{tab:elastic_continuum}.  We obtain $|M_{{\rm{sf}}}|^2$ for each of the modes (TA, LA, TO and LO).\cite{footnote_trans_mode} The final expressions are given in Table~\ref{tab:intra_g_M_Sn} (the case of $\mathbf{n}\|[001]$). To examine the accuracy of the $\mathbf{q}$ dependence, Fig.~\ref{fig:intra_M_q} shows the analytical intravalley results of Table~\ref{tab:intra_g_M_Sn} next to numerically calculated matrix elements along inequivalent high-symmetry scattering angle directions. The numerical procedure will be discussed in Sec.~\ref{sec:lifetime} together with application of the derived matrix elements to (analytically) find the spin lifetime expression.

\subsection{Intervalley $g$-process spin flips}\label{sec:g-process}

$g$-process scattering angles are strongly directional ($\mathbf{q} = \mathbf{k}_2 -\mathbf{k}_1\approx -2\mathbf{k}_0$ ). The leading spin-flip matrix element of a $g$-process depends on the mean wavevector $\mathbf{K}=(\mathbf{k}_1+\mathbf{k}_2)/2$ (and not on $\mathbf{q}$ as in the intravalley case; See Table~\ref{tab:symmetry_argument}). When $\mathbf{k}_2=-\mathbf{k}_1$ the opposite spin states are Kramers conjugates and their coupling via scattering with any type of phonon vanishes.\cite{Yafet_SSP63} As will be shown below, the LA mode leads the $g$-process spin relaxation where its matrix element is linear in $\mathbf{K}$. The scattering constant, however, is not as concise as in the intravalley case. This is conceivable because the translational factors, $e^{\pm i\mathbf{k}'_0\cdot\mathbf{r}}$, of the initial and final states are to be expanded around a common $X$ point ($\mathbf{k}'=0$). Other than the dominant LA phonon mode, we also discuss the general shape of the matrix element due to scattering with TA phonon modes. Their contribution to $g$-process spin relaxation at room temperature will be shown to be non-negligible in comparison with the LA mode. This property is analogous to $g$-process momentum scattering, where TA phonon modes with a higher-order matrix element but lower energy (compared to the leading LO phonon) are important  in describing the charge transport.\cite{Ferry_PRB76} In what follows we continue to work with the $X$-point basis states which can be readily related to the $\Delta$-axis basis states via $\psi_{\Delta_i,k_z}\simeq e^{ik'_z z}\psi_{X_i}$.

For $\mathbf{k}_1$ and $\mathbf{k}_2$ at the vicinity of the $\pm z$ valley center, we expand the matrix elements around $\mathbf{K}=0$,
\begin{eqnarray}
&&M_g(\mathbf{\mathbf{k}_1},\Uparrow;\mathbf{k}_2,\Downarrow)\nonumber\\
&=&\sum_{\pm}\langle \tfrac{\mathbf{q}}{2}+\mathbf{K},\Downarrow\left|\bm\xi^{\pm}(\mathbf{q})\cdot\bm \nabla\mathcal{V}_{\pm}\right| \mbox{-}\tfrac{\mathbf{q}}{2}+\mathbf{K},\Uparrow\rangle \nonumber\\
&=&\bcancel{M}^{(0)}_{{\rm{sf}}}+\bcancel{M}^{(1)}_{{\rm{sf}},1}+M^{(1)}_{{\rm{sf}},2}\nonumber\\
&&+\bcancel{\mathcal{O}(K^2)}+\mathcal{O}(K)\mathcal{O}(|\mathbf{q}/2+\mathbf{k}_0|)+ \mathcal{O}(K^3),\label{eq:M_g_order}
\end{eqnarray}
and explain each term separately. In a $g$-process $\xi^{-}(\mathbf{q})$ is not treated as a small quantity compared to 1. The in-phase and out-of-phase interaction are treated on an equal footing. However, we will find that the out-of-phase contribution drops.

Applying time reversal symmetry on a spinor state and $\mathbf{k}$-derivative operator $\bm{\mathcal{L}}$ [Eq.~(\ref{eq:TR_Prop})], it is readily seen that the zeroth-order term ($\mathbf{K}=0$) vanishes
\begin{eqnarray}
M^{(0)}_{{\rm{sf}}}=\sum_{\pm}\langle \tfrac{\mathbf{q}}{2},\Downarrow\left|\bm\xi^{\pm}(\mathbf{q})\cdot\bm\nabla\mathcal{V}_{\pm}\right|  \mbox{-}\tfrac{\mathbf{q}}{2},\Uparrow\rangle=0,
\end{eqnarray}
and that the two linear-in-$\mathbf{K}$ terms are equal to each other,
\begin{eqnarray}
M^{(1)}_{{\rm{sf}}}=\sum_{\pm}\langle \tfrac{\mathbf{q}}{2},\Downarrow\left|\left(\bm\xi^{\pm}\cdot\bm\nabla\mathcal{V}_{\pm}\right)\left(\mathbf{K}\cdot \bm{\mathcal{L}}\right) \right|  \mbox{-}\tfrac{\mathbf{q}}{2},\Uparrow\rangle\nonumber\\
+\sum_{\pm}\langle \tfrac{\mathbf{q}}{2},\Downarrow\left|(\bm{\mathcal{L}}^\dag\cdot\mathbf{K})\left(\bm\xi^{\pm}\cdot\bm \nabla\mathbf{\mathcal{V}}_{\pm} \right) \right|  \mbox{-}\tfrac{\mathbf{q}}{2},\Uparrow\rangle. \label{eq:g_lin}
\end{eqnarray}
When $\bm{\mathcal{L}}$ acts on the translational part, $e^{i\mathbf{k}'\cdot\mathbf{r}}$, the resulting matrix element becomes
\begin{eqnarray}
M^{(1)}_{{\rm{sf}},1}=\sum_{\pm}\langle \tfrac{\mathbf{q}}{2},\Downarrow\left| [\bm\xi^{\pm}\cdot\bm\nabla\mathcal{V}_{\pm},\, i\mathbf{K}\cdot\mathbf{r}]\right| \mbox{-}\tfrac{\mathbf{q}}{2},\Uparrow\rangle\approx0 ,
\end{eqnarray}
where $[U,W]=UW-WU$. The Elliott part of this term naturally drops since $\bm{\nabla}V$ and $\mathbf{r}$ commute. The Yafet part results in a small `$\alpha \mathbf{K}$' factor and a weak coupling between basis states (i.e., non-dominant coefficient products). Thus, $M^{(1)}_{{\rm{sf}},1}$ can be safely discarded.

The second part of the first-order term, $M^{(1)}_{{\rm{sf}},2}$, results from operating $\bm{\mathcal{L}}$ on the $\mathbf{k}$-dependent coefficient $\mathbf{C}(\mathbf{k})$ and $\tilde{\mathbf{C}}(\mathbf{k})$ [Eq.~(\ref{eq:C_k})] of the bra and ket states in Eq.~(\ref{eq:g_lin}). Time reversal symmetry is utilized again where following Eq.~(\ref{eq:TR_Prop}) we get that $\langle\tfrac{\mathbf{q}}{2},\Downarrow| = (\mathcal{T}|-\tfrac{\mathbf{q}}{2},\Uparrow\rangle)^\dag$, and the action of $\mathcal{T}$ on individual $X$-point basis states follows Eq.~(\ref{eq:X_TR}). All in all, the second part of the first-order term reads
\begin{eqnarray}
M^{(1)}_{{\rm{sf}},2}&=&2\sum_{\mu,\nu}\sum_{\pm} \left\langle X_{\mu}\left|\mathbf{K}\cdot\bm\nabla_{\mathbf{k}}\tilde{C}_{\mu'}(\mathbf{k}_0) e^{ik'_0z} \right.\right.\nonumber\\
&&\left.\left.(\bm\xi^{\pm}\cdot\bm\nabla\mathcal{V}_{\pm})C_{\nu}(\mathbf{k}_0) e^{ik'_0z} \right| X_{\nu}\right\rangle+\mathcal{O}(|\mathbf{q}/2+\mathbf{k}_0|^2)\nonumber\\
&=& I^{E}+I^{Y}+\mathcal{O}(|\mathbf{q}/2+\mathbf{k}_0|^2) \label{eq:M_g}
\end{eqnarray}
where both $\tilde{C}_{\mu'}$ and $C_{\nu}$ [Eq.~(\ref{eq:C_k})]  are evaluated at $\mathbf{k}_0\approx-\mathbf{q}/2$ and for $\mu=\{1,2,3,4\}$, $\mu'=\{2,1,4,3\}$. The Yafet part contribution $I^Y$ is not negligible for $g$-process. The detailed derivation of $I^E$ and $I^Y$ (Appendix \ref{app:details_intra_g}) bears some similarity with that of intravalley spin flip, and invokes the selection rules governing the opposite points of $\Delta$ star.

\begin{figure}
\includegraphics[width=4.5cm, height=4.2cm]{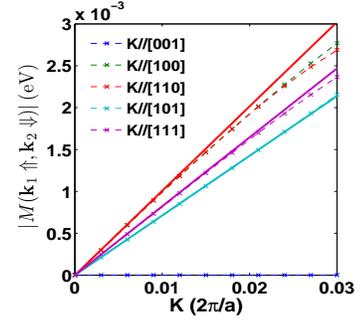}
\caption{(Color online)  Analytical (solid lines) and numerical results (-x- lines) of $g$-process spin-flip matrix elements as a function of $\mathbf{K}=\tfrac{1}{2}(\mathbf{k}_1+\mathbf{k}_2)$ due to scattering with LA phonon modes. Results are shown for five typical scattering directions where each is represented by a different line color. Analytical expressions are given in Table ~\ref{tab:intra_g_M_Sn} and their results for $\mathbf{K}\|[001]$ and $[100]$  overlap each other. Numerical results rely on empirical pseudopotential and adiabatic bond charge models (to be discussed in Sec.~\ref{sec:numerical_procedures})} \label{fig:g_M}
\end{figure}

We obtain the leading $g$-process matrix element,
\begin{eqnarray}
M_{g,{\rm{LA}}}(\mathbf{k}_1,\Uparrow; \mathbf{k}_2, \Downarrow)&\approx&I^E+I^Y \nonumber \\
&=& D_{gs} (K_x -i K_y), \label{eq:Mg}
\end{eqnarray}
where
\begin{eqnarray}
D_{gs}&=&2P/E_{\scriptscriptstyle{g,X}} [D_{\rm{so}} + 4 \Delta_X k'_0D_{zz}/\Delta_C] + \mathcal{O}(k'^3_0) \nonumber\\
&\approx& 0.1\,\rm{eV}.\label{eq:Dgs}
\end{eqnarray}
$D_{zz}=\Xi_d+\Xi_u$ denotes the sum of dilation and uniaxial deformation potential parameters [see Eq.~(\ref{eq:Djk}) and discussion after Eq.~(\ref{eq:projection})] and $D_{\rm{so}}\approx 6.7$ meV is a spin-dependent scattering constant [see Eq.~(\ref{eq:D_so})].

Squared matrix elements of the $g$-process are summarized in Table~\ref{tab:intra_g_M_Sn}, and Fig.~\ref{fig:g_M} compares these analytical expressions with numerical results along typical $\mathbf{K}$ directions. The scattering-angle dependence is predicted correctly where the linear relation holds best when both states are at the vicinity of the valleys centers. From numerical results we note that $\mathcal{O}(k'_0)^3$ terms from Elliott interaction yield a negative $10\%$ correction.

Before concluding this part we briefly discuss spin-flips matrix elements of higher order. We only mention the general nature of these matrix elements without deriving explicit forms. The reason is that the $g$-process has the weakest contribution to spin relaxation in unstrained bulk silicon. Among the higher-order matrix elements [last line of Eq.~(\ref{eq:M_g_order})], the $\mathcal{O}(K^2)$ terms vanish by time reversal symmetry. This property  can be proven by using Eq.~(\ref{eq:TR_Prop}) and the matrix elements expansion to quadratic-$\mathbf{K}$ terms. On the other hand, the non-vanishing second-order term has a wavevector dependence of the type $\mathcal{O}(K)\mathcal{O}(|\mathbf{q}/2+\mathbf{k}_0|)$.  Such a symmetry allowed term is more important than $\mathcal{O}(K^3)$ terms and it is governed by matrix elements of TA phonon modes. Its contribution to spin relaxation time from numerical calculations will be discussed in Sec.~\ref{sec:lifetime}. Here it is noted that compared with intravalley processes the relative contribution to spin relaxation from high-order terms is larger. In addition, the signature of further bands (outside the $X_1$ and $X_4$ basis states) is larger in $g$-process spin flips.

\subsection{Spin orientation dependence}\label{sec:valley-spin dependence}

We relax the restriction of a fixed spin orientation ($\mathbf{n}\| \mathbf{z}$) and explore this degree of freedom. The spin orientation dependence originates from the anisotropy of the conduction band. This anisotropy suggests that only the relative direction of spin orientation to a valley axis is important. Accordingly, we choose the $+z$ valley and express $\mathbf{n}$ in terms of polar and azimuthal angles ($\theta,\phi$) with respect to the $+z$ and $+x$ directions. This choice is convenient since selection rules between $X$-point basis states of the $+z$ valley have been already derived in Secs.~\ref{sec:kp hamilton}-\ref{sec:intravalley}.

Derivation of spin-flip matrix elements when $\mathbf{n} \nparallel \mathbf{z}$ relies on the specific form of $|\mathbf{k}, \Uparrow(\Downarrow)_\mathbf{n}\rangle$. These states satisfy the conditions of Eq.~(\ref{eq:spin_up_down_states}) and their forms can be found by solving the Hamiltonian matrix when written in terms of the new basis $\{|\mathbf{X}\rangle \otimes |\uparrow(\downarrow)_\mathbf{n} \rangle\}$. However, we can avoid this labor and find these states using a two-step procedure. First, the previously derived forms of $|\mathbf{k}, \Uparrow(\Downarrow)_\mathbf{z}\rangle$ [Eq.~(\ref{eq:spin_up_state})] are re-expressed in terms of the new basis $\{|\mathbf{X}\rangle \otimes |\uparrow(\downarrow)_\mathbf{n} \rangle\}$. This change of basis amounts to a rotation of the spin coordinate by $-\theta$ about the axis $\bm{\omega}=\hat{\mathbf{n}} \times \hat{\mathbf{z}}$. Specifically, the operator
\begin{eqnarray}
U=\displaystyle\exp\left(\frac{-i\bm\sigma \!\cdot\! \hat{\bm{\omega}} \,\theta}{2}\right),\label{eq:rot_matrix2}
\end{eqnarray}
is applied on $each$ of the four spinors $[A_{X_i}(\mathbf{k}), B_{X_i}(\mathbf{k})]^T$ that Eq.~(\ref{eq:state_structure}) is comprised of.\cite{footnote_rotation} In the second step, we find $|\mathbf{k}, \Uparrow(\Downarrow)_\mathbf{n}\rangle$ by forming linear combinations of the re-expressed states (still oriented along $\mathbf{z}$) such that Eq.~(\ref{eq:spin_up_down_states}) is satisfied. Technical details of finding this superposition are summarized in  Appendix~\ref{app:spin_align}.  The states can be expressed in the original $\{|\mathbf{X}\rangle \otimes |\uparrow(\downarrow)_z \rangle\}$ basis using $U^\dagger$.

Having the forms of $|\mathbf{k}, \Uparrow(\Downarrow)_\mathbf{n}\rangle$, we repeat the procedure of Sec.~\ref{sec:intravalley} and derive the dominant intravalley spin-flip matrix elements. Scattering with acoustic phonon modes reads
\begin{eqnarray}
&& \!\!\!\!\!\!M_{\lambda}^{\text{intra}} \left(\mathbf{k}_1,\Uparrow_{\mathbf{n}} \;\; ;\;\;  \mathbf{k}_2,\Downarrow_{\mathbf{n}} \right) \nonumber\\
&=&i\eta/\Delta_C D'_{xy}\left(q_x\xi^+_{y,\lambda}(\mathbf{q})+q_y\xi^+_{x,\lambda}(\mathbf{q})\right)
\nonumber\\
&&\times \left[\cos^2\frac{\theta}{2} (q_x-i q_y)- \sin^2\frac{\theta}{2}e^{2i\phi} (q_x+i q_y) \right],\quad \label{eq:ac_Sn}
\end{eqnarray}
where $\lambda$=$\{$TA$_1$, TA$_2$, LA$\}$. Scattering with optical phonon modes reads
\begin{eqnarray}
&& \!\!\!\!\!\!M_{\chi}^{\text{intra}} \!\left(\mathbf{k}_1,\Uparrow_{\mathbf{n}} \; ;\;  \mathbf{k}_2,\Downarrow_{\mathbf{n}} \right) \nonumber\\
&=&-\eta/\Delta_C \,D_{\rm{op}}\,\,\xi^-_{z,\chi}(\mathbf{q})\nonumber\\
&&\times\left[\cos^2\frac{\theta}{2} (q_x- iq_y) - \sin^2\frac{\theta}{2}e^{2i\phi} (q_x+ i q_y)\right].\quad \label{eq:op_Sn}
\end{eqnarray}
where $\chi$=$\{$TO$_1$, TO$_2$, LO$\}$. Table~\ref{tab:intra_g_M_Sn} lists the squared matrix element expressions of all phonon modes using elastic continuum approximation for diamond crystal structures. These spin-flip expressions are specified for spin orientations along all inequivalent high-symmetry crystal directions. This chosen set of directions is important for two reasons. First, the invoked elastic continuum approximation is accurate along these directions. Second, the oriented spins in typical spin injection experiments point along these directions. Using the results in Table~\ref{tab:intra_g_M_Sn}, we have shown in Fig.~\ref{fig:intra_M_q} that the analytical expression for $\mathbf{n \| \mathbf{z}}$  ($\theta$=0) agrees with independent numerical calculations. This agreement is also true (not shown) for $\mathbf{n \nparallel \mathbf{z}}$  cases in Table~\ref{tab:intra_g_M_Sn}. As will be explained in the next section, the numerical models automatically incorporate time reversal and space-group symmetries of the crystal. The agreement between these independent calculation approaches manifest the robustness of the major terms we  have kept in intravalley spin flips.

Deriving the $g$-process spin-flip matrix elements is similar. We study scattering from $z$ to $-z$ valley for an arbitrary direction of $\mathbf{n}$. Repeating the procedure of Sec.~\ref{sec:g-process}, the dominant spin-flip matrix element reads
\begin{eqnarray}
&&\!\!\!\!\!\!M_{g,{\rm{LA}}}(\mathbf{k}_1,\Uparrow_{\mathbf{n}}\; \mathbf{k}_2, \Downarrow_{\mathbf{n}}) =\nonumber\\
&&\!\!\!\!\!\!D_{gs}\left[ \cos^2\frac{\theta}{2} (K_x- iK_y) - \sin^2\frac{\theta}{2}e^{2i\phi} (K_x+ i K_y) \right],\quad\quad\label{eq:g_Sn}
\end{eqnarray}
and it originates from scattering with LA phonon modes. Using this expression, Table~\ref{tab:intra_g_M_Sn} lists squared matrix elements along high-symmetry crystal directions. As can be seen from the square brackets terms of Eqs.~(\ref{eq:ac_Sn})-(\ref{eq:g_Sn}), the $g$-process shares the same angular dependence as in the intravalley case but with replacing $\mathbf{q}=\mathbf{k}_1-\mathbf{k}_2$ with $\mathbf{K}=\tfrac{1}{2}(\mathbf{k}_1+\mathbf{k}_2)$. The similar angular dependence is not surprising for two related reasons. First, the spin orientation only affects the electron states while the phonon properties play no role in setting the angular dependance. Second, the electron states that we use in deriving intravalley or $g$-process matrix elements are all expanded around the same $X$ point. The replacement of $\mathbf{q}$ with $\mathbf{K}$ can be understood by time reversal symmetry.

Analysis of the spin relaxation time due to scattering within the $+z$ valley and between $\pm z$ valleys will be provided in Sec.~\ref{sec:lifetime_oritentaion} for various directions of $\mathbf{n}$.

\section{Interplay between analytical derivation and EPM}\label{sec:EPM}
In this work we compare our results with an empirical method in which the electronic states and phonon polarization vectors are calculated, respectively, via empirical pseudopotential and adiabatic bond charge models (EPM and ABCM).\cite{Chelikowsky_PRB76,Weber_PRB77} The states and polarization vectors  are used in calculating the electron-phonon interaction following a rigid-ion approximation.\cite{Allen_PRB81} The EPM and ABCM provide very accurate symmetry-related results and trends of contributions from high-order wavevector components. Given that a sufficiently large plane-wave basis is employed, then in addition to time reversal symmetry these models capture the symmetries of the Bravais lattice and of the primitive cell. This ability is independent of the specific chosen values of empirical parameters (e.g, form factors of the pseudopotential). On the other hand, intensive numerical calculations do not automatically guarantee an accurate spin-flip matrix element result. Our theory provides a clear insight to the identity of critical empirical parameters that are relevant for spin relaxation. In this section we will elaborate on  fundamental aspects in understanding the application of EPM in spin-flip processes.

The usual way of finding the energy band structure by adjusting the pseudopotential form factors is not sufficient for scattering problems.  Specific derivative values of the pseudopotential at the first few reciprocal lattice vectors [$d^mV(k)/dk^m$ at $k=g_n$] are additional necessary conditions. In momentum scattering, a correctly interpolated pseudopotential is capable of  reproducing the energy shifts of the conduction band in response to applied stress. Spin scattering is more than momentum scattering in the sense that the leading-order matrix element is of higher order in the wavevector (e.g., intravalley and $g$-process scattering in silicon). One consequence is that other deformation potential constants may come into play (e.g., $D'_{xy}$ as was shown in the previous section). In fact, our spin-dependent EPM is matched not only with energy band structure but also with different deformation potential quantities ($\mathcal{E}_1+a, b, d, \mathcal{E}_2$ and $\mathcal{E}^*_2$ in Ref.~[\onlinecite{Laude_PRB71}]).

The analytical derivation shows that the two-band degeneracy at the $X$ point plays an important role in silicon due to its proximity to the valley center. This proximity results in
$\Delta_C/E_{\scriptscriptstyle{g,X}}\ll1$ which allowed us to discard interaction terms of the type $r_i r_j \partial V_+/\partial r_k$ [Eq.~(\ref{eq:Msf2_in4})]. Considering the Fourier transform of the pseudopotential, this simplification means that intravalley spin flips are only sensitive to values of $V(g_n)$ and $dV(g_n)/d k$ while high-order derivatives can be discarded. However, silicon is a specific case. In germanium, for example, intravalley spin flips have a cubic dependence on the wavevector,\cite{footnote_Ge} and they may require additional information on $d^2V(g_n)/d k^2$ and $d^3V(g_n)/d k^3$. In the framework of deformation potential theory, this amounts to expanding the strained crystal potential with quadratic or higher-order strain-tensor components. Notably, a comprehensive experimental analysis of the intravalley spin relaxation time in germanium may provide new information on its crystal potential.

The $f$-process spin-flip matrix elements in silicon depend on various parts of the pseudopotential curve. An {\it a priori} and independent determination of $V(|\mathbf{k}|)$ is difficult. To empirically interpolate the pseudopotential curve from  momentum and spin relaxation experiments, one has to know the relaxation times dependencies on temperature, electric field, stress and related `knobs'. This information can resolve the values of individual scattering constants rather than their combined effect. In addition, knowledge of the intravalley scattering parameters allows one to reduce the uncertainties in interpolating the pseudopotential.

All in all, the $\mathbf{k}\cdot \mathbf{p}$ method and group theory provide an unambiguous guidance in relating EPM parameters with different experimental measurements and in determining the relevant parts for spin relaxation. These considerations rationalize the investigation of spin-flip problems by joining analytical and numerical approaches.

\section{spin lifetime}\label{sec:lifetime}

The spin relaxation time is an experimentally accessible quantity.
With a specific electron distribution $\mathcal{F}$, detailed expressions for spin-flip matrix elements can in principle provide a transparent physical picture of the spin relaxation under a variety of conditions. The spin relaxation rate has the form,\cite{Yafet_SSP63}
\begin{eqnarray}
\frac{1}{\tau_{{\rm{sf}}}}\!=\!\frac{4\pi}{\hbar}\!\left\langle\int\!\! \frac {d^3k_2}{(2\pi)^3} |\langle \mathbf{k}_2 | \mathcal{H}^{{\rm{sf}}}_{\rm{ep}} |\mathbf{k}_1\rangle|^2\delta(E_{\mathbf{k}_2}\!-\!E_{\mathbf{k}_1}\!\pm\!\hbar \omega_\mathbf{q})\! \right\rangle_{\!\mathbf{k}_1}\!,\nonumber\\ \label{eq:tau_sf}
\end{eqnarray}
where $\langle\mathbf{k}_2 | \mathcal{H}^{{\rm{sf}}}_{\rm{ep}} |\mathbf{k}_1\rangle$ denotes the expression of Eq.~(\ref{eq:k2Hepk1}) with opposite spins $\mathbf{s}_1$ and $\mathbf{s}_2$. The material volume $N a^3/4$ is chosen as the unit volume. $+(-)$ corresponds to phonon emission (absorption). The average over $\mathbf{k}_1$ represents $\partial \mathcal{F}/\partial E_{\mathbf{k}_1}$ weighted integration over $\mathbf{k}_1$, which is exact at the limit of infinitesimal spin-dependent chemical potential splitting. The prefactor of $4\pi/\hbar$ instead of $2\pi/\hbar$ denotes the fact that the net number of spin-polarized electrons ($N_{\Uparrow}$-$N_{\Downarrow}$) changes by two with each spin flip.

A few applications will be shown in this section. After briefly describing our numerical integration effort, we present the commonly used $\tau_{{\rm{sf}}}$ under a normal condition both analytically and numerically. Here `normal' stands for a non-degenerate bulk silicon without strong fields (i.e., Boltzmann distribution $\mathcal{F}_{\rm{MB}}$ for electrons), and with spin orientation along the valley axis. The average over $\mathbf{k}_1$ in Eq.~(\ref{eq:tau_sf}) then becomes $\mathcal{F}_{\rm{MB}}$ weighted integration over $\mathbf{k}_1$. This weighted integration is valid thanks to the relation $d \mathcal{F}_{\rm{MB}}/dE\propto \mathcal{F}_{\rm{MB}}$. In the last part of this section, some essential relations between the spin orientation and $\tau_{{\rm{sf}}}$ are derived.

\subsection{Numerical integrations and approximations} \label{sec:numerical_procedures}
We have performed numerical integrations of Eq.~(\ref{eq:tau_sf}) at different levels of approximation. These calculations are presented in decreasing order of their computation time. `EPM+ABCM' denotes the full numerical results from EPM and ABCM program codes.\cite{footnote_EPM_ABCM} The calculated electron states and phonon polarization vectors are then incorporated into a rigid-ion model following the procedure in Ref.~[\onlinecite{Cheng_PRL10}].  `kp+ABCM' replaces the EPM results with $\mathbf{k}\cdot \mathbf{p}$ energy band [Eq.~(\ref{eq:eigen_energy})]. It employs the general analytical form of $|M_{\rm{sf}}|$ from Eqs.~(\ref{eq:intra_AC}), (\ref{eq:intra_OP}) and (\ref{eq:Mg}). `ellip+$\omega(q)$' further employs a spheroidal energy dispersion and replaces the phonon frequency from ABCM with $\omega=qv_{\rm{TA/LA}}$ for intravalley and $g$-process scattering with acoustic phonon modes ($v_{\rm{TA/LA}}$ are the phonon velocities). For intravalley scattering with long-wavelength optical phonon modes, it replaces the phonon energy with a constant $\hbar \omega_{\rm{op}}$=63.5~meV. Finally, it employs the elastic continuum approximation for phonon polarizations (Table~\ref{tab:intra_g_M_Sn} with $\mathbf{n} \|[001]$).

Numerical integrations of Eq.~(\ref{eq:tau_sf}) were performed using a grid spacing of $0.01\times 2\pi/a$ in $\mathbf{k}$ space. This grid leads to converged results for intermediate and high temperatures ($T>50 K$). We take advantage of the eight-fold symmetry of the $\Delta$ axis valleys and reduce the intensity of computation whenever possible. The edges of the irreducible wedge are weighted to prevent overlap with their neighbors.  Also helpful is the strict equality $\mathcal{F}_{\rm{MB}}(\mathbf{k}_1)n(\mathbf{q}) = \mathcal{F}_{\rm{MB}}(\mathbf{k}_2)[n(\mathbf{q})+1]$ when $E(\mathbf{k}_2)-E(\mathbf{k}_1) = \hbar \omega(\mathbf{q})$. Typical execution times are as follows: tens of seconds for `ellip+$\omega(q)$', tens of minutes for `kp+ABCM' and a few days (with 64 CPU cores) for `EPM+ABCM'. The bottleneck of the `EPM+ABCM' computation speed lies in calculation of individual matrix elements. This calculation involves the product of states and interactions written, respectively, in vector and matrix forms with a basis of hundreds of plane waves.\cite{footnote_computing_time}

\subsection{Fixed spin orientation along the valley axis}

\begin{figure}
\centering
\includegraphics[width=8.7cm, height=4.0cm]{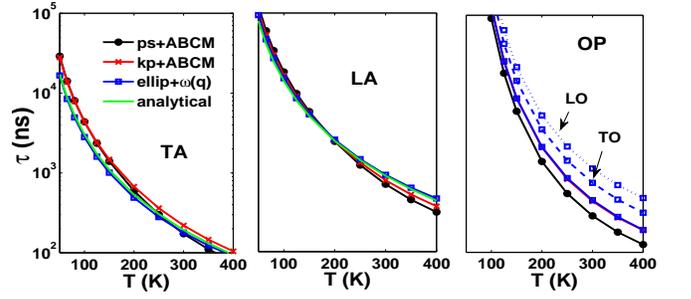}
\caption[intravalley tau vs T] {(Color online) Intravalley $\tau_{\rm{sf}}(T)$ induced by scattering with TA, LA and optical (OP) phonon modes. The relaxation time is contributed from all valleys and the spin orientation is set along one of the valley axes ($\Delta$ axis). Analytical curves of the acoustic modes follow Eq.~(\ref{eq:tau_ac}). Other curves refer to numerical approaches with calculation intensities that depend on details of the band structure, phonon energy, and polarization vectors (see text). Due to the crossing of TO and LO dispersion curves, we do not separate their contributions in the ABCM calculation.} \label{fig:intra_tau}
\end{figure}

We separately study the spin relaxation time due to intravalley, $g$-process, and $f$-process spin flips. Figure~\ref{fig:intra_tau} shows results of the intravalley spin relaxation time by integration of Eq.~(\ref{eq:tau_sf}) using the aforementioned numerical procedures. The figure also includes analytical curves for scattering with acoustic phonon modes (left and middle panels). These analytical integrations are carried out by employing an elastic scattering approximation and a high temperature limit for phonon population $n(q)\approx k_{\rm{B}} T/\hbar \omega(q)$. Together with the mentioned simplifications in numerical integrations, these common practices allow us to accurately calculate the relaxation time by employing the relevant spin-flip matrix elements in Table~\ref{tab:intra_g_M_Sn} ($|M_{\rm{TA/LA}}^{\rm{intra}}|^2$ with $\mathbf{n}\| [001]$ and $\mathbf{n}\| [100]$). The integrated spin relaxation rate reads
\begin{eqnarray}
\!\!\!\!\frac{1}{\tau^{\scriptscriptstyle{\rm{TA(LA)}}}_{{\rm{sf}},i}}\!\!&=&\!\!\!\!\zeta_{\rm{TA(LA)}}\frac{16\sqrt{2}m_d^{5/2}}{3\pi^{3/2}\hbar^6\rho }\!\! \!\left(\frac{|\eta|}{\Delta_C} D'_{xy}\right)^2\!\!\!\!\! (k_{\rm{B}} T)^{\frac{5}{2}},\nonumber\\
\zeta_{\rm{TA}}&=&\frac{1}{v^2_{\rm{TA}}}\left[-\frac{r^{4/3}(23-12r+4r^2)}{3(1-r)^3}\right.\nonumber\\
&&\qquad\left.+\frac{r^{5/6}(3+2r)}{(1-r)^{7/2}}\arcsin(\sqrt{1-r})\right]\nonumber\\
\zeta_{\rm{LA}}&=&\frac{1}{v^2_{\rm{LA}}}\left[\frac{r^{4/3}(3+16r-4r^2)}{3(1-r)^3}\right.\nonumber\\
&&\qquad\left.+\frac{r^{5/6}(1-6r)}{(1-r)^{7/2}}\arcsin(\sqrt{1-r})\right]\label{eq:tau_ac}
\end{eqnarray}
where $\tau^{\scriptscriptstyle{\rm{TA}}}_{{\rm{sf}},i}$ and $\tau^{\scriptscriptstyle{\rm{LA}}}_{{\rm{sf}},i}$ are, respectively, the intravalley spin relaxation times due to scattering with long-wavelength TA and LA phonon modes. Table~\ref{tab:params_2} lists the values of all parameters in the above expression. To enable an accurate analytical integration in comparison with the full numerical integration, the explicit dependence on the band structure anisotropy has been considered.\cite{footnote_analytical_integration} This anisotropy is expressed in Eq.~(\ref{eq:tau_ac}) via $r$$\,$$=$$\,$$m_t$/$m_l$ which denotes the ratio between the longitudinal and transverse effective masses of the electron (with respect to the valley axis).
As seen from the middle and left panels of Fig.~\ref{fig:intra_tau}, the analytical integrations  match very well with the most detailed numerical integration. These figures also show that $1/\tau_{\rm{sf}}(T)$ of the `EPM+ABCM' numerical results decrease slightly faster than $T^{-5/2}$ for both scattering with TA and LA phonon modes. It indicates the dependence of $|M_{\rm{sf}}|$ on higher-order wavevector components when $|\mathbf{q}|$ gradually increases.\cite{Cheng_PRL10} The figure shows that only minor changes are introduced in all intravalley processes when replacing the detailed conduction band structure with spheroid dispersion and the numerical phonon data with the analytical approximation. This behavior supports the validity of the invoked approximations, including the use of an elastic continuum approximation. The stronger deviation of the  `EPM+ABCM' curve in the case of optical phonon modes shows the effect of higher-order matrix element terms.  Although the contribution to spin relaxation from LA and optical phonon modes is negligible at low temperatures (compared with TA modes), their effect should be considered at room temperature (especially the optical modes).

\begin{table}
\caption{\label{tab:params_2}
Parameter values in Eqs.~(\ref{eq:tau_ac}) and (\ref{eq:tau_g}).}
\renewcommand{\arraystretch}{1.2}
\tabcolsep=0.2 cm
\begin{tabular}{c|ll|l}
\hline \hline 
$\rho$              & 2.33                & gr/cm$^3$                    &                                                                  \\
$m_t$               & 0.19                & $m_0$                        &   $(m_0^{-1}+2P^2/\hbar^2 E_{g,\scriptscriptstyle{X}})^{-1}$     \\
$m_l$               & 0.92                & $m_0$                        &                                                                  \\
$m_d$               & 0.32                & $m_0$                        &           $(m_lm_t^2)^{1/3}$                                     \\ \hline
$r$                 & 0.2                 & $m_t/m_l$                    &                                                   \\
$v_{\rm{TA}}$       & 5$\times$10$^5$     & cm/s                         &                                                   \\
$v_{\rm{LA}}$       & 8.7$\times$10$^5$   & cm/s                         &                                                   \\
$\Delta_C$          & 0.5                 & eV                           &   Table~\ref{tab:params}                          \\
$|\eta|$            & 16.7                & meV$\cdot$$a/2\pi$           &   Table~\ref{tab:params}                          \\
$D'_{xy}$           & 6                   & eV                           &   Eq.~(\ref{eq:D_xy})                             \\ \hline
$E^g_q$             & 21                  & meV                          &                                                   \\
$D_{gs}$            & 0.1                 & eV                           &   Eq.~(\ref{eq:Dgs})                              \\
\hline \hline
\end{tabular}
\end{table}

\begin{figure}
\includegraphics[width=4.5cm, height=4.4cm]{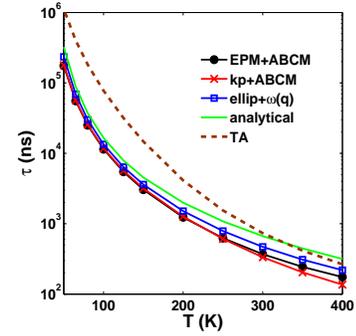}
\caption{(Color online) $g$-process $\tau_{\rm{sf}}(T)$ from LA phonon at different levels of approximation (see text). The `TA' curve is calculated only by `EPM+ABCM'. The relaxation time is contributed from all valleys and the spin orientation is set along one of the valley axes. In `ellip+$\omega(q)$', the phonon velocity is the same as that of a long-wavelength LA phonon mode. In the analytical curve the phonon energy is constant (21~meV). } \label{fig:g_tau}
\end{figure}

We study the spin relaxation time due to $g$-process spin flips in a similar way. Results of the analytical and numerical integrations are presented in Fig.~\ref{fig:g_tau}. The analytical relaxation rate due to scattering with LA phonon modes (dominant effect) is reached by integrating Eq.~(\ref{eq:tau_sf}) with spin-flip matrix elements taken from Table~\ref{tab:intra_g_M_Sn} (for the  case of $\mathbf{n}\| [001]$ and $\mathbf{n}\| [100]$),
\begin{eqnarray}
\frac{1}{\tau^{\scriptscriptstyle{\rm{LA}}}_{{\rm{sf}},g}}= \frac{\sqrt{2}D^2_{gs} m_t m^{\frac{3}{2}}_d E^g_q } {16 \pi^{\frac{3}{2}} \hbar^4 \rho} \frac{K_2(E^g_q/2k_{\rm{B}} T)}{\sqrt{k_{\rm{B}} T} \sinh(E^g_q/2k_{\rm{B}} T)}. \label{eq:tau_g}
\end{eqnarray}
Values of the scattering constant, phonon energy and effective masses are listed in Table~\ref{tab:params_2}. $K_i$ is the $i^{th}$ order modified Bessel function of the second kind.  Figure~\ref{fig:g_tau} also shows a full numerical curve due to scattering with TA phonon modes. The relative weight of the TA part clearly increases with temperature. Also can be seen at the high temperature end, is a relatively large deviation between the analytical and `ellip+$\omega(q)$' curves. This deviation is caused by using a constant phonon energy instead of a linear dispersion relation. The energy difference affects the phonon population.

\begin{figure}
\includegraphics[width=8.7cm, height=4.1cm]{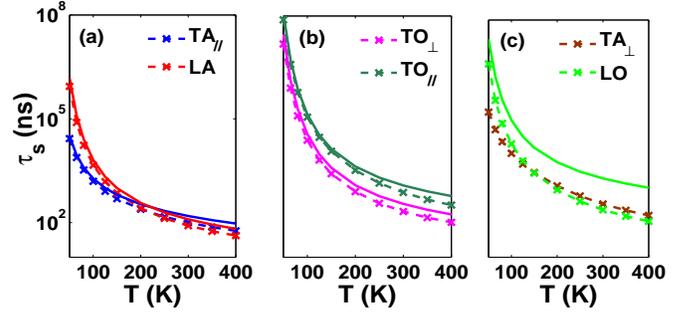}
\caption[f_tau] {(Color online) Analytical (solid lines) and numerical (-x- lines) results of $\tau_{\rm{sf}}(T)$ in $f$-process spin flips where each phonon mode is represented by a different color.  The relaxation time is from all valleys and the spin orientation is set along one of the valley axes. $\perp$ and $\|$ are taken with respect to the cross product of valley centers (direction of $\mathbf{k}_{0,1} \times\mathbf{k}_{0,2}$). The analytical contribution from $\rm{TA}_{\bot}$ modes vanishes (at the zeroth order).} \label{fig:f_tau}
\end{figure}

Finally, we study the spin relaxation due to $f$-process spin flips. Analytical and `EPM+ABCM' numerical integrations of Eq.~(\ref{eq:tau_sf}) are presented in Fig.~\ref{fig:f_tau} for each of the phonon modes. In the analytical integration, we have used the approximations of a spheroidal energy dispersion in the conduction band, and of wavevector independent spin-flip matrix elements and phonon energies (Sec.~\ref{sec:f-process}). The analytical integration becomes relatively simple for each of the nonvanishing modes and for spin orientation along the valley axis,
\begin{eqnarray}
\frac{1}{\tau^{\scriptscriptstyle{{\Sigma_i}}}_{{\rm{sf}}}}= \frac{\sqrt{2}m_d^{\frac{3}{2}}  } {3 \pi^{\frac{3}{2}} \hbar^2 \rho }A_i D^2_{\Sigma_i} \frac{K_1(E^f_q/2k_{\rm{B}} T)}{\sqrt{k_{\rm{B}} T} \sinh(E^f_q/2k_{\rm{B}} T)}. \label{eq:tau_f}
\end{eqnarray}
Values of the scattering constants and phonon energies ($D_{\Sigma_i}$ and $E^f_q$) are listed in Table~\ref{tab:f_character} where $D_{\Sigma_1}$ stands for $D_{\Sigma_1,s}$. The value of $A_i$ is obtained by summing $|M_{{\rm{sf}}}|^2$ over 12 pairs of valleys. Following the results of Sec.~\ref{sec:f_para}, we get that $A_1=8$ and $A_{2/3}=16$. In calculating the values of $A_i$ we have used the facts that for $\Sigma_1$, there is no coupling between valleys of the $x$-$y$ plane whereas the coupling is $D^2_{\Sigma_1}$ between any of the remaining eight pairs of valleys (assuming that $\mathbf{n}$$\,$$\|$$\,$$\mathbf{z}$). For $\Sigma_{2/3}$, the coupling is $|(1+i)D_{\Sigma_{2/3}} |^2$ between each of the 4 pairs of the $x$-$y$ plane and $D^2_{\Sigma_{2/3}}$ between each of the remaining 8 pairs.

Figure~\ref{fig:f_tau} shows that the dominant scattering is with phonon modes of the LA and $\rm{TA}_{\|}$ ($\Sigma_1$ and $\Sigma_3$ symmetries; see Table~\ref{tab:f_character}). Their temperature trends are correctly predicted by the wavevector-independent analytical analysis. The same applies for scattering with phonon modes of the $\rm{TO}_{\bot}$ and $\rm{TO}_{\|}$. After a qualitative analysis we find that the big difference between analytical and numerical results in the LO phonon case is caused by quadratic-wavevector terms. This wavevector dependence is further complicated by even higher-order terms when the electron states are further away from valley centers. For the $\rm{TA}_{\bot}$ phonon case, we found in Sec.~\ref{sec:f-process} that there is no wavevector independent term. The numerical result is attributed to linear terms and it is non-negligible due to the higher phonon population of this mode [lowest energy; see Fig.~\ref{fig:BZ_phonon}(a)]. In general, wavevector dependent terms contribute at all modes when the temperature increases. The dependence on the electron wavevectors ($\mathbf{k}_{1,2}$) can be similarly analyzed between decomposed $\mathbf{k}\cdot\mathbf{p}$ basis states (as in the intravalley and $g$-process cases). We do not make an explicit derivation of these terms since the focus is on the leading-order term contribution (wavevector independent in the $f$-process case).

\subsection{Spin orientation anisotropy coupled with symmetry breaking mechanisms}\label{sec:lifetime_oritentaion}

The spin relaxation time is in general a function of spin orientation [$\tau_{\rm{sf}}(\mathbf{n})$]. It can be obtained by integration with the general matrix element expressions derived in
Eqs.~(\ref{eq:Mm_Sigma1_Sn})-(\ref{eq:Ms_Sigma3_Sn}) and (\ref{eq:ac_Sn})-(\ref{eq:g_Sn}). We can firstly calculate $\tau_{\rm{sf}}(\mathbf{n})$ from one valley (intravalley) or a pair of valleys (intervalley). The total $\tau_{\rm{sf}}(\mathbf{n})$ is a summation from all of the involved valleys, whose individual $\tau_{\rm{sf}}(\mathbf{n})$ can be related to the calculated one by a proper coordination rotation.

\begin{figure}
\includegraphics[width=8cm, height=6.8cm]{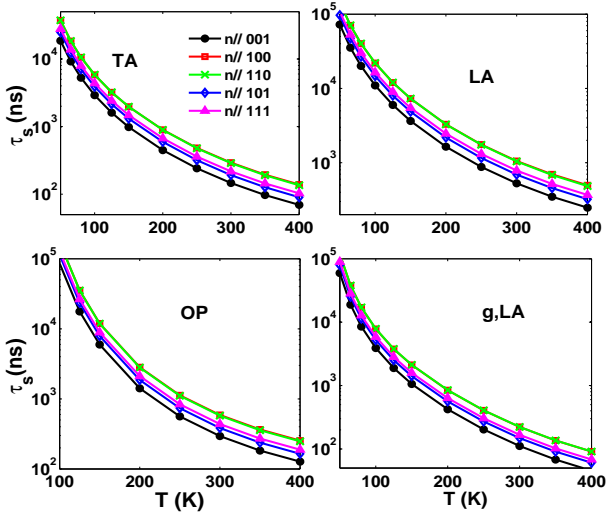}
\caption {(Color online) $\tau_{\rm{sf}}(T, \mathbf{n})$ of electrons in the $z$ valley (intravalley) and $\pm z$ valleys ($g$-process) for spin orientation $\mathbf{n}$ along high symmetry directions.  The curves of $\mathbf{n}\|[001]$ and $[110]$ overlap each other since integrations of their respective spin-flip matrix elements over the azimuthal angle of $\mathbf{q}$ yield the same result.  In this sense, only the projection $\hat{n}_z$ on the valley axis is relevant.} \label{fig:intra_g_tau_zv_Sn}
\end{figure}

We first discuss the integrated effect of changing $\mathbf{n}$ on intravalley and $g$-process spin flips. Figure~\ref{fig:intra_g_tau_zv_Sn} shows the temperature dependence of $\tau_{\rm{sf}}(\mathbf{n})$ due to intravalley scattering of electrons in the $z$-valley with TA, LA, and optical modes. It also shows $\tau_{\rm{sf}}(\mathbf{n})$ due to the dominant intervalley scattering between the $\pm z$ valleys ($g$-process). These results were calculated using the `kp+ABCM' integration procedure with matrix elements taken from Table~\ref{tab:intra_g_M_Sn}. Each panel shows results of $\mathbf{n}$ along all of the inequivalent high-symmetry crystallographic directions. The spin lifetime in a given valley increases with decreasing the projection of the spin orientation on the valley axis. This effect implies that suppression of the spin relaxation in one valley is compensated by enhanced relaxation at perpendicular valleys. As a result, the spin lifetime due to intravalley and $g$-process scattering from all valleys is expected to have a diminished dependence on the spin orientation. Nonetheless, Fig.~\ref{fig:intra_g_tau_zv_Sn} can be seen as a simplified example of how different spin orientations, coupled with a symmetry-breaking mechanism, can lead to different experimentally measurable quantities. For example, stress or electrical fields can selectively change the electron distribution of different valleys leading to a significant electron population only at valleys along a certain axis. In that case, the total spin lifetime will present a pronounced degree of anisotropy when changing the spin orientation.

For $f$-process spin flips, changing the spin orientation results in slightly more involved relations between spin relaxation times of different pairs of valleys. The wavevector integration of Eq.~(\ref{eq:tau_sf}) is not affected by the spin orientation ($M_{\rm{sf}}^f$ are wavevector independent). Therefore, one only needs to obtain the values of $A_i$ in Eq.~(\ref{eq:tau_f}) for each of the $\Sigma_i$ symmetries. In this application, the 12 pairs of involved valleys can be divided into 6 groups. Each group consists of two pairs related by space inversion operation and they always have the same value of $|M_{\Sigma_i}(\mathbf{n})|=\langle \mathbf{k}_2,\Downarrow_{\mathbf{n}} |H_{\Sigma_i}| \mathbf{k}_1,\Uparrow_{\mathbf{n}} \rangle$. For $\pm x\leftrightarrow \pm y$ pairs, we use Eqs.~(\ref{eq:Ms_Sigma1_Sn}), (\ref{eq:Ms_Sigma2_Sn}) and (\ref{eq:Ms_Sigma3_Sn}), and denote them with $M_{0,\Sigma_i}(\mathbf{n})$. Using crystal symmetry, the matrix elements of all other pairs relate to $M_{0,\Sigma_i}(\mathbf{n})$ by,
\begin{eqnarray}
\pm x\leftrightarrow \mp y &:& M_{\Sigma_i}(\mathbf{n})=M_{0,\Sigma_i}({n}_x, -{n}_y,{n}_z),\nonumber\\
\pm x\leftrightarrow \pm z &:& M_{\Sigma_i}(\mathbf{n})=M_{0,\Sigma_i}({n}_x, {n}_z,{n}_y),\nonumber\\
\pm x\leftrightarrow \mp z &:& M_{\Sigma_i}(\mathbf{n})=M_{0,\Sigma_i}(n_x, -n_z,-n_y), \nonumber\\
\pm y\leftrightarrow \pm z &:& M_{\Sigma_i}(\mathbf{n})=M_{0,\Sigma_i}(n_z, n_y,n_x),\nonumber\\
\pm y\leftrightarrow \mp z &:& M_{\Sigma_i}(\mathbf{n})=M_{0,\Sigma_i}(-n_z, n_y,-n_x).
\end{eqnarray}
After summing all of the contributions, one finds that $A_1=8$ and $A_{2/3}=16$ [the same as those obtained in Eq.~(\ref{eq:tau_f})] with the direction of $\mathbf{n}$ first expressed in terms of $(\theta, \phi)$  and then substituted into expressions of $M_{0,\Sigma_i}$. Therefore, when all valleys are equally populated then the total $\tau_{{\rm{sf}}}$ is invariant of $\mathbf{n}$ (to the leading order). On the other hand, when symmetry breaking effects are introduced then valley repopulation brings in a dependence of the spin relaxation time on the spin orientation. For example, consider an internally strained structure in which the $f$-process scattering with $\pm z$ valleys is suppressed (i.e., the $\pm x$ and $\pm y$ valleys are equivalent and have sufficiently lower energy). In such a structure, the above analysis reveals that if the spin is oriented along $z$ then one should assign in Eq.~(\ref{eq:tau_f}) values of $A_1=0$ and $A_{2,3}=8$. Similarly, for orientation along $x$ or $y$ one should assign $A_1=4$ and $A_{2,3}=4$. Using the parameters of this stressed configuration, the $f$-process spin relaxation time is estimated to be shortened by $\sim$50\% when changing the spin orientation from the $z$-axis to the perpendicular plane.

\section{summary and outlook}\label{sec:conclusion}

We have presented a comprehensive analysis of all phonon-induced spin relaxation processes in bulk silicon. The applied temperature and doping regime of this mechanism has been identified, among other mechanisms (Sec.~\ref{sec:phases}). In decreasing order of contributions to spin relaxation, detailed expressions of $f$-process, intravalley and $g$-process matrix elements have been derived and their dependence on the spin orientation are unveiled. We have elaborated on the wavevector dependence and symmetry properties of each spin-flip process. In analogy to Herring and Vogt theory on momentum relaxation in silicon,\cite{Herring_PR56} this work unravels the magnitudes and symmetries of all phonon-induced spin relaxation processes in silicon.

In studying the $f$-process spin flips, double group selection rules are used to obtain the wavevector-independent (leading order) matrix elements. Spin orientation dependent spin-flip (and spin-conserving) matrix elements were expressed in terms of scattering constants $D_{\Sigma_i s}$ (and also $D_{\Sigma_1 m}$), for electron-phonon interaction with $\Sigma_i$ symmetry.

Intravalley spin flips were studied by using a combination of single group theory, $\mathbf{k}\cdot \mathbf{p}$ perturbation method and rigid-ion model. The spin-dependent coupling between the expanded basis functions and symmetrized interactions is  formulated via selection rules. This approach allowed us to derive the leading order intravalley matrix elements [Eqs.~(\ref{eq:ac_Sn}) and (\ref{eq:op_Sn})] and to resolve their exact dependence on the phonon polarization ($\bm{\xi}$), its wavevector ($\mathbf{q}$), and on the spin orientation with respect to the valley axis [$\mathbf{n}(\theta,\phi)$]. By incorporating the diamond crystal structure into an elastic continuum model (expressing $\bm{\xi}$ in terms of $\mathbf{q}$), we have derived appealing forms of these matrix elements (Table~\ref{tab:intra_g_M_Sn}). Finally, the analysis identifies the important band structure parameters. The $\eta$ parameter is a measure of the \textit{wavevector independent} spin-orbit coupling between conduction and valence states at the $X$ point (Table~\ref{tab:params}). The deformation potential parameter due to scattering with long-wavelength acoustic (optical) phonon modes is $D_{\rm{op}}$ ($D'_{xy}$), and it corresponds to interband coupling between the lowest pair of conduction bands [Eq.~(\ref{eq:D_xy}) and (\ref{eq:Dop})]. This coupling is brought by the proximity of the valley to the two-band degeneracy at the $X$ point. The $\Delta_C$ parameter denotes the energy gap between the lowest pair of conduction bands at the valley center.

A complete picture of the relation between intravalley spin-conserving and spin-flip processes has been provided (together with the detailed derivations in Appendix \ref{app:momentum} and \ref{app:details_intra_g}). This comparison reveals important physical aspects that are being overlooked when relating spin and momentum relaxation times via the shift of the $g$-factor (conventional approach in quantifying the spin lifetime due to the Elliott-Yafet relaxation mechanism).

$g$-process spin flips are studied in a similar way to the intravalley case. In spite of the opposite valley positions of an electron before and after scattering, Kramers conjugation relation allows us to expand the electronic states by basis functions of the same $X$ point. One result of this relation is that the matrix elements depend on the average between the initial and final electron wavevectors [$\mathbf{K}=\tfrac{1}{2}(\mathbf{k}_1+\mathbf{k}_2)$] rather than on their difference ($\mathbf{q}=\mathbf{k}_1-\mathbf{k}_2$). The spin orientation dependent matrix element of the $g$-process is provided in Eq.~(\ref{eq:g_Sn}). The involved scattering constant has a large contribution from dilation and uniaxial deformation potential constants [Eq.~(\ref{eq:Dgs})]. Comparing the derived matrix elements with respective results of independent numerical calculations shows that our analytical approach provides accurate spin-flip amplitudes at all scattering angles for both intravalley and $g$-process cases (Figs.~\ref{fig:intra_M_q} and \ref{fig:g_M}).

Our analysis provides insights into which parts of the interaction (spin-independent `Elliott' or spin orbit coupling `Yafet') dominate the phonon-induced spin relaxation in silicon.  In silicon, the sum of Elliott and Yafet contributions vanishes at the zero and first order of intravalley scattering with acoustic phonon modes.\cite{Yafet_SSP63} At the leading order of this scattering (quadratic-in-$\mathbf{q}$), we have shown that the Elliott part dominates the spin relaxation. The Elliott contribution is also shown to dominate the spin relaxation due to intravalley scattering with optical phonon modes and $g$-process intervalley scattering. The latter two process are, respectively, linear in $\mathbf{q}$ and $\mathbf{K}$. Yafet contributions, however, cannot be completely ignored in $g$-process scattering where the deformation potential constant is affected by the spin-orbit coupling [Eq.~(\ref{eq:D_so})]. In silicon, Elliott and Yafet contributions are comparable only in the $f$-process.

We have derived the spin lifetime due to each of the spin-flip processes by integrating its leading matrix elements. Analytical forms are given for intravalley scattering with acoustic phonon modes [Eq.~(\ref{eq:tau_ac})] and for both types of intervalley scattering [Eqs.~(\ref{eq:tau_g}) and (\ref{eq:tau_f})]. Comparison of these results with numerical calculations at different levels of approximation show good agreements (Figs.~\ref{fig:intra_tau}, \ref{fig:g_tau} and \ref{fig:f_tau}). The analysis also identifies the phonon modes which lead to strongest spin relaxation. The $f$-process is led by scattering with $\Sigma_1$ and $\Sigma_3$ symmetry (LA and TA$_{\rm{\|}}$) phonon. Intravalley spin relaxation is led by scattering with TA phonon. Intravalley contributions from scattering with LA and optical phonon become comparable to the TA's at room temperatures. $g$-process is led by scattering with LA phonon. We have also considered the secondary contribution to $g$-process spin relaxation from scattering with TA phonon (quadratic-wavevector dependence but a larger phonon population).

\subsection*{Outlook}\label{sec:outlook}

Results of this work shed light on new research directions in group IV spintronics. By having a thorough understanding of the underlying physics, one can devise a means to enhance the spin lifetime of room temperature silicon spintronic devices. Quenching of the dominant $f$-process by certain stress configurations is the first step in this direction.\cite{Dery_APL11,Tang_PRB12}  In this case, only the valleys along one crystallographic axis are practically populated with electrons. To further suppress the remaining relaxation processes (intravalley and $g$-process spin-flips),  one can make use of the slower relaxation when the spin orientation is perpendicular to this crystallographic axis (Fig.~\ref{fig:intra_g_tau_zv_Sn}).

In addition to quenching the $f$-process and further optimization by spin orientation, one can also impose geometrical constraints on the transport. For example, promising candidates seem to be stressed silicon wires with a cross-section area that is large enough to prevent detrimental surface effects but is small enough to restrict the phase-space for scattering. Having the wire axis parallel to the axis of populated valleys would allow one to achieve significantly longer spin lifetimes. This fact can be seen from the detailed intravalley and $g$-process matrix elements: while in these structures forward and backward scattering with respect to the axis of the wire (and populated valleys) dominate the transport, these types of scattering would not be accompanied by spin flips (e.g., for $z$ valley electrons, assign $q_x \approx q_y \approx 0$ in Table~\ref{tab:intra_g_M_Sn}). This example shows the insights one can gain from understanding the symmetries of the matrix elements rather than only having a knowledge of the integrated effect.

The complete set of matrix elements is also instrumental in calculating the spin relaxation in the presence of large electric fields. Due to the mass anisotropy, the valley population depends on the direction of the field. In addition, the field can lead to a large departure of the electron distribution from equilibrium conditions.\cite{Canali_PRB75} As a result of these effects, intervalley processes are enabled already at low temperatures and certain scattering processes are enhanced.\cite{Li_PRL12}  Using the dependence of the relaxation on spin orientation and scattering directions, one can accurately model and understand the spin relaxation in these conditions.

The presented theory identifies a handful of scattering and band structure constants which have not been experimentally determined yet. Evidently, the most important constants are $D_{\Sigma_{1(3)}s}$ of the $f$-process scattering with phonon modes along the LA (TA$_\|$) branches, and $\Delta_X$ which denotes the spin-orbit coupling between conduction and valence states at the $X$ point.  In the absence of experimental data, we have used the empirical pseudopotential method to calculate their values (Tables~\ref{tab:f_character} and \ref{tab:params}). To determine these constants experimentally one should resolve various contributions to the measured spin lifetime. Intravalley, $g$ and $f$-processes have different dependencies on the wavevector components, phonon polarization and spin orientation. In addition, the energies of the respective phonon modes are different. As a result, the measured spin lifetime of each of these processes has a unique dependence on temperature [Eqs.~(\ref{eq:tau_ac})-(\ref{eq:tau_f})], and it can be clearly resolved by application of a symmetry breaking mechanism (e.g., stress or electric fields as mentioned before).

Finally, the theoretical approaches presented in this paper can be used to study the spin relaxation of materials with different symmetry groups and consequently different wavevector-order analysis (e.g., germanium and graphene with respective utilization of the space groups at the $L$ and $K$ points of their Brillouin zones). Results of such a study provide a clear picture of preferred scattering angles, spin orientation and dominant spin relaxation mechanisms. As in the case of silicon, having this information provides guidance in tailoring the spin relaxation by application of stress, external fields or geometrical constraints. When such external influences become too large, one can repeat the steps of the presented procedure after adding the external perturbation explicitly in the Hamiltonian and interaction terms.

This work is supported by DOD/AF/AFOSR FA9550-09-1-0493 and by NSF ECCS-0824075. 

\appendix

\section{detailed application of double group theory for $f$-process matrix elements}\label{app:f}
In this appendix, we intend to express interaction matrix elements $\langle \mathbf{k}_2,\mathbf{s}_2 |H_{\Sigma_i}| \mathbf{k}_1,\mathbf{s}_1 \rangle$  in terms of $\mathcal{N}_{\Sigma_i}$ independent constants. First, by the general time reversal and space inversion symmetries we can connect different matrix elements of each phonon mode [Eq.~(\ref{eq:TS})].

For $\Sigma_1$ mode, the $(\rho_z|\tau)$ operation equates the spin-flip matrix element to negative of itself (seen from the character of $\Sigma_1$ and the IR matrix of $D_{\Delta_6}=D_{\Delta_1} \times D_{1/2}$ in Table~\ref{tab:f_character}). We show this example explicitly,
\begin{eqnarray}
{H}_{\Sigma_1}&\rightarrow&{H}_{\Sigma_1},\nonumber\\
| \mathbf{k}, \Uparrow_z \rangle&\rightarrow& e^{ \frac{-ik_0 a}{4}}\times(-i)| \mathbf{k}, \Uparrow_z \rangle,\nonumber\\
 | \mathbf{k}, \Downarrow_z \rangle&\rightarrow& e^{ \frac{-ik_0 a}{4}}\times i| \mathbf{k}, \Downarrow_z \rangle, \nonumber
\end{eqnarray}
remembering that basis states in $D_{\Delta_6}$ are $|\mathbf{k},\Uparrow_z(\Downarrow_z)\rangle$. $(\rho_z|\tau)$ and other symmetry operations do not provide constraints on the spin-conserving matrix element. Since $\mathcal{N}_{\Sigma_1}=2$ [Eq.~(\ref{eq:N_0_Sigma})], there are two real constants $D_{\Sigma_1,m}$ and $D_{\Sigma_1,s}$. The physical significance of these two constants will become clear later. They are defined such that
\begin{eqnarray}
\langle \mathbf{k}_2, \Uparrow_z |H_{\Sigma_1}| \mathbf{k}_1, \Uparrow_z \rangle = D_{\Sigma_1,m} + i D_{\Sigma_1,s}. \label{eq:Sigma_1_app}
\end{eqnarray}

For $\Sigma_2$, spin-conserving matrix element vanishes (can be seen, for example, by applying the $T\sigma_z$ operation). The spin-flip matrix element is
\begin{eqnarray}
\langle \mathbf{k}_2, \Downarrow_z |{H}_{\Sigma_2}| \mathbf{k}_1, \Uparrow_z \rangle & \stackrel{\rm{TR}}{=} & -\langle \mbox{-}\mathbf{k}_1, \Downarrow_z |{H}_{\Sigma_2}| \mbox{-}\mathbf{k}_2, \Uparrow_z \rangle\nonumber\\
&\stackrel{(\rho_{xy}|\tau)}{=} & i\langle \mathbf{k}_2, \Uparrow_z |{H}_{\Sigma_2}| \mathbf{k}_1, \Downarrow_z \rangle, \nonumber
\end{eqnarray}
by time reversal and $(\rho_{xy}|\tau)$ sequentially. Combined with Eq.~(\ref{eq:TS_s}), we are led to
\begin{eqnarray}
\langle \mathbf{k}_2, \Downarrow_z |{H}_{\Sigma_2}| \mathbf{k}_1, \Uparrow_z \rangle = D_{\Sigma_2} - i D_{\Sigma_2},\label{eq:Sigma_2_app}
\end{eqnarray}
where $D_{\Sigma_2}$ is the independent real constant.

Applying the same operations for $\Sigma_3$ and noting that the only difference from $\Sigma_2$ mode is the sign of $\chi_{\scriptscriptstyle\Sigma_3}(\rho_{xy}|\tau)$, we get
\begin{eqnarray}
\langle \mathbf{k}_2, \Downarrow_z |{H}_{\Sigma_3}| \mathbf{k}_1, \Uparrow_z \rangle = D_{\Sigma_3} + i D_{\Sigma_3},\label{eq:Sigma_3_app}
\end{eqnarray}
where $D_{\Sigma_3}$ is the independent real constant. As shown in Eq.~(\ref{eq:N_0_Sigma}), the $\Sigma_4$ phonon symmetry does not couple electrons of any spin species ($\mathcal{N}_{\Sigma_4}=0$).

\section{Group $G_{32}^2$}\label{app:G32}

Symmetry operations of the $G_{32}^2$ group are listed in Table~\ref{tab:OperatorPolarVector}. The table refers to the $X$ point in the $z$ direction. Tables of $X$ points in the $x$ and $y$ directions are derived by cyclic permutations. From Table~\ref{tab:OperatorPolarVector} one can study how vectors and axial-vectors are transformed under the 32 group elements. Character table of the $G_{32}^2$ group is listed in Table~\ref{tab:CharactorTable}. From the character table one can decompose the direct products of $\{X_1,X_4\}$ into direct sums of IRs [Eq.~(\ref{eq:XX})]. One can also construct all sorts of direct product rules from IRs that present components of vectors, axial-vectors or (anti)symmetric potentials [Eq.~(\ref{eq:operator_IR})].

\begin{table} [h]
{ 
\caption{List of operations of the $G_{32}^2$ group. Notations follow Ref.\onlinecite{Lax_PR61}.}
\label{tab:OperatorPolarVector}
\renewcommand{\arraystretch}{1.2}
\begin{tabular}{c|l}
\hline \hline
Class symbol  & Operations \\ \hline
$C_1$  & $(\epsilon|0)$\\
$C_2$  & $(\delta_{2x}|0)$, $(\delta_{2y}|0)$, $(\delta_{2x}|t_{xy})$, $(\delta_{2y}|t_{xy})$ \\
$C_3$  & $\delta_{2z}$\\
$C_4$  & $(\delta_{2xy}|\tau)$,$(\delta_{2\bar{x}y}|\tau+t_{xy})$\\
$C_5$  & $(\delta_{4x}|\tau)$,$(\delta^{-1}_{4x}|\tau)$,$(\delta_{4x}|\tau+t_{xy})$,$(\delta^{-1}_{4x}|\tau+t_{xy})$\\
$C_6$  &  $(\rho_z|\tau)$, $(\rho_z|\tau+t_{xy})$\\
$C_7$  &  $(\rho_y|\tau)$, $(\rho_x|\tau)$,$(\rho_y|\tau+t_{xy})$, $(\rho_x|\tau+t_{xy})$\\
$C_8$  &  $(i|\tau)$, $(i|\tau+t_{xy})$\\
$C_9$  &  $(\rho_{xy}|0)$,$(\rho_{\bar{x}y}|0)$\\
$C_{10}$  &  $(\sigma_{4z}|0)$, $(\sigma^{-1}_{4z}|0)$, $(\sigma_{4z}|t_{xy})$, $(\sigma^{-1}_{4z}|t_{xy})$\\
$C_{11}$  &  $(\rho_{xy}|t_{xy})$,$(\rho_{\bar{x}y}|t_{xy})$\\
$C_{12}$  &  $(\delta_{2\bar{x}y}|\tau)$,$(\delta_{2xy}|\tau+t_{xy})$\\
$C_{13}$  & $(\delta_{2z}|t_{xy})$ \\
$C_{14}$  & $(\epsilon|t_{xy})$ \\\hline
\end{tabular}
}
\end{table} $\,$
\begin{table}
{ 
\caption{Character table of the $G_{32}^2$ group}
\label{tab:CharactorTable}
\renewcommand{\arraystretch}{1.2}
\begin{tabular}{c|cccccccccccccc}
\hline \hline
$h_i$  & 1   &4    & 1   & 2   &4    &2    &4    &2    &2    &4     &2     &2 &1 &1  \\ \hline
       &$C_1$&$C_2$&$C_3$&$C_4$&$C_5$&$C_6$&$C_7$&$C_8$&$C_9$&$C_{10}$&$C_{11}$&$C_{12}$&$C_{13}$&$C_{14}$ \\ \hline
$M_1$  & 1 &1& 1& 1 &1 &1 &1 &1 &1 &1 &1 &1 &1 &1\\
$M_2$  & 1 &1& 1&-1 &-1 &1 &1 &1 &-1 &-1 &-1 &-1 &1 &1\\
$M_3$  & 1 &-1& 1& -1 &1 &1 &-1 &1 &-1 &1 &-1 &-1 &1 &1\\
$M_4$  & 1 &-1& 1& 1 &-1 &1 &-1 &1 &1 &-1 &1 &1 &1 &1\\
$M_5$  & 2 &0& -2& 0 &0 &2 &0 &-2 &0 &0 &0 &0 &-2 &2\\
$M_1'$  & 1 &1& 1& 1 &1 &-1 &-1 &-1 &-1 &-1 &-1 &1 &1 &1\\
$M_2'$  & 1 &1& 1&-1 &-1 &-1 &-1 &-1 &1 &1 &1 &-1 &1 &1\\
$M_3'$  & 1 &-1& 1& -1 &1 &-1 &1 &-1 &1 &-1 &1 &-1 &1 &1\\
$M_4'$  & 1 &-1& 1& 1 &-1 &-1 &1 &-1 &-1 &1 &-1 &1 &1 &1\\
$M_5'$  & 2 &0& -2& 0 &0 &-2 &0 &2 &0 &0 &0 &0 &-2 &2\\ \hline
$X_1$  & 2 &0& 2& 0 &0 &0 &0 &0 &2 &0 &-2 &0 &-2 &-2\\
$X_2$  & 2 &0& 2& 0 &0 &0 &0 &0 &-2 &0 &2 &0 &-2 &-2\\
$X_3$  & 2 &0& -2& 2 &0 &0 &0 &0 &0 &0 &0 &-2 &2 &-2\\
$X_4$  & 2 &0& -2& -2 &0 &0 &0 &0 &0 &0 &0 &2 &2 &-2 \\\hline
\end{tabular} }
\end{table}

\section{$\mathbf{k}\cdot\mathbf{p}$ Hamiltonian parameters and the Partitioning}\label{app:partitioning}

Comparing directly the matrices form of Eq.~(\ref{eq:H_block}) and the Hamiltonian $H_1$ in Eq.~(\ref{eq:kp_perturb}), the four independent integral constants readily follow,
\begin{subequations}
\begin{align}
\hbar k_0 &=  \left\langle X_1^{2'}\left| p_z \right|X_1^{2'}\right\rangle,  \label{eq:k0} \\
m_0 P &= \hbar \left\langle X_4^x\left| p_y\right| X_1^{2'} \right\rangle,  \label{eq:P} \\
4m_0^2c^2 \Delta_{\scriptscriptstyle{X}} &=i \hbar \left \langle X_4^x\left| (\mbox{\boldmath$\nabla$}V\times\mathbf{p})_y\right|X_1^{2'}\right\rangle,  \label{eq:Delta_X} \\
4m_0^2c^2 \alpha &=  -i\hbar^2  \left\langle X_4^x\left|\nabla_{i} V \right|X_1^{2'}\right\rangle. \label{eq:alpha}
\end{align}
\end{subequations}
These constant can also be expressed as other equivalent integrals (e.g., $\hbar k_0 =  -\left\langle X_1^{1}\left| p_z \right|X_1^{1}\right\rangle$).

To analytically diagonalize the Hamiltonian matrix in Eq.~(\ref{eq:8 by 8}) we note that the $X$-point energy gap, $E_{g,\scriptscriptstyle{X}} \approx 4.3$~eV, is significantly larger than other energy scales. Therefore, we use degenerate second-order perturbation theory and lump the valence band effect onto the conduction band (L\"{o}wdin partitioning).\cite{Lowdin_JCM51} We get a reduced $4\times 4$ matrix,
\begin{eqnarray}
\bar{H}_{cc} &=& H_{cc} + \frac{H_{vc}^\dag H_{vc}}{E_{g,\scriptscriptstyle{X}}} \,\,,\label{eq:Hccbar}
\end{eqnarray}
whose four spin-dependent basis states are
\begin{eqnarray}
\bar{X}_L = \bar{X}_1 + \frac{H_{vc}}{E_{g,\scriptscriptstyle{X}}} \bar{X}_4 \,.  \label{eq:reduced_basis}
\end{eqnarray}
$\bar{X}_1=[|X_1^{2'},\uparrow\rangle, |X_1^{2'},\downarrow\rangle, |X_1^1,\uparrow\rangle, |X_1^1,\downarrow\rangle]^T$ and $\bar{X}_4=[|X_4^x,\uparrow\rangle, |X_4^x,\downarrow\rangle, |X_4^y,\uparrow\rangle, |X_4^y,\downarrow\rangle]^T$. Higher-order perturbation does not bring  dominant terms up to quadratic $k$ dependence. It has been explicitly checked.
The eigenvalues of $\bar{H}_{cc}$ that pertain to the energies of the upper and lower conduction bands read
\begin{eqnarray}
&&E_{\pm}(\mathbf{k}) =  \frac{\hbar^2 \!k_z^{'2}}{2m_0}\! + \! \frac{\hbar^2 (k_x^2 + k_y^2)}{2m_t}\!  \pm \frac{\Delta E_c(\mathbf{k})}{2},  \label{eq:eigen_energy} \\
&&\Delta E_c(\mathbf{k}) = 2\times \\ &&\sqrt{\!\left(\frac{\hbar^2k_0k_z'}{m_0}\!\right)^2\!\!\!\!-\!\left(\frac{\hbar^2k_x k_y}{m_{cv}}\!\right)^2\!\!\!\!+\! |\eta|^2(k_x^2+k_y^2) \!\!-\!\!\frac{4\Delta_C\Delta_{\scriptscriptstyle{X}}\alpha k_z'}{E_{g,\scriptscriptstyle{X}}}}, \nonumber \label{eq:delta_energy}
\end{eqnarray}
where $k_z'=k_z-2\pi/a$. The energy gap between the conduction bands is $\Delta E_c(\mathbf{k})$. Values of the parameters are provided in Table~\ref{tab:params}. The $\eta$ related term\cite{Pengke_PRL11} as well as the $\alpha k_z'$ related term in Eq.~(\ref{eq:eigen_energy}) are the leading spin-orbit effect on the energy. Along the $Z$-symmetry axis connecting the $X$ and $W$ points [$k_z'=k_y=0$ or $k_z'=k_x=0$; see Fig.~\ref{fig:BZ_phonon}(a)] as well as the part of $\Delta$ axis very close to the $X$ point, the splitting between the lower and upper conduction bands is induced by the spin-orbit coupling. This feature corresponds to the celebrated spin hot-spot at the edge of the Brillouin zone.\cite{Cheng_PRL10,Pengke_PRL11} Figures~\ref{fig:band_structure}(b) and (c) show the energy dispersion of the two conduction bands along the $Z$ axis.

\section{Spin alignment}\label{app:spin_align}

In this appendix, we present the routine for linearly combining doubly degenerate eigenvectors such that the resulting eigenvectors spins are aligned along a desired direction of $\mathbf{n}$ [Eq.~(\ref{eq:spin_up_down_states})]. This routine is for Hamiltonians that include spin-orbit coupling and when the crystals have a space inversion symmetry. Particularly, it is tailored for basis states which go back to themselves after sequential space inversion and time reversal operations. The chosen X-point basis as well as general plain-wave basis belong to this category. Otherwise the routine can be readily modified and made applicable.

In general, we express the double degenerate eigenvectors in basis $|\mathbf{X}\rangle \otimes|\uparrow(\downarrow)_{\mathbf{n}}\rangle$, where $|\mathbf{X}\rangle$ is the spin-independent part  (Secs.~\ref{sec:kp hamilton} and \ref{sec:valley-spin dependence}). $\bm{\sigma}\cdot \hat{\mathbf{n}}$ in Eq.~(\ref{eq:spin_up_down_states}) is written as [1,0;0,-1] in this basis. Suppose one of the double degenerate eigenvectors is $\mathbf{c}^a(\mathbf{k})$, and it has $2m$ elements, where $m$ is the number of spin independent basis states ($m=4$ in our $\mathbf{k}\cdot\mathbf{p}$ Hamiltonian). $c^a_{2i\mbox{-}1(2i)}(\mathbf{k})$ are coefficients of pure spin up (down) basis states. Then, from the general consideration of time reversal and space inversion symmetries of the Hamiltonian and the basis states, we know that components of the other eigenvector $\mathbf{c}^b(\mathbf{k})$ can be written as $c^b_{2i}=(c^a_{2i-1})^*$ and $c^b_{2i-1}=-(c^a_{2i})^*$. This property satisfies the first equality of Eq.~(\ref{eq:spin_up_down_states}).  To satisfy the other spin alignment definition in Eq.~(\ref{eq:spin_up_down_states}), we write a normalized linear combination
\begin{eqnarray}
\mathbf{c}^{\Uparrow}=(\mathbf{c}^a+w\mathbf{c}^b)/\sqrt{1+w^2} \label{eq:w}
\end{eqnarray}
such that,
\begin{eqnarray}
 (\mathbf{c}^{\Downarrow})^\dag \left[I_m\otimes \left(
 \begin{array}{cc}
 1 &0\\
 0 &-1
 \end{array}
 \right)\right]
  \mathbf{c}^{\Uparrow}=0,\nonumber
\end{eqnarray}
where $c^{\Downarrow}_{2i}=(c^{\Uparrow}_{2i-1})^*$, $c^{\Downarrow}_{2i-1}=-(c^{\Uparrow}_{2i})^*$, and $I_m$ is a $m$-dimensional identity matrix. We are led to
\begin{eqnarray}
-d-wb+w^2d^*=0,\nonumber
\end{eqnarray}
where
\begin{eqnarray}
b=\sum_{i=1}^m(|c^a_{2i-1}|^2-|c^a_{2i}|^2),\, d=\sum_{i=1}^m(c^a_{2i-1}c^a_{2i}).\nonumber
\end{eqnarray}
Thus, we get the combination parameter in Eq.~(\ref{eq:w})
\begin{eqnarray}
w=(b-\sqrt{b^2+ 4|d|^2})/(2d^*).
\end{eqnarray}
We further have $w\approx-d/b$ if $b\gg d$. In the case of EPM states, this general procedure requires to replace the $|\mathbf{X}\rangle$ basis ($m$=4) with a plane-wave basis (where typically $m$$>$100).

\section{Intravalley momentum scattering and deformation potential}\label{app:momentum}

In this appendix, we derive results of intravalley momentum scattering by selection rules with $\mathbf{k}\cdot\mathbf{p}$ Hamiltonian eigenstates. The procedure paves the way for analyzing more involved intravalley (Sec. \ref{sec:intravalley}) and $g$-process (Sec. \ref{sec:g-process}) {\it{spin-flip}} scattering in a similar approach.

The leading terms in momentum scattering matrix elements depend linearly on the phonon wavevector. We show it by a wavevector-order analysis.

\textbf{Zeroth-order}: At $\mathbf{k}_1=\mathbf{k}_2$, the in-phase atomic vibration does not depend on lattice sites. Thus, the phonon-induced interaction in Eq.~(\ref{eq:Hep1}) reduces to displacement of the entire crystal, $\sum_{j} \bm\nabla_{\mathbf{r}} \mathcal{V}_+(\mathbf{r}-\mathbf{R}_j)=  \bm\nabla \mathcal{V}_{\rm{crystal}}$. Based on the relation
\begin{eqnarray}
\bm{\nabla}\mathcal{V}_{\rm{crystal}}= i[\mathbf{p},H]/\hbar,\label{eq:commutator_2}
\end{eqnarray}
the coupling of the in-phase part between spin-degenerate eigenstates of $H$ vanishes,
\begin{eqnarray}
M_{\rm{i}}(\mathbf{k},\mathbf{s};\mathbf{k},\mathbf{s})= M_{\rm{i}}(\mathbf{k},\mathbf{s};\mathbf{k},\mbox{-}\mathbf{s}) =0.\label{eq:zeroth_in}
\end{eqnarray}
For the out-of-phase part, $\bm{\xi}^-(\mathbf{q})$ is linear with $\mathbf{q}$,
\begin{eqnarray}
M_{\rm{o}}(\mathbf{k},\mathbf{s};\mathbf{k},\mathbf{s})= M_{\rm{o}}(\mathbf{k},\mathbf{s};\mathbf{k},\mbox{-}\mathbf{s}) =0.\label{eq:zeroth_out}
\end{eqnarray}
Therefore, all the zero-order terms vanish.

\textbf{First-order}: We write the matrix elements of momentum scattering using the basis states of the $X$ point. Here, the effect of spin-orbit coupling can be safely neglected. As a result, in the expansion of $|\mathbf{k},\mathbf{s}\rangle$  [Eqs.~(\ref{eq:spin_up_state})-(\ref{eq:state_structure})] we omit $\mathbf{B}(\mathbf{k})$ and keep only the coefficients of $\mathbf{A}(\mathbf{k})$. These coefficients and the translational part are then linearized around $\mathbf{K}=(\mathbf{k}_1+\mathbf{k}_2)/2$ providing
\begin{eqnarray}
\!\!\! \mathbf{A}(\mathbf{k}_{1,2})&=&\mathbf{A}(\mathbf{k}_0)\mp \mathbf{q}/2\cdot\bm\nabla_k \mathbf{A}(\mathbf{k})\Big|_{\mathbf{k}_0}+...\,\nonumber \\
&=& \left[0,1,\pm\frac{Pq_x}{2E_{\scriptscriptstyle{g,X}}},\pm\frac{Pq_y}{2E_{\scriptscriptstyle{g,X}}}\right]+...\,, \label{eq:A_linear} \\ \!\!\! \exp(i\mathbf{k}'_{1,2}\!\cdot\mathbf{r})&=& \exp(i\mathbf{k}'_0\!\cdot\mathbf{r}) (1\mp i\mathbf{q}\cdot\mathbf{r}/2+...\,,
\end{eqnarray}
where $\mathbf{K}$ was replaced by the wavevector of the valley center ($\mathbf{k}_0$). The error brought into the matrix element with this replacement is of higher order (quadratic) in $\mathbf{q}$. We substitute these linearized forms into Eq.~(\ref{eq:spin_up_state}) and the resulting states are then plugged into the in-phase and out-of-phase part of Eq.~(\ref{eq:M}). This procedure identifies three linear terms with relatively large coefficients, and the matrix element up to leading order reads
\begin{eqnarray}
&&M(\mathbf{k}_1,\mathbf{s};\mathbf{k}_2,\mathbf{s})=\bcancel{M}_{\rm{m}}^{(0)}+{M}_{\rm{m}}^{(1)} = \sum_{j=1}^3 I_{{\rm{m}},j}  \nonumber \\
&&I_{\rm{m,1}} = \bm{\xi}^-(\mathbf{q})\cdot \langle X_1^1| \bm{\nabla} V_{-}|X_1^1 \rangle\, \nonumber \\
&&I_{\rm{m,2}} = \frac{P}{2E_{\scriptscriptstyle{g,X}}} \sum_{\ell=x,y} \left[ q_{\ell} \bm{\xi}^+(\mathbf{q})\cdot \langle X_4^{\ell}|  \bm{\nabla} V_+ |X_1^1 \rangle-\rm{c.c.}\right]\nonumber\\
&&I_{\rm{m,3}} = - \bm{\xi}^+(\mathbf{q})\cdot \langle X_1^1| (i\mathbf{q}\cdot\mathbf{r}) \bm{\nabla} V_+ |X_1^1 \rangle \;,\label{eq:momentum}
\end{eqnarray}
where c.c. denotes complex conjugate. These conceivably dominant terms are further examined to see if they are kept by selection rules.

The integral forms of intravalley momentum scattering are further restricted by group theory.  Both the basis states couplings and interactions are identified as parts belonging to IRs of $G^2_{32}$. Using Eqs.~(\ref{eq:XX})-(\ref{eq:potentialIR}) and the discussion that follows, we write the following decompositions,
\begin{subequations}
\begin{align}
\!\!\!\!\! \langle X_1^1 | ... | X_1^1\rangle & \rightarrow  X_1 \otimes X_1 =  M_1 \oplus M_3' \oplus \bcancel{M_4} \oplus \bcancel{M_2}'    \label{eq:x11x11}\\
\langle X_4^\ell | ... | X_1^1 \rangle & \rightarrow  X_4\otimes X_1= M_5\oplus M_5'\;, \label{eq:x11x4} \\
\bm{\nabla}V_+ \;\;\;\; & \rightarrow   (M'_3\oplus M_5)\otimes M_1 = M'_3\oplus M_5\;, \label{eq:vs_sel} \\
\bm{\nabla}V_- \,\;\;\; & \rightarrow   (M'_3\oplus M_5)\otimes M'_2 = M_4 \oplus M_5'\;. \label{eq:vas_sel}
\end{align}
\end{subequations}
In writing the first line, we have used Eq.~(\ref{eq:rep_cc}) to cross-out IRs in which only off-diagonal coupling between basis states is possible (i.e, between $X_1^{2'}$ and $X_1^1$). The first integral, $I_{\rm{m,1}}$ in Eq.~(\ref{eq:momentum}), couples $X_1^1$ states via a vector-type operation on the antisymmetrical potential part ($\bm{\nabla}V_{-}$). Since the respective decompositions in Eqs.~(\ref{eq:x11x11}) and (\ref{eq:vas_sel}) have no mutual IR, this type of coupling vanishes. We get
\begin{eqnarray}
I_{\rm{m,1}} = 0 \;. \label{eq:Im1}
\end{eqnarray}
The second integral, $I_{\rm{m,2}}$ in Eq.~(\ref{eq:momentum}), couples $X_1$ and $X_4$ states via a vector-type operation on the symmetrical potential part ($\bm{\nabla}V_+$). $M_5$ appears in the respective  decompositions [Eqs.~(\ref{eq:x11x4}) and (\ref{eq:vs_sel})]. Following the discussion that precedes Eq.~(\ref{eq:possible_couplings}) we can find the invariant integrand form. Then, by noting that $V_+$ can be replaced by the crystal potential (set $\mathbf{k}_{1,2}=(0,0,1)2\pi/a$ in Ref.~[\onlinecite{footnote_translational}]), the integral is analytically solved using Eq.~(\ref{eq:commutator_2}),
\begin{eqnarray}
I_{\rm{m,2}} &=&  \frac{im_0P^2}{\hbar^2} \left[ q_{x}\xi^+_x(\mathbf{q}) + q_{y}\xi^+_y(\mathbf{q}) \right]\;. \label{eq:Im2}
\end{eqnarray}
The third integral, $I_{{\rm{m}},3}$ in Eq.~(\ref{eq:momentum}), couples $X_1^1$ states via a second-rank tensor [$\mathbf{r}\!\otimes\!\bm{\nabla}V_+(\mathbf{r})$]. To write a decomposition expression, it should first be casted into a symmetrized form,
\begin{eqnarray}
\mathbf{r}\!\otimes\!\bm{\nabla}V_+ \!=\!  \sum_{\alpha} (\mathbf{r}-\bm{\tau}_{\alpha})\!\otimes\!\bm{\nabla}V_{\rm{at}}(\mathbf{r}-\bm{\tau}_{\alpha}) \!+\!\bm{\tau}\!\otimes\!\bm{\nabla}V_{-}\,.\quad \label{eq:Im4_casting}
\end{eqnarray}
As shown before, $\bm{\nabla}V_{-}(\mathbf{r})$ cannot couple between $X_1^1$ states. The symmetrized sum term on the right-hand side transforms as
\begin{eqnarray}
\!\!\! (M_3' \!\! &\oplus& \!\! M_5)^{\otimes 2} = (M_5\otimes M_5)\oplus (M_3'\otimes M_3') \oplus 2(M_3'\otimes M_5)   \nonumber \\
&=& (M_1 \oplus M_2 \oplus M_3 \oplus M_4) \oplus (M_1) \oplus 2(M_5').  \label{eq:Im4_decomp}
\end{eqnarray}
Combined with Eq.~(\ref{eq:x11x11}), we see that the two $M_1$ in Eq.~(\ref{eq:Im4_decomp}) contribute to $I_{{\rm{m}},3}$ in Eq.~(\ref{eq:momentum}). There should be two independent parameters associated with the two $M_1$'s. Out of the nine tensor components of $\sum_{\alpha}(r_i-\tau_{\alpha,i})\partial V_{\rm{at}}(\mathbf{r}-\bm{\tau}_\alpha)/\partial r_j$, one independent parameter originates from the product of longitudinal components ($i=j=z$) and belongs to $M'_3\otimes M'_3=M_1$, and the other from the sum of transverse component products ($i=j=x$ plus $i=j=y$) and belongs to the $M_1$ out of $M_5\otimes M_5$. Putting all of the pieces together, the overall matrix element of intravalley momentum scattering reads
\begin{eqnarray}
\!\! && \!\! M(\mathbf{k}_1,\mathbf{s};\mathbf{k}_2,\mathbf{s}) = \label{eq:mom_matrix_element} \\ && \,\,\, \frac{i m_0P^2}{\hbar^2} \left[ q_{x}\xi^+_x(\mathbf{q}) + q_{y}\xi^+_y(\mathbf{q}) \right] \nonumber \\ && -i\sum^{x,y,z}_{j} q_j \xi^+_j \sum^{A,B}_{\alpha} \left\langle X^1_1\right|   (r_j\!-\!\tau_{\alpha\!j})\frac{\partial V_{\rm{at}}(\mathbf{r}\!-\!\bm{\tau}_{\alpha})}{\partial r_j}\left| X^1_1\right\rangle \;.\nonumber
\end{eqnarray}

Deformation potential theory provides a concise appearance for the matrix element of intravalley momentum scattering,\cite{Herring_PR56,Bir_Pikus_Book,Rieger_PRB93}
\begin{eqnarray}
\sum^{x,y,z}_{j}\langle\,\mathbf{k}_0\left|\mathcal{D}_{jj}\right|\mathbf{k}_0\, \rangle \,\epsilon_{jj}(\mathbf{q}),\label{eq:deformation_theory}
\end{eqnarray}
where deformation potential operators and strain tensor ($\bar{\bm\epsilon}$) elements at the long-wavelength regime are
\begin{eqnarray}
\mathcal{D}_{jk}&=&-\frac{p_jp_k}{m_0} + \lim_{\bar{\bm{\epsilon}}\rightarrow 0} \frac{\partial V_{\epsilon}[(1+\bar{\bm{\epsilon}})\cdot \mathbf{r}]}{\partial \epsilon_{jk}}, \label{eq:Djk}\\
\epsilon_{jk}(\mathbf{q})&=&i (q_j \xi^+_k(\mathbf{q}) + q_k \xi^+_j(\mathbf{q}))/2. \label{eq:strain_jk}
\end{eqnarray}
$V_{\epsilon}$ is the crystal potential under strain $\bar{\bm{\epsilon}}$. Here $\epsilon_{jk}(\mathbf{q})$ in Eq.~(\ref{eq:strain_jk}) is stripped out of the amplitude factor $\sqrt{\hbar/[2\rho V \omega(\mathbf{q})]}\sqrt{n(\mathbf{q})+1/2 \pm 1/2} $ in order to compare it with the expression of Eq.~(\ref{eq:M}).  To relate with Eq.~(\ref{eq:mom_matrix_element}) we substitute Eq.~(\ref{eq:Djk}) and (\ref{eq:strain_jk}) into Eq.~(\ref{eq:deformation_theory}) and use $\psi_{\mathbf{k}_0}(\mathbf{r}) \simeq e^{i k'_0 z}\psi_{X^1_1}(\mathbf{r})$. The second term in $\mathcal{D}_{jj}$ leads to exactly the last term of Eq.~(\ref{eq:mom_matrix_element}) with an opposite sign. For the kinetic term in $\mathcal{D}_{jj}$, we write
\begin{eqnarray}
\langle X^1_1| p^2_j | X^1_1\rangle \!=\! \sum_n \langle X^1_1| p_j  | X_n \rangle \langle X_n| p_j | X^1_1\rangle \!\simeq \!\frac{m_0^2P^2}{\hbar^2}, \quad \label{eq:projection}
\end{eqnarray}
where $j=\{x,y\}$ considering the dominant coupling of $p_{x,y}$ between $X^1_1$ and valence $X_4$ states.\cite{footnote_projection} $p^2_z$ does not have this dominant coupling. Therefore, the first term in $\mathcal{D}_{jj}$ is exactly the second line of Eq.~(\ref{eq:mom_matrix_element}) with a minus sign. Physical results do not change upon this global minus sign. Of the diagonal deformation potential constants $D_{jj}=\langle\,\mathbf{k}_0\left|\mathcal{D}_{jj}\right|\mathbf{k}_0\rangle$, there are two independent values $D_{zz}$ and $D_{xx}=D_{yy}$ (in $z$ valley). Dilation and uniaxial deformation potentials are related to them via $\Xi_d=D_{xx}$ and $\Xi_u=D_{zz}-D_{xx}$.

All in all, we have shown the equivalency of our procedure with known deformation potential theory for intravalley momentum scattering.
\section{some details in intravalley $\&$ $g$-process spin flip matrix elements}\label{app:details_intra_g}
\subsection{intravalley}\label{sec:details_intra}

We invoke group theory and evaluate the integrals in  Eqs.~(\ref{eq:Msf2_out2})-(\ref{eq:Msf2_in3}). The out-of-phase matrix element,  $\sum_{\mu,\nu}I_{\mu,\nu;2}$,  includes a single dominant Elliott process. Its coefficient product, $i\eta/\Delta_C$, comes from $A_{X^1_1}$ and $\partial B_{X^{2'}_1}/\partial k_{x(y)}$. Its integral reads $\langle X_1^1 | \bm\nabla V_{-} | X_1^{2'}\rangle$.  Using  appropriate decompositions
\begin{subequations} \label{eq:Msf2_selection}
\begin{align}
\!\!\!\!\!\langle X_1^1 | ... | X_1^{2'}\rangle & \rightarrow  X_1 \otimes X_1 =  \bcancel{M_1} \oplus \bcancel{M_3}' \oplus M_4 \oplus M_2'    \label{eq:Msf2_decom_out2A}\\
\bm{\nabla}V_- \,\;\;\; & \rightarrow   (M'_3\oplus M_5)\otimes M'_2 = M_4 \oplus M_5'\;, \label{eq:Msf2_decom_out2B}
\end{align}
\end{subequations}
we see that these states can be coupled by the longitudinal component of $\bm{\nabla}V_-$ (transforms as $M_4$). We have crossed out IRs in which only diagonal coupling is possible (e.g, between $X_1^1$ and $X_1^1$). All in all, one combination is kept
\begin{eqnarray}
\sum_{\mu,\nu} I_{\mu,\nu;2} = \frac{i\eta}{\Delta_C}\left\langle X^{2'}_1\left|\frac{\partial V_{-}}{\partial z} \right|X^1_1\right\rangle (iq_x+q_y)\xi^-_z(\mathbf{q}).\label{eq:spin-out term2}
\end{eqnarray}

The remaining non-vanishing matrix elements relate to the in-phase potential. The sum $\sum_{\mu,\nu}I_{\mu,\nu;3}$ includes a single dominant Elliott process. Its coefficient product, $-iP\eta/E_{\scriptscriptstyle{g,X}}\Delta_C$, comes from $\partial B_{X^{2'}_1}/\partial k_{x,y}$ and $\partial A_{X^{x(y)}_4}/\partial k_{x(y)}$. Its integral reads $\langle X_1^{2'} | \bm\nabla V_{+} | X_4^{x,y}\rangle$. Using  appropriate decompositions
\begin{subequations}
\begin{align}
\langle X_1^{2'}  | ... | X_4^{\ell} \rangle & \rightarrow  X_4\otimes X_1= M_5\oplus M_5'\;,  \label{eq:Msf2_decom_in3A} \\
\bm{\nabla}V_+ \;\;\;\; & \rightarrow   (M'_3\oplus M_5)\otimes M_1 = M'_3\oplus M_5\;,  \label{eq:Msf2_decom_in3B}
\end{align}
\end{subequations}
we see that these states can be coupled by the transverse components of $\bm{\nabla}V_+$ (transform as $M_5$). Similar to the derivation of the second momentum integral [Eq.~(\ref{eq:Im2})], we reach at an analytical result for the third spin integral,
\begin{eqnarray}
\sum_{\mu,\nu} I_{\mu,\nu;3} =\frac{i \eta P^2 m_0 }{2 \Delta_C \hbar^2}(q_x-iq_y)\!\left[q_x \xi^+_y(\mathbf{q}) \!+\! q_y \xi^+_x(\mathbf{q})\right]\!.\,\,\, \label{eq:spin-in term3}
\end{eqnarray}

The sum $\sum_{\mu,\nu}I_{\mu,\nu;4}$ also includes a single dominant Elliott process. Its coefficient product, $-i\eta/\Delta_C$, comes from $A_{X^1_1}$ and  $\partial B_{X^{2'}_1}/\partial k_{x,y}$. Its integral reads $\langle X_1^{2'} | \mathbf{r}\!\otimes\! \bm\nabla V_{+} | X_1^1\rangle$. The interaction is first casted into Eq.~(\ref{eq:Im4_casting}) where both parts (tensor and antisymmetric potential) can couple between the states. Each of the respective decompositions in Eqs.~(\ref{eq:Im4_decomp}) and (\ref{eq:Msf2_decom_out2B}) share a common $M_4$ IR with Eq.~(\ref{eq:Msf2_decom_out2A}). Using the transformation properties of $M_4$, the resulting integral reads
\begin{eqnarray}
&&\!\!\!\!\!\!\!\!\!\!\!\!\sum_{\mu,\nu} I_{\mu,\nu;4} =\frac{i\eta(q_x-iq_y)}{ \Delta_C} \left\langle X^{2'}_1\left|  \frac{\partial V_-}{\partial z} \mathbf{q}\cdot\bm\tau \xi^+_z(\mathbf{q})\right.\right.\label{eq:spin-in term5} \\
&&\!\!\!\!\!\!\!\!\!\!\!\!\left.\left.+  \sum^{A,B}_\alpha (y-\tau_{\alpha,y}) \frac{\partial V_{\!\rm{at}} (\mathbf{r}-\bm\tau_\alpha)}{\partial x}  \left[q_x \xi^+_y(\mathbf{q}) + q_y \xi^+_x(\mathbf{q})\right]\right|X^1_1\right\rangle . \nonumber
\end{eqnarray}

The sum $\sum_{\mu,\nu}I_{\mu,\nu;5}$ includes dominant Elliott processes coupled by $\bm\nabla V_{+}$. One of the dominant products comes from $A_{X^1_1}$ and $2\partial^2 B_{X^{2'}_1}/\partial k_z\partial k_{x(y)}$. The corresponding integral $\langle X_1^{2'} | \bm\nabla V_{+} | X_1^1\rangle$ vanishes, for that there are no common IRs between Eqs.~(\ref{eq:Msf2_selection}) and (\ref{eq:Msf2_decom_in3B}). Other dominant products come from $A_{X^1_1}$ and $\partial^2 B_{X^{x(y)}_4} /\partial k^2_{y(x)}$, and from $B_{X^{x(y)}_4}$ and $2\partial A_{X^{2'}_1}/\partial k_x \partial k_y$. Repeating the analysis that led to Eqs.~(\ref{eq:Im2}) and (\ref{eq:spin-in term3}), we reach at
\begin{eqnarray}
\sum_{\mu,\nu} I^E_{\mu,\nu;5}&=&\frac{-iP^2 m_0 }{2\Delta_C \hbar^2}\left\{ \eta \left[iq^2_y \xi^+_x(\mathbf{q}) - q^2_x \xi^+_y(\mathbf{q})\right]\right. \nonumber\\
&&\left.- (2\eta'-\eta) q_x q_y \left[ \xi^+_x(\mathbf{q}) -i \xi^+_y(\mathbf{q})\right]\right\}\,.\quad\label{eq:spin-in term4}
\end{eqnarray}

The only Yafet process is included in $I_{\mu,\nu;5}$ with coefficient product $4P^2/(E_{\scriptscriptstyle{g,X}}E_C)$ that comes from $A_{X^1_1}$ and $2 \partial^2 A_{X^{2'}_1}/\partial k_x \partial k_y$. Its integral reads $\langle X_1^{2'} | \bm\nabla V^{\rm{so}}_{+,\mathbf{k}_0} | X_1^1\rangle$. The $k$-independent part of $V^{\rm{so}}_{+,\mathbf{k}_0}$ does not contribute to this matrix element: $\langle X_1^{2'} |\bm\nabla (\bm\nabla V_+\times\mathbf{p})| X_1^1\rangle$ vanishes by time reversal symmetry [Eqs.~(\ref{eq:vectorTR}) and (\ref{eq:XX_TR})]. The relatively small interaction $\bm\nabla (\bm\nabla V_+\times\hbar\mathbf{k}_0)$ can couple $X_1^1$ with $X_1^{2'}$ states. The interaction transforms as second-rank tensor and belongs to $(M'_3\oplus M_5)^{\otimes2}$. From Eqs.~(\ref{eq:Im4_decomp}) and (\ref{eq:Msf2_decom_out2A}), we see that $M_4$ is the common IR. To be specific, it is the component $\partial^2 V_+/\partial x \partial y$ that belongs to $M_4$. In spite of the relatively small magnitude, we still give its expression explicitly,
\begin{eqnarray}
\sum_{\mu,\nu} I^Y_{\mu,\nu;5}=&&\frac{P^2 q_x q_y (\xi^+_x-i \xi^+_y)}{E_{\scriptscriptstyle{g,X}} \Delta_C}\nonumber\\
&& \times\frac{\hbar^2 k'_0}{4m_0^2c^2}\left\langle X^{2'}_1\left| \frac{\partial^2 V_+}{\partial x \partial y} \right|X^1_1\right\rangle, \label{eq:spin-in termY}
\end{eqnarray}
and shall find it a compensating part in leading to a concise result.

Given these dominant contributions for spin relaxation, we may attempt to relate them to some known physical quantities. First of all, we find out that the terms in Eq.~(\ref{eq:spin-in term3}) and Eq.~(\ref{eq:spin-in term4}) (excluding small $\eta'-\eta$ part) can be combined and they share a common factor $i\eta(q_x-iq_y)(q_x\xi^+_y+q_y\xi^+_x)/\Delta_C$ with the second term in Eq.~(\ref{eq:spin-in term5}). Secondly, the left-out $\eta'-\eta$ part in Eq.~(\ref{eq:spin-in term4}) is found to be exactly compensated by the Yafet part [Eq.~(\ref{eq:spin-in termY})]. It is shown with the help of \cite{footnote_projection2}
\begin{eqnarray}
\frac{\hbar^2}{4m_0^2c^2}\left\langle X^{2'}_1\left| \frac{\partial^2 V_+}{\partial x \partial y} \right|X^1_1\right\rangle\simeq\frac{2P m_0 \alpha}{\hbar^2}.
\end{eqnarray}
At this phase, the total of Eqs.~(\ref{eq:spin-in term3})-(\ref{eq:spin-in termY}) closely resemble to a deformation potential constant $D'_{xy}=\langle X^{2'}_1|\mathcal{D}_{xy}|X^1_1\rangle$ defined in Eq.~(\ref{eq:Djk}), up to a constant prefactor. However, care should be used due to its off-diagonal nature. Contrary to diagonal deformation potential that appeared in the momentum scattering (e.g., dilation and uniaxial), $\partial V_{\epsilon}[(1+\bar{\bm{\epsilon}})\cdot \mathbf{r}]/\partial \epsilon_{xy}$ contains a part induced by internal displacement.\cite{Bir_Pikus_Book} Thus, the sum of Eqs.~(\ref{eq:spin-in term3})-(\ref{eq:spin-in termY}) alone is not sufficient to form a complete deformation potential constant. Out-of-phase phonon polarization vector $\bm{\xi}^-$, on the other hand, can be expressed in terms of $\bm{\xi}^+$ and internal displacement for small $\mathbf{q}$ (see Table~\ref{tab:elastic_continuum} and its discussion). Replacing $\bm{\xi}^-$ in Eq.~(\ref{eq:spin-out term2}) with equivalent $\bm{\xi}^+$ terms (from the fourth row in Table~\ref{tab:elastic_continuum}), we obtain the final intravalley spin-flip matrix element expression in Eq.~(\ref{eq:intra_AC}) with Eq.~(\ref{eq:D_xy}) of the main text.

Finally, we mention an alternative choice to derive these results with a basis states of the $\Delta$ axis at $\mathbf{k}_0$. In this case, $\it{third}$-order perturbation theory expanded around  $\mathbf{k}=\mathbf{k}_0$ gives a similar result (with the help of wavevector-order analysis and the appropriate adjustments to the space group of $\Delta$-axis). From our results, we are able to conclude that the leading contributing term with this alternative approach is
\begin{eqnarray}
&&\frac{\hbar}{m_0 E_{g,\mathbf{k}_0} \Delta_C}\sum_{i,j} \langle \Delta_1\!\!\downarrow\!\!|\mathbf{q}\!\cdot\!\mathbf{p} |\Delta^i_5\!\downarrow\rangle \langle\Delta^i_5\!\downarrow \!\!|V_{\rm{so}} |\Delta_{2'}\!\!\uparrow \rangle \! \nonumber\\
&&\quad\times\Big\langle\Delta_{2'}\!\! \uparrow\!\!\Big| \frac{\hbar}{m_0 E_{g,\mathbf{k}_0}}\left[ \mathbf{q}\!\cdot\!\mathbf{p} |\Delta^j_5\uparrow\rangle \langle \Delta^j_5\uparrow\!\!| \bm{\xi}^+(\mathbf{q})\! \cdot\!\!\! \bm\nabla V_0 \right]\nonumber\\
&&\qquad\qquad\qquad\qquad\qquad\qquad\qquad+ \epsilon_{xy}(\mathbf{q}) V_{xy}  \Big|\Delta_1\!\!\uparrow\!\Big \rangle ,\label{eq:intra_Delta}\nonumber
\end{eqnarray}
where $i, j=\{x,y\}$.  $\mathbf{p}$, $|\Delta \rangle$ and $E_{g,\mathbf{k}_0}$ denote, respectively, the momentum operator, the spin-independent state and the energy gap between conduction and valence bands at $\mathbf{k}_0$. $\epsilon_{xy}(\mathbf{q}) V_{xy}$, with  $V_{jk}$ defined by the last term in Eq.~(\ref{eq:Djk}),  is a partial combination of in-phase and out-of-part interactions.

\subsection{$g$-process}\label{sec:details_g}
We analyze the terms $I^E$ and $I^Y$ in Eq.~(\ref{eq:M_g}) and derive their expressions in detail. Following the reasoning that led to observation 2 of the intravalley case, the Elliott part ($I^{E}$) has a dominant coefficient $i\eta/\Delta_C$ from $\partial B_{X^{2'}_1}/\partial k_{x,y}$ and $A_{X^1_1}$. However, the basis states associated with this coefficients are $\langle X_1^1|$ and $|X_1^1\rangle$, as can be inferred by the index arrangement of $\mu'$ [below Eq.~(\ref{eq:M_g})]. The Yafet part ($I^{Y}$) has a dominant coefficient $-P/E_{\scriptscriptstyle{g,X}}$ from $\partial A_{X^{x(y)}_4}/\partial k_{x(y)}$ and $A_{X^1_1}$ between $\langle X_4^{y(x)}|$ and $|X_1^1\rangle$. In spite of this coefficient disobeying observation 2, the Yafet part is kept and will be shown not too small compared with the Elliott part.

Having identified the important coupling between basis states, we can determine which of the phonon modes dominate the $g$-process spin relaxation. It is determined by applying selection rules connecting opposite points of the $\Delta$ star. The Elliott part is analyzed first. From Eq.~(\ref{eq:M_g}), a first Brillouin zone phonon wavevector $\mathbf{q}=-2\mathbf{k}'_0 \approx (0,0,0.3)2\pi/a$ is needed to conserve the crystal momentum. With $\psi^*_{X_1^1}=\psi_{X_1^{2'}}$, the character of the state product has the following identity, $\chi^{k_0}_{\Delta'_2}\chi^{k_0}_{\Delta_1} =\chi^{2k_0}_{\Delta'_2} =(\chi^{-2k_0}_{\Delta'_2})^* =(\chi^{-2k'_0}_{\Delta_1})^*$. The last equality is obvious if one refers to the character table of the $\Delta$ group (e.g., Table V of Ref.~[\onlinecite{Lax_PR61}]). Thus, the Elliott part involves a phonon with a character $\chi^{-2k'_0}_{\Delta_1}$. IR of $\Delta_1$ with a first Brillouin zone wavevector is identified with the LA mode. Similarly, the coupling of the Yafet part is between $\psi_{\Delta_1,k_0}$ and $\psi_{\Delta_5,k_0}$, and it leads to $\chi^{-2k'_0}_{\Delta_5}$ for the interaction. The spin-orbit potential transforms as $\Delta_5$, and $\chi_{\Delta_1(\Delta'_2)}\chi_{\Delta_5} =\chi_{\Delta_5}$. Thus, the Yafet part involves a phonon mode that transforms as $\Delta_1$ or $\Delta'_2$, which corresponds to LA or LO mode, respectively. The LO phonon has much larger energy than that of the LA phonon and we drop it from the leading Yafet contribution. In conclusion, the LA mode dominates both the Elliott and Yafet coupling.

The leading-order matrix element of the $g$-process [Eq.~(\ref{eq:M_g})] can be related to some form of a deformation potential parameter. It can be done since the coupling can be expressed between basis states of the same $X$ point. The Elliott part is explicitly written as
\begin{eqnarray}
I^{E}&=&\frac{-2\eta}{\Delta_C}  \left\langle X^1_1\left| e^{2 i k'_0 z}
\left( \xi^+_{LA,z}\frac{\partial V_+}{\partial z}+ \xi^-_{LA,z}\frac{\partial V_-}{\partial z} \right)\right|X^1_1 \right\rangle\nonumber \\
&& \times( K_x- i K_y)\,,
\end{eqnarray}
where we have used the approximation that $\bm\xi^{\pm}_{\rm{LA}}$ has only a nonzero longitudinal component (it is exact when the wavevector is on the $\Delta$ axis).
To find the nonvanishing part of this integral, the integrand is converted into parts belonging to IRs of the $X$-point space group. The new feature that emerges in a $g$-process is the phase factor $e^{2i k'_0z}=1+ 2ik'_0 z -2k'^2_0 z^2 +\mathcal{O}((k'_0 z)^3)$. For the in-phase part, $\partial V_+/\partial z$ and $z \partial V_+/\partial z$ belong to $M'_3$ and $M_1\oplus M_4$. The higher-order term is
\begin{eqnarray}
&&\!\!\!\!\!\!\!\!z^2\frac{\partial V_+}{\partial z}=\sum\limits_{\alpha}(z-\tau_{\alpha,z})^2 \frac{\partial V_{\!\rm{at}} (\mathbf{r}-\bm{\tau}_\alpha)}{\partial z} +\tau^2_z \frac{\partial V_+}{\partial z} \nonumber\\
&&\!\!\!\!\!\!+ 2\tau_z \left[ (z-\tau_{A,z}) \frac{\partial V_{\!\rm{at}} (\mathbf{r}-\bm{\tau}_A)}{\partial z} - (z-\tau_{B,z})\frac{\partial V_{\!\rm{at}} (\mathbf{r}-\bm{\tau}_B)}{\partial z}\right]\nonumber,
\end{eqnarray}
which belongs to
\begin{eqnarray}
M'^{\otimes2}_3\otimes M'_3\oplus M'_3 \oplus M'_3\otimes M'_3\otimes M'_2\,.\nonumber
\end{eqnarray}
As shown by Eq.~(\ref{eq:x11x11}), operators that couple $X_1^1$ states belong either to $M_1$ or $M_3$. Operators that are all even under time reversal [Eq.~(\ref{eq:vectorTR})] are further restricted to $M_1$ for even-parity states [Eqs.~(\ref{eq:rep_cc}) and (\ref{eq:XX_TR})]. Thus, the allowed coupling interaction is $\sum_{\alpha} (z-\tau_{\alpha,z}) \partial V_{\!\rm{at}} (\mathbf{r}-\bm{\tau}_\alpha)/\partial z$. Similarly, we find that there is no comparable out-of-phase contribution. The resulting Elliott part of the matrix element reads
\begin{eqnarray}
I^{E} \approx \frac{(-i\eta) 4 k'_0}{\Delta_C} D_{zz} (K_x-i K_y),\label{eq:g_Ell}
\end{eqnarray}
where we have used $\xi^{+}_{{\rm{LA}},z}(-2\mathbf{k}'_0)\approx 1$ and $D_{zz}$ in Eq.~(\ref{eq:Djk}).

For Yafet part, the unity leading term in the expansion of $e^{2i k'_0 z}$ results in a nonvanishing integral. It is the reason that we have kept this part in spite of disobeying observation 2. The symmetry properties of $\partial(\bm\nabla V_+\times \mathbf{p})_{x(y)}/\partial z$ follow $M'_3\otimes M'_5=M_5$ which also appears in the decomposition of the states product [$X^{y(x)}_4 \otimes X^1_1$; see Eq.~(\ref{eq:X1X4})]. Together with the leading coefficient of the state expansion, $-P/E_{\scriptscriptstyle{g,X}}$, the matrix element of the Yafet part reads
\begin{eqnarray}
I^Y=\frac{2P}{E_{\scriptscriptstyle{g,X}}}D_{\rm{so}}(K_x -i K_y),\label{eq:g_Yaf}
\end{eqnarray}
where the scattering constant integral is denoted by
\begin{eqnarray}
D_{\rm{so}}&=&\frac{\hbar}{4 c^2 m_0^2}\left \langle X^{x}_4\left | \frac{\partial (\bm\nabla V_+ \times \mathbf{p})_{y}} {\partial z}\right|X^1_1 \right\rangle \nonumber\\
&\approx& 6.7\,\rm{meV}\cdot2\pi/a\label{eq:D_so}
\end{eqnarray}
with its value calculated from EPM. The out-of-phase part leads to a slightly smaller coupling integral ($\approx$$\,$4~$\rm{meV}\cdot 2\pi/a$), but with a small polarization $\xi^-_{{\rm{LA}},z}(-2k'_0)\approx0.2$. This property renders its contribution too small compared with the leading Elliott part. It is therefore neglected.

All together, the leading $g$-process matrix element is Eq.~(\ref{eq:Mg}) with Eq.~(\ref{eq:Dgs}) in the main text.


\begin{thebibliography}{99}
\bibitem{footnote_translational} After variable substitution $\mathbf{r}-\mathbf{R}_j\rightarrow \mathbf{r}$, invoking the periodicity in $u_k(\mathbf{r})$ and summing over $N$ primitive cells, we have $ \sum_j\int u^{*}_{k_2}(\mathbf{r})\exp(-i\mathbf{k}_2\cdot\mathbf{r}) \exp(i\mathbf{q}\cdot\mathbf{R}_j) \bm{\nabla}\mathcal{V}_{\pm}(\mathbf{r}-\mathbf{R}_j) \exp(i\mathbf{k}_1\cdot\mathbf{r})u_{k_1}(\mathbf{r}) d\mathbf{r}
    = N \int u^{*}_{k_2}(\mathbf{r})\exp(-i\mathbf{k}_2\cdot\mathbf{r}) \bm{\nabla}\mathcal{V}_{\pm}(\mathbf{r})
    \exp(i\mathbf{k}_1\cdot\mathbf{r}) u_{k_1}(\mathbf{r})d\mathbf{r} \delta_{\mathbf{q}+\mathbf{k}_1-\mathbf{k}_2,\mathbf{g}}$, where $\mathbf{g}$ is a reciprocal lattice vector.
\bibitem{footnote_rot_spin} The transformation matrix $D_{1/2}$ for a proper rotation operation is the same as the corresponding rotation matrix for spin, and is unit for the inversion operation since the later keeps the spin angular momentum invariant.
\bibitem{footnote_rot_Delta6} This transformation matrix effectively treats the two basis states of $\Delta_6$ as pure spins. Strictly speaking, it is an approximation that involves a negligible error of the order of $\Delta_{so}/E_g\sim0.1\%$ for the state at the valley center. In intravalley spin flips (Sec.~\ref{sec:valley-spin dependence}), on the other hand, it will induce inaccurate quadratic-in-$\mathbf{q}$ terms which participate in the leading matrix element terms and we will use a strict and more tedious procedure.
\bibitem{footnote_Xpoint_choice} With $X$-point ($\Delta$-axis) basis states the scattering can be viewed as second-order degenerate (third-order non-degenerate) perturbation, seen later in the paper. Expansion with $X$-point basis states naturally incorporates the evolution away from the $X$ point. It elucidats the nature of the spin hot-spot \cite{Cheng_PRL10,Pengke_PRL11} and renders a smoother transition to the case of shear strained silicon.
\bibitem{footnote_realIR}  We omitted the complex conjugation in these  decompositions since the character table is real (e.g., $X_1^*\otimes X_1$ $\rightarrow$ $X_1\otimes X_1$).
\bibitem{footnote_check_state}  Spin-independent EPM states at the $X$ point are symmetrized into $X_1^{2'}$, $X_1^{1}$, and $X_4^x$ and $X_4^y$ states. Then the coefficients in $\mathbf{A}(\mathbf{k})$ and $\mathbf{B}(\mathbf{k})$ are validated, term by term, via projection of spin-dependent EPM states at the vicinity of the valley bottom onto these $X$-point states.
\bibitem{footnote_trans_mode}  The transverse mode results are a sum of two modes and do not depend on the specific choice of orthogonal polarizations.
\bibitem{footnote_rotation}  The rotation in Eq.~(\ref{eq:rot_matrix2}) can be regarded as a `passive' rotation, as opposed to the `active' rotation that we have utilized in the $f$-process case [Eq.~(\ref{eq:k_n})]. In the latter, the rotated two-dimentional $\Delta_6$ state points at the $\mathbf{n}$ direction. On the other hand, after applying Eq.~(\ref{eq:rot_matrix2}) on the components of Eq.~(\ref{eq:state_structure}), the state still points along the $\mathbf{z}$ direction but it is expressed with the basis of $\{|\mathbf{X}\rangle \otimes |\uparrow(\downarrow)_\mathbf{n} \rangle\}$.
\bibitem{footnote_Ge} The cubic dependence of intravalley spin-flips in germanium is due to the high-symmetry at its valley center ($L$ point) and due to the relatively large and comparable energy-gap of the valley center from nearby bands.
\bibitem{footnote_computing_time} Eigenenergies, electron states and phonon polarization vectors are pre-prepared in $1/48$ of the irreducable Brillouin zone wedge. Their computation consumes a relatively small amount of time. Similarly, transformation of electron states and phonon polarization vectors under symmetry operations as well as the search algorithm to satisfy energy conservation consume a relatively negligible computation time.
\bibitem{footnote_analytical_integration} The analytical integration difficulty mainly lies in the $|M_{\rm{sf}}|^2$'s denominator $q^2_z m_l/m_d+ q^2_{\perp} m_t/m_d$ after a spheroid-spherical coordinate transformation,\cite{Herring_PR56} where $m_d=(m_l m^2_t)^{1/3}$. We overcome it by first integrating over symmetrized azimuthal angles $\phi_{\pm}=(\phi_2\pm\phi_1)/2$ and then over symmetrized polar angles $\theta_{\pm}=(\theta_2\pm\theta_1)/2$. In this way, integrations of different variables are decoupled.
\bibitem{footnote_projection} $\langle X_n| p_j | X^1_1\rangle=  \langle X_n| [p_j, H_0]| X^1_1\rangle /(E_{X^1_1}-E_{n} ) = N \hbar\langle X^{x,y}_4| \partial V_+/\partial r_j | X^1_1\rangle /(E_{X^1_1}-E_{n} )$, and we keep only the upper valence band $X_4$ for $X_n$ since $E_{\scriptscriptstyle{g,X}}$ is much smaller than other energy gaps ($|E_{X^1_1}-E_{n}|$). When taking further bands explicitly into consideration, the kinetic term in Eq.~(\ref{eq:projection}) remains in agreement with $I_{{\rm{m}},2}$ in Eq.~(\ref{eq:Im2}) if our state basis includes these further bands.
\bibitem{footnote_projection2} In deriving this equation we performed integration by parts and related $\partial^2 V_+(\mathbf{r})/\partial x \partial y$ with $[\partial V_+/\partial x,p_y]/i\hbar$. Then we have inserted a complete sum of projection operators $|X_i\rangle\langle X_i|$ between $\partial V_+/\partial x$ and $p_y$. Finally, we keep only $|X_4\rangle\langle X_4|$ of the valence band following energy proximity as in Eq.~(\ref{eq:projection}).
\end{thebibliography}
\end{document}